\documentstyle[amssymb,aps,preprint,epsfig]{revtex}
\newcommand{\vev}[1]{\left\langle #1\right\rangle}

\newcommand{\goes}{\rightarrow} 
\newcommand{\MeV}{\; \mathrm{MeV}} 
\newcommand{\GeV}{\; \mathrm{GeV}} 
\newcommand{\TeV}{\; \mathrm{TeV}}

\newcommand{\lapproxeq}{\lower .7ex\hbox{$\;\stackrel{\textstyle  
<}{\sim}\;$}} 
\newcommand{\gapproxeq}{\lower .7ex\hbox{$\;\stackrel{\textstyle  
>}{\sim}\;$}} 
\newcommand{\stackdown}[2]{\lower 1.4ex\hbox{$\;\stackrel{\textstyle{#1}}  
{\scriptstyle{#2}}\;$}}
\newcommand{\beq}{\begin{equation}} 
\newcommand{\eeq}{\end{equation}} 
\newcommand{\bea}{\begin{eqnarray}} 
\newcommand{\eea}{\end{eqnarray}}
\newcommand{\lsp}{\tilde{\chi}}
\newcommand{\bfactor}{{\bar{\beta}}_f}
\newcommand{\ttchi}{\tilde{\chi}}

\newcommand{\ori}{\hspace{.1in}}
\newcommand{\fp}{f^{\prime}}

\newcommand{\relic}{\Omega_{\lsp}\,h_0^2}
\newcommand{\xsec}{\vev{ \sigma v_{rel} }} 
\newcommand{\etal}{\textit{et. al.}}
\newcommand{\reaction}{\lsp\,\lsp \goes X\,Y}  
\newcommand{\mlsp}{m_{\lsp}} 

\makeatletter 
\def\slash{\@ifnextchar[{\fmsl@sh}{\fmsl@sh[0mu]}} 
\def\fmsl@sh[#1]#2{% 
  \mathchoice 
    {\@fmsl@sh\displaystyle{#1}{#2}}% 
    {\@fmsl@sh\textstyle{#1}{#2}}% 
    {\@fmsl@sh\scriptstyle{#1}{#2}}% 
    {\@fmsl@sh\scriptscriptstyle{#1}{#2}}} 
\def\@fmsl@sh#1#2#3{\m@th\ooalign{$\hfil#1\mkern#2/\hfil$\crcr$#1#3$}} 
\makeatother 
\begin{document}
\draft
\tightenlines
\preprint{\parbox{7cm}{ ACT-8/99, CTP-TAMU-32/99,\\ 
                         UA/NPPS-04-99, hep-ph/9909497 } }

\title{Neutralino Relic Density in a Universe with a non-vanishing
 Cosmological Constant}

\author{{\bf A.~B.~Lahanas} $^{1}$, \, 
{\bf D.~V.~Nanopoulos} $^{2}$  \, and \, {\bf V.~C.~Spanos} $^{1}$ }

\maketitle
\begin{center}
$^{1}$ {\it University of Athens, Physics Department,  
Nuclear and Particle Physics Section,\\  
GR--15771  Athens, Greece}\\ 
 
$^{2}$ {\it Department of Physics,  
         Texas A \& M University, College Station,  
         TX~77843-4242, USA, 
         Astroparticle Physics Group, Houston 
         Advanced Research Center (HARC), Mitchell Campus, 
         Woodlands, TX 77381, USA, and \\ 
         Academy of Athens,  
         Chair of Theoretical Physics,  
         Division of Natural Sciences, 28~Panepistimiou Avenue,  
         Athens 10679, Greece }  \\ 
\end{center}

%%%%%%%%%%%%%%%%
\begin{abstract}
We discuss the relic density of the lightest of the supersymmetric particles
in view of new cosmological data, which favour the concept of an accelerating
Universe with a non-vanishing cosmological constant. 
Recent astrophysical observations provide us with very precise values of the
relevant cosmological parameters. Certain of these parameters have direct
implications on particle physics, e.g., the value of matter density, which in
conjunction with electroweak precision data put severe constraints on the
supersymmetry breaking scale. In the context of the 
Constrained Minimal Supersymmetric
Standard Model ({\small CMSSM}) such limits read  as:
$M_{1/2} \simeq 300 \GeV - 340 \GeV$,
$m_0 \simeq 80 \GeV - 130 \GeV$.
Within the context of the {\small CMSSM}
a way to avoid these constraints is either to go to the large
$\tan \beta$ and $\mu > 0$ region, or make ${\tilde \tau}_R$, the next to
lightest supersymmetric particle 
({\small LSP}), be almost degenerate in mass with {\small LSP}.
\end{abstract}


%%%%%%%%%%%%%%%%%%%% S E C T I O N --> I  %%%%%%%%%%%%%%%%%%%%%%%%%%%%%%%%%%%
%%%%%%%%%%%%%%%%%%%%%%%%%%%%%%%%%%%%%%%%%%%%%%%%%%%%%%%%%%%%%%%%%%%%%%%%%%%%%
\baselineskip=20pt
\section{Introduction}
%%%%%%%%%%%%%%%%%%%%%%%%%%%%%%%%%%%%%%%%%%%%

During the last few years the 
knowledge of the cosmological parameters has started entering 
an era of high precision with far reaching consequences
not only for cosmology,
but for particle physics as well. The cosmic microwave background 
temperature is accurately known, $T_0 = 2.7277 \pm 0.002\;^0\mathrm{K}$,
the Hubble parameter is determined with a relatively small
error, $H_0 = 65 \pm 5 \;
\mathrm{Km /sec /Mpc}$,
the baryonic mass density is precisely determined by big-bang
nucleosynthesis, $\Omega_{\mathrm{B}}  h_0^2 = 0.019 \pm 0.001$, while the
determination of the age of
the Universe from the oldest stars, as well as other sources, yield 
$t_U  = 14 \pm 1.5\; \mathrm{Gyr}$ \cite{Turner}.
Very recent observations of type Ia
supernovae ({\small SNI}a), 
as well as measurements of the anisotropy of the Cosmic
Background Radiation ({\small CBR}) 
provide additional information favouring  
an almost {\em flat} and {\em accelerating} Universe,
where the acceleration mainly is driven by a {\em non-vanishing
cosmological constant}  \cite{Lineweaver,Perlmutter,Riess}.

There is a growing consensus that the anisotropy of
the {\small CBR} offers the best
way to determine the curvature of the Universe and hence the total
matter-energy density $\Omega_0$ \cite{Turner}.
The data are consistent
with a flat Universe, ${\Omega}_0=1.0 \pm 0.2$, 
while we are confident that 
the radiation component of the matter-energy density,
that is the contribution from {\small CBR} and/or
ultra relativistic neutrinos, is very small
\cite{Turner,Lineweaver}. 
Therefore the present matter-energy density
can be decomposed principally into matter density
$\Omega_{\mathrm{M}}$ and vacuum energy $\Omega_{\Lambda}$:
\beq
\Omega_0=\Omega_{\mathrm{M}}+\Omega_{\Lambda}\,. \label{omega0}
\eeq

There is also supporting evidence, coming 
from many independent astrophysical observations, 
that the matter density weighs $\Omega_{\mathrm{M}}=0.4 \pm 0.1$
(see for instance Ref.~\cite{Turner} and references therein).
The $\Omega_{\mathrm{M}},\; \Omega_{\Lambda}$
values are then restricted by the age
of the Universe and by the value of the Hubble parameter through 
%%%%%
\beq
t_{U}={\frac{1}{H_{0}}}\;{\int_{0}^{1}}\;dy\;
\sqrt{y \over {{\Omega}_{\mathrm{M}} (1-y) 
+ {\Omega}_{\Lambda} (y^3-y) + y} } \,.
\label{age}
\eeq
%%%%%%%%%%%
The constraints stemming from Eq.~(\ref{age}) are however less restrictive
than those coming from the supernovae {\small SNI}a data.
Recently two groups, the Supernova Cosmology Project \cite{Perlmutter}
and the High-$z$ Supernova Search Team \cite{Riess},
using different methods of analysis, each found evidence
for accelerated expansion, driven by a vacuum energy contribution:
\beq
\Omega_{\Lambda}={4\over{3}}\Omega_{\mathrm{M}}+{1\over{3}}\pm {1\over{6}}\,.
\eeq
So, for $\Omega_{\mathrm{M}}=0.4 \pm 0.1$ this relation implies that
the vacuum energy is non-vanishing,
$\Omega_{\Lambda}=0.85\pm 0.2$, 
a value which is compatible with
a flat Universe, as the anisotropy of 
{\small CBR} measurements indicate.
Taking into account the fact that 
the baryonic contribution to the matter density
is small, ${\Omega}_{\mathrm{B}}=0.05 \pm 0.005 $,
the values for matter energy density $\Omega_{\mathrm{M}}$ 
result to a Cold Dark Matter ({\small CDM}) density 
${\Omega}_{\mathrm{CDM}} \simeq 0.35 \pm 0.1$, which combined with 
more recent measurements \cite{Turner,Freedman}
of the scaled Hubble parameter
$h_0=0.65 \pm 0.05$,  result to small
{\small CDM} relic densities:  
\beq 
\Omega_{\mathrm{CDM}} \, {h_0}^2 \simeq 0.15 \pm 0.07\,.
\label{bound0}
\eeq 

From measurements of the ratio of the baryonic to total mass
in rich clusters, smaller values for the mass density are
obtained. This ratio is found to be 
$\Omega_{\mathrm B}/ \Omega_{\mathrm M} \thickapprox 0.15$ \cite{white,fabian}
which entails to even tighter limits  
$\Omega_{\mathrm{CDM}} \, {h_0}^2 = 0.12\pm 0.04$ \cite{ostriker}.

Such stringent bounds for the {\small CDM} relic density
affect supersymmetric predictions and may lower the limits of
the effective supersymmetry breaking scale, and hence
the masses of the supersymmetric particles.
In Ref.~\cite{Lopez0} within the framework 
of the string inspired no-scale
$SU(5)\times U(1)$ supergravity model, 
by relaxing the cosmological constant, regions of the parameter
space compatible with 
$\Omega_{\mathrm{CDM}} \, {h_0}^2 \thickapprox 0.2$ were delineated, and
phenomenological predictions for the sparticle spectrum were given.
The relevance of the high precision cosmology to constrained supersymmetry was
addressed
 in Refs.~\cite{gondolo,nath}.
More recently the {\small CDM} relic 
abundance with non-vanishing cosmological
constant, in the framework of the Minimal Supersymmetric Standard Model
({\small MSSM}), was shown to put limits on supersymmetric mass spectrum 
\cite{Wells}.
In fact it was shown that gauginos can be within LHC reach, 
if the recent cosmological data are used.
As stated in Ref.~\cite{LNS1} it is worth pointing out 
that while electroweak ({\small EW}) precision data are in perfect
agreement with Standard Model ({\small SM}) predictions, and
hence in agreement with
supersymmetric models which are characterized by a large
supersymmetry breaking scale 
${M}_{\mathrm{SUSY}}$ \cite{Lahanas}, the data on
${\Omega}_{\mathrm{CDM}} \, {h_0}^2$
push ${M}_{\mathrm{SUSY}}$
to the opposite direction preferring  small values of
${M}_{\mathrm{SUSY}}$.
Therefore {\small EW} precision data may be hard to reconcile with
the assumption that the lightest supersymmetric
particle ({\small LSP} or $\lsp$), is a candidate for
{\small CDM} \cite{LNS1}. 

The method to calculate the relic
abundance of a Dark Matter ({\small DM}) 
candidate particle in the Universe is outlined
in Ref.~\cite{weinb}. In 
$R$-parity conserving supersymmetric theories the {\small LSP} may be a 
neutralino, which is a good candidate to play the 
role of {\small DM} \cite{first}.
Many authors 
\cite{old1,Sred,griest2,old2,old3,%
Mizuta,Lopez2,Lopez3,arno,Lopez,Drees,%
recent,reports,Drees2,gondolo,nath,Wells,Falk1,Falk2,kawamura,khalil}
have since calculated the relic neutralino density.  
In the early works, only the most important neutralino annihilation
channels were considered, but later works \cite{Drees,recent} included all
annihilation channels.
Also more refined calculations of thermal averages of cross sections 
were employed, which took into account threshold effects and integration over
Breit-Wigner poles \cite{Griest,Gelmini}.

Our study in this paper is based on the {\small CMSSM}, which is
motivated by Supergravity, assuming
universal boundary conditions for the soft supersymmetry breaking parameters,
and in which the {\small EW} symmetry is radiatively
broken \cite{report}. Our strategy of 
calculating the neutralino relic density follows three steps: 
First the {\small SUSY} particle spectrum and
the relevant couplings are generated,
according to the supersymmetric scenario mentioned above.
Then the thermally averaged cross sections $\vev{\sigma v}$
are calculated in their non-relativistic limit, using analytic expressions. 
Finally we numerically solve the Boltzmann equation, which governs the
evolution
of the neutralino relic density, by using very accurate routines able to
handle stiff problems of differential equations.
Regarding the calculation of the relic density, we solve the Boltzmann
equation
numerically by finding a proper boundary condition along the lines described
in Ref.~\cite{Lopez}. This is reminiscent of the 
{\small WKB} approximation; it yields
very accurate results and differs from the standard approaches used in most
works. 
We want to emphasize that
for the sake of the effectiveness of our computational
code we have chosen to use analytic results in order to calculate
the amplitudes of the processes contributing to
thermally averaged cross section $\vev{\sigma v}$ \cite{Drees}.
The price one pays, is that these analytic
results break down in the vicinity of the poles or thresholds 
of the cross section.
However the comparison of our results with those of other
studies \cite{Lopez3,recent}, 
which treat the problem of poles and thresholds
in a more accurate manner by calculating numerically the thermally averaged
cross section \cite{Griest,Gelmini}, shows that they
are in striking agreement. This occurs, at least, in regions of the parameter
space of the {\small CMSSM} where this comparison is feasible.

The effect of the coannihilation between the {\small LSP} and the
next-to-lightest supersymmetric particle ({\small NLSP})
is quite important and should be duly taken into account
%%%\cite{binetruy,Drees2,gondolo,Mizuta,Griest}.
\cite{binetruy,Griest,Mizuta,Drees2,gondolo}.
The importance of coannihilation of the lightest of the neutralinos
$\lsp$, which in most of the parameter space of the {\small CMSSM}
is a bino, with $\tilde{\tau}_R$ has been pointed out in 
Refs.~\cite{Falk1,Lazarides}. $\lsp-\tilde{\tau}_R$ coannihilation
are of relevance
for values of the parameters near the edge where $\lsp$ and
${\tilde \tau}_R$ are almost degenerate in mass. 
In such regions of the parameter space
the results reached using the ordinary methods, 
in which these effects are neglected, 
have to be properly modified to correctly account
for the effect of the coannihilation.

As a preview of our results: \\
We have found that within the context of the {\small CMSSM} the recent cosmological
data, in combination with {\small EW} precision measurements, lead to rather
tight
limits for the relevant supersymmetric breaking parameters $m_0$, $M_{1/2}$,
provided the next to the {\small LSP} particle (${\tilde \tau}_R$) is not nearly
degenerate in mass with the {\small LSP}. In this regime the only option to avoid
these limits is to move to the large $\tan \beta$ region, where acceptable
relic densities can be obtained if the pseudoscalar Higgs mass is
approaching twice the mass of the {\small LSP}. This case is consistent with
$b \goes s \gamma$ and may be of relevance for
models in which Yukawa coupling unification is enforced.

In regions of the parameter space in which ${\tilde \tau}_R$'s mass is close
to that of the {\small LSP}, 
where coannihilation processes need be taken into account
for the calculation of the actual neutralino relic abundance, 
such limits can be evaded.

This paper is organized as follows: \\
In the first section we give the basic formalism and discuss various details
of our calculations. In section II and III we discuss the methodology
we follow in solving the Boltzmann equation and give details of our numerical
computation. In section IV our results for the {\small LSP} relic density are
presented and regions of the parameter space consistent with the new
astrophysical data are delineated. Towards the end of this section
a discussion is devoted to the coannihilation effects.
Finally we end up with the conclusions.
To facilitate the reader 
the supersymmetric conventions used throughout this paper are presented in
the Appendix.

%%%%%%%%%%%%%%%%%%%% S E C T I O N --> II  %%%%%%%%%%%%%%%%%%%%%%%%%%%%%%
%%%%%%%%%%%%%%%%%%%%%%%%%%%%%%%%%%%%%%%%%%%%%%%%%%%%%%%%%%%%%%%%%%%%%%%%%

\section{Supersymmetric relic density}

Our aim is to calculate the cosmological relic density of the lightest of the
supersymmetric particles, which will be denoted by $\lsp$ throughout this
paper. This we assume is one of the four neutralinos
states. In supersymmetric models with $R$-parity conservation this particle
is stable. The cosmological constraints on  
${\Omega}_{\mathrm{CDM}}$  discussed
previously may impose stringent constraints on its mass, as well as on the
masses
of other supersymmetric particles which are exchanged in graphs, 
contributing to pair annihilation reactions 
%%%%%%%%%%%%%%%%%%%%%%%%%%%%%%%%%%%%%%%%%
\bea
   \lsp \; \lsp \goes X \; Y \,, \nonumber 
\eea
%%%%%%%%%%%%%%%%%%%%%%%%%%%%%%%%%%%%%%%%%
constraining the predictions of supersymmetry. 

The basic ingredient in calculating the {\small LSP} relic abundance is the
calculation of the thermally averaged cross sections
$ \xsec$ for the annihilation processes
$\reaction $, 
which enter into the Boltzmann transport equation whose solution yields the
mass density of the $\lsp$ particles at present 
epoch\footnote{We neglect at this stage slepton--$\lsp$ coannihilations
and slepton-slepton annihilations.}.
$v_{rel}$ denotes the relative velocity of the two annihilating $\lsp$'s. 
Although these issues have been covered
in numerous articles we will briefly repeat them from this stand too,
in order to pave the ground for the discussion in the remainder of this 
paper.

Our principal objective is to calculate the present {\small LSP} mass density
%%%%%%%%%%
\beq
\rho_{\lsp} = m_{\lsp}\, n(T_0)\,,
\eeq
%%%%%%%%%%%%%%%%%
where $T_0 \approx 2.7 \;^0\mathrm{K} $ is today's Universe temperature.
This determines the {\small LSP} energy density
$\Omega_{\lsp} = \frac{\rho_{\lsp}}{\rho_{crit}}$, where ${\rho_{crit}}$
is the critical density of Universe.
$\rho_{\lsp}$ is calculated by solving the Boltzmann equation given by
%%%%%%%%%%
\beq
\frac{dq}{dx} = \lambda (x) \, (q^2 - q_0^2)\,, \label{bol}
\eeq
%%%%%%%%%%%%%%%%%
where $x = T/m_{\lsp}$ and
%%%%%%%%%%%%%%%
\beq
q \equiv \frac{n}{T^3 h(T)} \;\; , \; q_0 \equiv \frac{n_0}{T^3 h(T)}\; .
\label{qs}
\eeq
%%%%%%%%%%%%%%%%%

In the equation above $n$ denotes the number density of $\lsp$'s
and $ n_0$ their density in thermal equilibrium. The latter is given by
%%%%%%%%%%%%%%%%%%%
\beq
n_0 = {\frac{k_{spin}}{2{\pi}^2}} \; {\frac{T^3}{x^3}} \;
    {\int_{1}^{\infty}}\; du \;
    {\frac{u \sqrt{(u^2-1)}}{e^{u/x}+1}}\,,
\eeq
%%%%%%%%%%%%%%%%
whose low temperature expansion (low $ x = T/m_{\lsp}$) is
%%%%%%%%%%%%%%%%%%%
\beq
n_0 =  \frac{k_{spin} \, e^{-1/x} \, T^3} { (2 \pi x)^{3/2} } \;
  (1 + \frac{15}{8} x + {\cal{O}} (x^2) ) \; .
\eeq
%%%%%%%%%%%%%%%%
In the equations above $k_{spin}$ is the number of the spin degrees of
freedom. The function $h(T)$  
counts the effective entropy degrees of freedom, determining 
the entropy density of the Universe 
%%%%%%%%%%
\beq
s = {\frac{2 \pi^2}{45}} \; T^3 \; h(T)\,,
\eeq
%%%%%%%%%%%%%%%%
which along with the effective energy degrees of freedom $ g(T)$,
which determines the energy density
%%%%%%%%%%%%%%%%%%%
\beq
\rho  =  {\frac{\pi^2}{30}} \; T^4 \; g(T) \,,
\eeq
%%%%%%%%%%%%%%
enter into the prefactor $\lambda (x) $ appearing on the right hand side of
Eq.~(\ref{bol})
%%%%%%%%%%%%%%%%%
\beq
\lambda (x) \equiv \left( {{\frac{4 \pi^3}{45}} G_N} \right)^{-1/2} \;
 {\frac{m_{\lsp}}{\sqrt{g(T)}}} \;
\left( h(T)+ {\frac{m_{\lsp}}{3}}{{h}^{\prime}}(T) \right) \; \xsec \; .\label{lala}
\eeq
%%%%%%%%%%%

Depending on the temperature $T$ the content of the particles in equilibrium
is different. In our analyses we use the expressions for $g(T)$, $h(T)$ as
given in Ref.~\cite{Lopez}. In the region $40 \MeV < T < 2.5 \GeV $, where
the quark--hadron phase transition takes place,
the values used for $g(T)$, $h(T)$ are those corresponding to 
a critical temperature $T_c = 150 \MeV $ as given in 
Ref.~\cite{Sred}. For a critical temperature 
$T_c = 400\MeV$, also quoted in Ref.~\cite{Sred}, we did not observe
a substantial change in our final results concerning the {\small LSP} relic density.
Recent lattice {\small QCD} results indicate 
that a first order phase transition takes place during the hadronization
\cite{Iwasaki}.
Using the corresponding data for the energy
and entropy densities \cite{Schwarz},
no significant change is observed in our final results, as it has been also 
noticed in Ref.~\cite{Sred}.

We postpone for later the details of the numerical scheme employed to
solving the Boltzmann equation (\ref{bol}) and pass to discuss the thermal
averages
$\xsec$ for the various processes involved. At this point we follow 
Ref.~\cite{Drees} and express the non-relativistic cross sections for the
annihilation processes $\reaction$ in terms of helicity
amplitudes as follows
%%%%%%%%%%%%%%%%%%%%%%%
\bea
v \; {\sigma (\reaction)} = 
{\frac{1}{4}}\;
{\frac{{\bar{\beta}}_f}{8 \pi s}}\; {\frac{1}{S_f}}\;   \hspace{5.9cm}
\nonumber   \\
\times {\sum_h}
\left( |{A^h}(^1 S_0)|^2 \; +\;
{\frac{1}{3}} ( |{A^h}(^3 P_0)|^2 + |{A^h}(^3 P_1)|^2 + |{A^h}(^3 P_2)|^2 )
  \right)\;,   \label{heli}
\eea
%%%%%%%%%%%%%%%%%%%%%
where $v$ is the relative velocity $v_{rel}$. 
In Eq.~(\ref{heli}) the amplitudes  ${A^h}(^{2\;S+1} L_J)$ depends on
the helicities of the final products denoted  collectively by $``h"$
and the total cross section is obtained as an incoherent sum over the
final helicity states. The cross section will be 
expanded up to ${\cal{O}} (v^2)$ terms
and for this reason only $S$ and $P$ waves in the initial state are of
relevance. The statistical factor $S_f$ appearing in the denominator in
Eq.~(\ref{heli}) equals to 2! when the final particles are identical. The
kinematical factor ${\bar{\beta}}_f$ is given by
%%%%%%%%%%%%%%%%
\beq
{{\bar{\beta}}_f} =
{ \left( 1- \frac{2 (m^2_X + m^2_Y)}{s} +
\frac{2 (m^2_X - m^2_Y)}{s^2}
\right)}^{1/2}   \; ,
\eeq
%%%%%%%%%%%%%%%%%%%%%%%%
where $s$ is the center of mass ({\small CM}) energy squared. 

Although our analysis in many respects resembles that pursued in
Ref.~\cite{Lopez} it differs in the particular method employed to calculate
the thermal
averaged cross sections, where we follow closely Ref.~\cite{Drees}.
The results of the two approaches ought to be identical if it were not for
the fact that some interference terms between graphs in
processes involving Higgs particles in the final state or one Higgs and
a $Z$-boson, were omitted. In our approach these terms are implicitly
included in Eq.~(\ref{heli}).

Since 
the r.h.s. of Eq.~(\ref{heli}) will be expanded up to terms ${\cal{O}}(v^2)$,
we need cast the helicity amplitudes into the following forms:
%%%%%%%%%%%%%
\bea
{A^h}&&(^1 S_0) = a^h_0  +  a^h_1 v^2 + ...    \\ 
 {A^h}&&(^3 P_{0,1,2}) = b^h (P_{0,1,2}) v + ...   \label{pwav}
\eea
%%%%%%%%%%%%%
The ellipses in the equations above include higher in $v$ terms.

Besides this 
the kinematical factor ${{\bar{\beta}}_f}$ has to be expanded, 
and also the {\small CM}
energy squared variable $s$ should be expressed in terms of the relative
velocity $v$ as given below
%%%%%%%%%%%%%%
\bea
{\bar{\beta}}_f = \beta_0 + \beta_1  v^2  + {\cal{O}}(v^4)\;\,,\;
s^{-1} = (1-\frac{v^2}{4}) /{4 {m_{\lsp}^2}  }  \,.
\eea
%%%%%%%%%
By using these, 
the cross section of Eq.~(\ref{heli}) can be brought into the form
%%%%%%%%
\beq
v \; { \sigma } = a + \frac{b}{6}  v^2   \label{xsec}
\eeq
%%%%%%%%%%%%%%%
with the constants $a , b \;$ defined by the following expressions
%%%%%%%%%%%%%%%%%
\bea
a &=&  k \; \sum_h  \; \beta_0 \; 
|  a_0^h |^2  \label{ab1}  \\ 
b &=& 6 \;k \; \sum_h \; \biggl( \;| a_0^h |^2 \;
(\beta_1-\frac{\beta_0}{4})\; + \;\beta_0 
\;( {a_0^h}^{*}  a_1^h +  h.c.)      \nonumber    \\ 
&+&\frac{\beta_0}{3} \; 
( \;| b^h (P_{0})|^2 + |b^h(P_{1})|^2+|b^h(P_{2})|^2\;)\; \biggr) \,. 
\label{ab2} 
\eea 
%%%%%%%%%
The prefactor $k$ appearing in the equations above is given by
%%%%%%%%%%
\bea
k^{-1} \;=\;128 \; \pi \; S_f \;{m_{\lsp}^2}\; .  \nonumber
\eea
%%%%%%%%%%%%%%%%%

It is well known that the expansion in the relative velocity $v$ breaks down
near thresholds or poles. Concerning the kinematical factor
$\bfactor$ we write
%%%%%%%%%%
\bea
\bfactor\;=\;\delta\; \sqrt{\epsilon }
\; \left(1 + \frac{v^2}{8\;\epsilon} +
{\cal{O}}\;(v^4)\right)  \; ,  \label{kine}
\eea
%%%%%%%%%%%%
where 
%%%%%%%%%%%%%%%%%
\bea
\delta \;=\; {\frac {\sqrt{4\;m_X\;m_Y\;}} {m_X\;+\;m_Y} } \;\,,\;
\epsilon \;=\; 1 \;-\; 
{\frac  {{(m_X\;+\;m_Y)}^2}  {4\;{{m_{\lsp}}^2}}} \; .  \label{epsi}
\eea
%%%%%%%%%%%%%%%
This expansion obviously breaks down when $\epsilon$ gets small, or
equivalently when we are near the threshold
%%%%%%%%%%%%%%%
\bea
2  {m_{\lsp}} = {m_X} + {m_Y} \,.    \label{thre}
\eea
%%%%%%%%%%%%%%%%%%

Also singular are the expansions (\ref{ab1}) and (\ref{ab2})  when we are
near an $s$-channel pole
of a particle of mass $m_I$ into which $\lsp \lsp$ are fused to. The
intermediate particle's propagator in this case is expanded as
%%%%%%%%%%%%%%%
\bea
\frac{1}{s-m_I^2+i\;m_I\; \Gamma_I}\;=\;
{\frac{1}{{m_{\lsp}}^2}} \; {\frac{1}{4-R_I^2+i\;G_I}} \;
\left(1\;-\; {\frac{v^2}{4-R_I^2+i\;G_I}} \right)   \; , \label{pole}
\eea
%%%%%%%%
where
%%%%%%%%%
\bea
R_I \;=\; \frac{m_I}{m_{\lsp}} \;\,,\;
G_I \;=\; \frac{m_I \; \Gamma_I}{{m_{\lsp}^2} } \; .
\eea
%%%%%%%

The expansion  (\ref{pole}) holds as long as we are away from poles, otherwise
the coefficient of the relative velocity squared gets large. The largeness
of this factor is dictated by the narrowness of the resonance and the
heaviness of the {\small LSP}.
For the
$Z$-boson resonance for instance, the corresponding 
rescaled width $G_Z$ is
$G_Z  \approx (230 / {m_{\lsp}^2})  {\GeV}^2 $, which for
$ {m_{\lsp}} \approx 100 \GeV  $  yields 
$G_Z  \approx  2.3 \times 10^{-2} $ invalidating the expansion
(\ref{pole}) on the resonance.

Therefore near poles
%%%%%%%%%%%%%%
\bea
m_{\lsp}  \;=\;  \frac{M_I}{2} \; ,
\eea
%%%%%%%%%%%%%%%
as well as near threshold, more  sophisticated methods should be used,
as those found in Refs.~\cite{Griest,Gelmini}, for the non-relativistic
expansion
of the cross section  in Eq.~(\ref{xsec}). 
We shall come back to this point later
when discussing the {\small LSP} relic density.

To make contact with the findings of Ref.~\cite{Sred}  we write the cross
section as
%%%%%%%%%
\bea
v  \sigma = {\frac{1}{{E_1}{E_2}}} \; w(s)\,,   \label{osw}
\eea
%%%%%%%%%
where  ${E_1}, {E_2}\;$  are the energies of the initial particles and $s$
the total {\small CM} energy squared. Eq.~(\ref{osw}) leads, up to
${\mathcal O}(x)$, to a thermal averaged
cross section (for details see  Ref.~\cite{Sred}) given by
%%%%%%%%%%%%%%%%%%%
\bea
\vev{v \sigma}  =   \frac{1}{{m_{\lsp}^2}} \; \left[ w_0 + \frac{3}{2}
( - 2 \; w_0 + w^{\prime}_0 ) \; x \right]   \,,  \label{expa}
\eea
%%%%%%%%
where $ x = T / {m_{\lsp}}$ and
%%%%%%%%%%%%%%%%%%%
\bea
w_0 \;=\; w(s_0) \; \; , \;   w^{\prime}_0 \;=\; 4 \; {{m_{\lsp}}^2} \;
%%(\frac{dw}{d{s_0}}) \quad  ( s_0 = 4 \; {{m_{\lsp}}^2} ) .
{\left(\frac{dw}{d{s}} \right)}_{s_0} \quad  ( s_0 = 4 \; {{m_{\lsp}}^2} )\; .
\eea
%%%%%%%%
By comparing Eqs. (\ref{heli}) and  (\ref{osw}) we can have
%%%%%%%%%%%%
\bea
a  =  {\frac{1}{{m_{\lsp}}^2}} \; w_0  \;\; , \; 
b  = {\frac{3}{2 {m_{\lsp}}^2}} \; ( w^{\prime}_0 - w_0 )\,,
\eea
%%%%%%%%%%%%%
which can be used to cast Eq.~(\ref{expa}) into the form
%%%%%%%%%
\bea
\vev{ v \sigma } \;=\; a + (b - \frac{3}{2} a ) \; x  \,.  \label{xab}
\eea
%%%%%%%%%%%%%%%

%%%%%%%%%%%%%%%%%%%% S E C T I O N --> III  %%%%%%%%%%%%%%%%%%%%%%%%%%%%%%%%%
%%%%%%%%%%%%%%%%%%%%%%%%%%%%%%%%%%%%%%%%%%%%%%%%%%%%%%%%%%%%%%%%%%%%%%%%%%%%%

\section{Solving the Boltzmann Equation}

The coefficients $a$ and $b$, appearing in Eq.~(\ref{xab}), are calculated for
each process
%%%%%%%
\beq
\lsp \; \lsp \goes X \; Y   \label{2body}
\eeq
%%%%%%%%%%%
where $\lsp$ is the lightest supersymmetric particle which we assume is one of
the four neutralinos as said in previous sections. At low temperatures
the particles in the final state may include ordinary fermions, gauge bosons
or Higgses.

The freeze out temperature $T_f$ usually 
occurs for values of $ x_f \equiv  \frac{T_f}{m_{\lsp}} \simeq 0.05$
and hence we can solve the Boltzmann equation  (\ref{bol}) in the regime
$x  \leq x_0$ by knowing the value of $q(x)$ at a properly chosen
point $x_0 \geq x_f$ 
which is not much beyond $x_f\;$  \footnote{ The choice 
of $x_0$ is related to the
particular method employed for solving the Boltzmann equation to be
discussed later in this chapter. The resulting values of $x_0$ turn out to
be around~{$\approx 0.1$}.}.
For temperatures $ T $ corresponding to
$ {x \leq {x_0}} $
contributions of sparticles other than the {\small LSP} to $ g(T)$, $h(T)$ are
negligible, relative to {\small LSP},
and can be safely ignored. The reason is that any sparticle's mass $m_{i}$
is larger than $m_{\lsp}$ and hence the relative Boltzmann factors 
$\exp \; [ - \frac{m_i - m_{\lsp}}{m_{\lsp} \;x} ] $ are suppressed in the
region
$x<x_0 \approx 0.1$. 
Hence only the contribution of the {\small LSP} is kept in the
effective energy and entropy degrees of freedom functions  $ g(T) $ and
$ h(T) $  respectively\footnote{ Obviously in regions where the
coannihilation effects are important this approximation does not hold
and the contributions of sparticles with masses close to mass of the $\lsp$
should be added to $g(T)$, $h(T)$.}.

Also, as stated previously, in the annihilation
process in Eq.~(\ref{2body}) only non-supersymmetric
particles are considered in the
final state. Although this is obviously correct at zero relative velocity of
the initial particles (at threshold), since
$\lsp$ is the {\small LSP}, it may not be the case at finite temperatures when
$\lsp$ are adequately thermalized acquiring kinetic energies sufficient
to produce heavier sparticles.
Therefore channels which are
forbidden at zero relative velocity may be activated at temperatures $T$.
%%This regards not only channels involving sparticles in the final state
%%but also other channels.
 In this work we will follow the standard treatment
and ignore contributions of all channels which are forbidden at zero relative
velocity. This is justified by the following argument.
The values of $x$ relevant
for our calculation are $x \leq 0.1 $ and  as a consequence the corresponding
temperatures are much smaller than $m_{\lsp}$. Therefore the initial state
particles $\lsp$ are not adequately thermalized to activate a forbidden
reaction. We can appeal to a more quantitative argument by recalling that
in the forbidden region the thermally averaged cross sections
are proportional to
%%%%%%%%%%%%%%
\bea
e^{- {{   {\mu_{-}^2}     } /x} } \; ,      \label{expon}
\eea
%%%%%%%%%%
see Ref.~\cite{Griest}, where ${\mu_{-}}$ depends on the masses of the final
products $X$, $Y$. When for instance these have equal masses, say $m_2$, and
$m_2 > m_{\lsp}$ this is given by
${\mu_{-}^2} = 1 - {\frac{m_{\lsp}^2}{m_2^2}}$. Therefore in the region
$x < 0.1$ the exponent in Eq.~(\ref{expon}) drops rapidly, 
unless $m_{\lsp}$ is close
to $m_2$. This is what is intuitively expected; at low temperatures
($ T \ll m_{\lsp}$) the thermal energies of the  $\lsp$'s in the initial state
are not sufficient to activate reactions in which the final products have 
masses well above their production threshold. Only when their masses are
very close to threshold even a small amount energy is adequate to furnish
enough kinetic energy to the initial particles to activate the reaction.
On these grounds we therefore ignore the contributions of forbidden channels.
This approximation is not expected to invalidate significantly
our results.

With this in mind the channels which contribute are 
(see also Refs.~\cite{Lopez,Drees,recent}),
%%%%%%%%%%%%%%%%%%%%%
\bea
q{\bar q}, \; l{\bar l}, \; 
W^{+}W^{-},\; ZZ,\; ZH,\; Zh,\; ZA,\; W^{\pm}H^{\mp},\;
HH,\; hh,\; Hh,\;AA,\; HA, \; hA,\; H^{+}H^{-} \; .   \nonumber
\eea
%%%%%%%%%%%%%%
$q$, $l$ denote quarks and leptons, $H$, $h$, $A$ denote the heavy, light and
pseudoscalar Higgses
respectively, while $H^{\pm}$ are the charged Higgses. The helicity amplitudes
for the above processes have been calculated in Ref.~\cite{Drees}
as we have already discussed.
Adjusting the results of 
that reference to conform with with our notation\footnote{Our notation
differs slightly from that used in Ref.~\cite{Drees}
(see Appendix).} we can calculate $\xsec$.

Our numerical procedure then goes as follows: 
\newcounter{bib} 
\begin{list} 
{(\roman{bib})}{\usecounter{bib}} 
\item Given the experimental inputs for {\small SM} 
fermion and gauge boson masses as
well as couplings and supersymmetry breaking parameters, 
we first run two-loop Renormalization Group  
Equations ({\small RGE}'s) in order to 
define physical masses and couplings of all particles
involved having as reference scale the physical $Z$-boson mass $M_Z$. 
\item We then calculate the coefficients $a$ and $b$ 
encountered in Eq.~(\ref{xab})
for each of the processes mentioned before.  
\item We solve the Boltzmann equation to define the relic density at today's
Universe temperature $T \simeq 2.7 \;^0\mathrm{K}$ . 
\end{list} 

Regarding point (i) we take as inputs the soft 
{\small SUSY} breaking parameters
namely squark, slepton, Higgs soft masses, trilinear scalar
couplings, gaugino masses  as well as the parameters 
$\tan \beta$ and $ \mathrm{sign}(\mu)$. $\mu$ is the
Higgsino mixing parameter.
We assume {\small CMSSM} with universal
boundary conditions at the unification scale $M_{\mathrm{GUT}}$.

Although in our analysis we have enforced unification on
gauge couplings at $M_{\mathrm{GUT}}$, the extracted values for the
relic density are insensitive to this 
assumption and can cover cases where one abandons the naive 
gauge coupling unification scenario. 
In those cases the unification scale $M_{\mathrm{GUT}}$ is defined as the
point where $g_1$ and $g_2$ meet. At this scale
$g_3=g_{1,2}(1+\Delta \epsilon)$. $\Delta \epsilon \neq 0$ signals
deviation from gauge coupling unification condition, which may be
attributed to the appearance of high energy thresholds.
Values of  $\Delta \epsilon$ of the order of 1\% produce 5\%
variation in $\alpha_s(M_Z)$, which however are not felt by the relic 
density. The reason is that the latter depends implicitly on 
$\alpha_s(M_Z)$, through its dependence on sparticle masses, and
therefore such small variations of $\alpha_s$ have negligible effect on
the relic density. Therefore our analysis can accommodate cases where
one allows for small departures from schemes where gauge couplings
unify at a common scale.

Running two-loop {\small RGE}'s for
all couplings and masses involved, in the usual manner, we determine the
parameters at the $Z$-pole mass which are  necessary to calculate masses and
couplings entering into the helicity amplitudes. Throughout
radiative breaking of the {\small EW} symmetry is assumed.
The magnitude of the $\mu$ parameter, but not its sign, as well
as the Higgs mixing soft parameter $m_3^2$ are both determined at $M_Z$ via
the minimization conditions of the one-loop corrected effective potential.

All couplings and running masses are calculated in the 
$\overline {\mathrm{\small DR}}$ scheme. 
Whenever needed these can be converted
to their corresponding $\overline {\mathrm{\small MS}}$ values.
In a mass independent renormalization scheme, as the 
$\overline {\mathrm{\small DR}}$, no theta functions enter into the
{\small RGE}'s to implement the decoupling of heavy sparticles at 
thresholds (see for instance Bagger {\etal} in Ref.~\cite{Lahanas}). 
Therefore corrections to physical masses, which are
calculated as the poles of propagators, receive contribution from
both light and heavy degrees of freedom.

The pole masses of the third generation fermions are taken
equal to $M_t^{pole}=175\GeV$, $M_b^{pole}=5\GeV$ and 
$M_{\tau}^{pole}=1.777\GeV$. From the pole masses we can have the values of
running masses at the pole, and then run the appropriate 
{\small RGE}'s to have the corresponding  running masses at the reference
scale $M_Z$. The $b$ and $\tau$ masses should evolve, according to the
$SU(3)_c \times U(1)_{em}$ group, since $M_b^{pole}$, $M_{\tau}^{pole}$
are below $M_Z$. Note that in the case of
$b$ and $t$-quarks, the 
two-loop {\small QCD} corrections relating 
pole and running mass are duly taken into account. In this way one obtains
the values of the running masses at $M_Z$, and from these
the corresponding Yukawa couplings at the same scale in the
$\overline {\mathrm{\small DR}}$ scheme as demanded.

Regarding Higgs boson masses, one-loop radiative correction
to their masses are assumed through out this paper. The effect
of the renormalization group improvement and leading two-loop
corrections although important for an accurate
determination of the Higgs masses does not significantly affect the
values of the relic density. Only the location of the Higgs $s$-channel
poles and the thresholds, whenever Higgses appear in the final state, 
are little affected.

Radiative corrections to the couplings of the {\small LSP} to Higgses are
not taken into account in this work. These can be important when  {\small LSP}
is a high purity Higgsino state \cite{Drees2}, since a pure Higgsino state
has no coupling to Higgs bosons. However in the {\small CMSSM}
with universal boundary conditions for the soft masses at the 
Unification scale a high purity Higgsino state is hardly
realized in view of negative results from {\small SUSY} particle searches,
and the aforementioned corrections are not of relevance.

We also assume that the {\small LSP} is the lightest of the neutralinos.
At the stage (i) of collecting inputs for the calculation of the coefficients
$a$, $b$ we do not impose all existing experimental bounds on sparticle masses,
especially those imposed on gluino and squark masses, some of which are
conditional and model dependent.
The reason for doing this relies on that we want to study the behaviour of 
the relic density $ \Omega_{\lsp} h_0^2 $ in as much enlarged portion of the 
parameter space as possible. 
Obviously the parameter space  will shrink even more if additional
experimental constraints are taken into account. We postpone a
discussion concerning the experimental bounds used in our analysis
for the following chapter.

Having all parameters at the scale $M_Z$ we pass to stage (ii)
and calculate the coefficients $a$ and $b$ (see Eqs.~(\ref{ab1},\ref{ab2}))
through which $\xsec$ are calculated. As discussed in the previous
section we have assumed non-relativistic approximation  and have expanded up
to ${\cal O}(v^2)$ in the relative velocity $v$. However such an expansion
breaks down near a threshold, or near a pole as discussed in the previous
section.
 In order to quantify the notion of nearness to
either a  threshold or to a pole we first consider the threshold case.
As is obvious from Eq.~(\ref{epsi}) we are on the threshold when the
parameter $\epsilon$ vanishes, in which case the expansion  
of Eq.~(\ref{kine})
breaks down. From this equation it is seen that the value of $\epsilon$
signalling departure from the validity of the expansion in powers of
$v$, is the one for which the coefficients of $v^2$ in 
Eq.~(\ref{kine}) is
unity. This occurs for $\epsilon_0 = 0.125$  which yields 
$z_0 \equiv {\frac{M_X + M_Y}{2\,m_{\lsp}}} \simeq 0.94$. Looking for a more
reliable criterion we invoke Ref.~\cite{Griest} where results relying on
more accurate analyses are compared against the standard schemes  which
we are using in this paper. From the figures displayed in the 
aforementioned reference
we find that $z_0 \simeq 0.95$ not very far from the value quoted above.
Therefore throughout our analysis we shall employ the following
``near threshold" criterion
%%%%%%%%%%%%%
\bea
0.95 \; \leq \; {\frac{M_X + M_Y}{2\,m_{\lsp}}} \leq \; 1  \; . \label{crit1}
\eea
%%%%%%%%%%%%

A similar analysis holds for the poles too. 
As is obvious from Eq.~(\ref{pole}) the expansion breaks down when
$R_I \; \equiv \;  {\frac{m_I}{m_{\lsp}}}$
is close to 2, unless the rescaled width $G_I$ (see Eq.~(\ref{pole}))
turns out to be large. The possible poles
encountered are the $Z$, $H$, $h$ and $A$ resonances which have small rescaled
widths unless the {\small LSP} is very 
light with mass ${\cal{O}} (10) \GeV $. In our
analysis we employ the following ``near pole" 
criterion\footnote{Outside this region 
the traditional series expansion, we use
in this paper, and exact results are almost identical. See for instance
Ref.~\cite{Lopez}.}
%%%%%%%%
\bea
   |\; 4 - R_I^2 \;| \; \leq \; 0.8 \,.    \label{crit2}
\eea
%%%%%%%%%%%%
For values of the parameters leading to either Eq.~(\ref{crit1}) 
or Eq.~(\ref{crit2})
the expansion of the cross section 
in powers of the relative velocity is untrustworthy and results based on
such an expansion are unreliable. In those cases other more accurate methods
should be used (see Refs.~\cite{Griest,Gelmini}).

In the final stage (iii) we solve the Boltzmann equation (\ref{bol}). Knowing
$\xsec$ from the procedure outlined previously, and by calculating
the functions $g(T)$, $h(T)$, $h^\prime (T)$, we can have the prefactor
$\lambda (x)$ appearing in Eq.~(\ref{bol}). At high temperatures, or same
large values of $ x=T/m_{\lsp}$, above the freeze-out temperature,
the function $q(x)$ approaches its
equilibrium value $q_0 (x)$ (see Eq.~(\ref{qs})). A convenient
and accurate method for
solving the Boltzmann equation is the 
{\small WKB} approximation employed in Ref.~\cite{Lopez}. 
This relies on the observation that $\lambda (x)$ is a rather
large number of the order of $10^8$ or so, or even larger in some cases. Due
to the largeness of $\lambda (x)$ an approximate solution is
%%%%%%%%%%%%
\bea
q  = {q_0} \; \left( 1\;+\;{\frac{q^{\prime}_0}{2 \lambda q^2_0}}\right) \;+\;
{\cal {O}}(1/{{\lambda}^2}) \; . \label{qappro}
\eea
%%%%%%%%%%%%
Obviously this holds for values of $x$ for which
${\frac{q^{\prime}_0}{2 \lambda q^2_0}} $ is smaller than unity. In our
numerical procedure we find a point $x_0$ for which
%%%%%%%%%
\bea
{\frac{q^{\prime}_0}{2 \lambda q^2_0}} (x_0) \simeq 0.1  \; .\label{bound}
\eea
%%%%%%%%%%%%%%%
For larger values of $x$, this ratio becomes even smaller while for smaller
values increases rapidly invalidating the approximation (\ref{qappro}).
This rapid change of the aforementioned ratio is mainly due to 
${\frac{q^{\prime}_0}{2 \;q^2_0}} $.
A typical sample  is shown
in Table I where for an {\small LSP} mass $\approx 100 \GeV$, for a top mass
equal to $175 \GeV$ and for masses
of the Higgses $h$, $H$, $A$, $H^{\pm}$ equal to
$100$, $250$, $270$ and $300\GeV$ respectively,
we list its values for
$x$ in the range 0.03--0.08.
One observes that  ${\frac{q^{\prime}_0}{2 \;q^2_0}} $  increases by almost
10 orders of magnitude from $x=0.08$ down to $x=0.03$ while $\lambda (x)$
remains almost constant in this interval. With a typical value of
$\lambda (x) \approx 10^{10}$  this ratio turns out to be 
$\approx {\frac{1}{10}} $ for values of $x $ around $0.06 $.

Given the point $x_0$, defined in Eq.~(\ref{bound}), we numerically
solve the Boltzmann equation, in order to obtain solutions
in the region $x \leq x_0$, having as boundary condition
%%%%%%%%%%%%%%%%%%%
\bea
q(x_0) \;=\; 1.1 \; q_0(x_0)  \,.   \label{boundary}
\eea
%%%%%%%%%%%%%
The omitted ${\cal {O}}(1/{{\lambda}^2})$ terms in Eq.~(\ref{qappro})
yield corrections which are less than $5 \%$. Therefore this scheme
yields very accurate results.
% We should remark that the
%solution (\ref{qappro} ) for values of $x < x_0 $, and
%especially for values of $x$ approaching zero,
%is not valid. This is the reason we have to appeal to numerical integration
%in order to know the relic density for values of $x$ in the vicinity of zero.
%This is due to the fact that the ratio  ${\frac{q^{\prime}_0}{2 \;q^2_0}} $
%appearing in the approximate solution  (\ref{qappro} ) is an increasing
%function of $x$ as $x$ decreases. For instance for an LSP mass of $100 \GeV$,
%for top mass equal to $175 \GeV$ and for Higgs masses .... 

The numerical solution is found by use of special routines found in {\small IMSL}
{\small FORTRAN} library, which are eligible to handle stiff differential
equations such as the Boltzmann equation.
Similar routines can be presumably found in other
libraries too. However it is important to stress that the choice of the
right routine and accuracy is of great importance. Due to the fact that
$q(x)$ varies rapidly for $x < x_0 $ a high degree of accuracy is demanded
which makes other routines being either extremely slow or unable to reach 
convergence.

To implement the numerical solution of the Boltzmann equation we need as
inputs
the function $\lambda (x)$, defined by Eq.~(\ref{lala}), whose values are
known provided $g$, $h$, $h^\prime$, as well as $\xsec$, are
calculated. The effective number and entropy degrees of freedom functions
$g$ and $h$ respectively are calculated in the way described in the first
section. The thermal integrals needed for their calculation are found by
invoking fast and reliable
integration routines found in the same Fortran library {\small IMSL}. 
Their correctness has been checked by comparing our findings against those
of other packages. The {\small LSP} mass and the masses of Higgses and the
remaining {\small SM} particles are 
needed in order to calculate the aforementioned
functions. Therefore we first run to get the values of all 
parameters involved at the physical scale $M_Z$, as well as all physical
masses among these the
radiatively corrected Higgs masses. These inputs are also used in
order to calculate all relevant sparticle couplings to other species 
necessary to calculate the coefficients $a$, $b$
and hence   $\xsec$, for each one of
the processes involved.

After this short description of our numerical procedure we pass to discuss
how this machinery is implemented to infer physics conclusions for the
{\small LSP} relic density.

%%%%%%%%%%%%%%%%%%%% S E C T I O N --> IV  %%%%%%%%%%%%%%%%%%%%%%%%%%%%%%
%%%%%%%%%%%%%%%%%%%%%%%%%%%%%%%%%%%%%%%%%%%%%%%%%%%%%%%%%%%%%%%%%%%%%%%%%

\section{The LSP relic density}

Following the numerical procedure outlined in the previous section we are
ready to embark on discussing 
the predictions for the {\small LSP} relic density. As discussed in
section I we have in mind minimal supersymmetry with universal boundary
conditions at the unification scale for the soft breaking parameters and
radiatively induced {\small EW} symmetry breaking. Therefore the arbitrary
parameters are $ m_0$, $M_{1/2}$, $ A_0$ and $\tan \beta$.
The value of $\mu$ is
determined from the minimization conditions of the one-loop corrected
effective potential. These also determine
the Higgs mixing parameter $m_3^2$. The sign of $\mu$ is undetermined
in this procedure and in our analysis both signs of $\mu$ are considered.
Therefore in this scheme the $\mu$ value as well as $m_3^2$
are not inputs.
%%%%%%%%%%%%%%%%%%%%%%%%%%%%
%%% Part transferred to where RG analysis is discussed
%%%%%%%%%%%%%%%%%%%%%%

At the first stage for each point in the parameter space we collect
outputs including all parameters relevant for the calculation of the relic
density, such as couplings and physical masses, in the way prescribed
in the previous section, without imposing
any experimental constraints. However we certainly exclude points that are
theoretically
forbidden, such as those leading to breaking of lepton and/or color number,
or points for which Landau poles are developed and so on. We also exclude
points for which the {\small LSP} is not a neutralino.
In subsequent runs the above  inputs are used to determine the $\lsp$ relic
density solving the Boltzmann equation as outlined in the previous
section.

In our analysis we should exclude points of the parameter space for
which violation of the experimental bounds on sparticle masses is
encountered. We use the bounds of Ref.~\cite{newdata} \\

%%%%%%%%%%%%%%%%%%%%%
\bea
Neutralinos &:& m_{{\ttchi}^0_1}\;>\;33\GeV \hspace{5.9cm} \nonumber \\
Charginos \; \; \; &:& m_{{\ttchi}^{+}_1}\;>\;95\GeV  \nonumber \\
Sleptons \; \; \; \; \; \; &:& m_{{\tilde{\tau}}_R} \;>\;71\GeV \,,\;
(m_{{\ttchi}^0_1}\;<\;20\GeV) \nonumber \\
&&m_{{\tilde{\mu}}_R}\,,\;m_{{\tilde{e}}_R}\;>\;84 \,,\; 89 \GeV \,,\;
(m_{{\tilde{\mu}}_R,{\tilde{e}}_R }\;>\;m_{{\ttchi}^0_1} \;+\;10)
\nonumber \\
&&m_{{\tilde{\nu}}_L} \;>\;43 \GeV  \nonumber \\
Higgses \; \; \; \; \; \; &:&m_{h_0}\;>\;81 \GeV \,,\; (light \;\; scalar)
\nonumber \\
&&m_{A}\;>\;81 \GeV \,,\; m_{H^{\pm}} \;>\;69 \GeV \,.
\eea
%%%%%%%%%%%%%%
At this stage we do not exclude yet points which violate the gluino
$\tilde{g}$ and squark $\tilde{q}$ mass bounds
%%%%%%%%%%%%%%%%%%%%%
\bea
m_{\tilde{g}} \;>\;173 \GeV \,,\;m_{\tilde{q}} \;>\;176 \GeV
\,.
%%%%\nonumber \\
\eea
Then for each point of the parameter space for which the above experimental
constraints are obeyed we calculate the $\lsp$ relic density.

From our outputs we have found that the chargino bound is the most stringent
of all listed above, in the parameter space examined.
The only exception is the light Higgs boson mass, which outstrips the
chargino bound for very low $\tan\beta$ values. 
 The gluino and
squark mass bounds quoted before, if subsequently imposed, 
are found to be weaker than the
chargino bound. Only a small region of the parameter space which is allowed
by the chargino mass constraint is excluded when one enforces the bound
$m_{{\tilde{t}}_1} > 176 \GeV$ on the lightest of the top squarks.

Before embarking to discuss our physics results we should stress that in our
scheme we have not committed ourselves to any particular
approximation concerning the masses or couplings
of sectors which are rather involved, such as neutralinos for instance, which
are crucial for our analysis. Therefore we do not only consider regions of the
parameter space in which the {\small LSP} is either purely a $ \tilde B$ (bino) or
purely a Higgsino, but also regions where in general the {\small LSP}
happens to be an admixture of the four available degrees of
freedom\footnote{ The case of a Higgsino-like LSP has 
been pursued in Refs.~\cite{Drees2,Falk2} where the 
dominant radiative corrections to neutralino
masses are considered. Analogous corrections to couplings of Higgsino-like
neutralinos to $Z$ and Higgs bosons are important and can change the relic
density by a factor 5 in regions of parameter space where LSP is a high purity
Higgsino state \cite{Drees2}. However this case is not
realized within the CMSSM with universal boundary
conditions for the soft masses.}.
%%In earlier sections we listed all two-body annihilation processes
%%contributing to the {\small LSP} relic density.
Regarding the LSP's mass we note that 
for large values of it many channels are
open but for small values ($m_{\lsp} < 40 \GeV$) only channels with 
fermions, except the top quark, in the final state are contributing. In these
processes the exchanged particles can be either
a $Z$-boson and a Higgs in the $s$-channel, 
as well as a sfermion $\tilde f $ in the $t$-channel. Higgs exchanges are
suppressed by their small couplings to light fermions,
and sfermion exchanges are suppressed when their masses are large. Then
the only surviving term, for large values of squark and slepton
masses is the $Z$-boson exchange. 
However in the parameter region where the {\small LSP} is a high purity 
bino, this is not coupled to the $Z$-boson resulting to very small cross
sections enhancing dramatically the {\small LSP} relic density. Therefore in 
considerations in which the {\small LSP} 
is a light bino\footnote{This happens 
when $|\mu| \gg M_W$, with $M_1$ small
$\approx M_W$.},
large squark or slepton masses are inevitably excluded since they lead to
large relic densities. 
If one relaxes this assumption and considers regions of the parameter space
in which the {\small LSP} is light but is not purely bino, 
heavy squarks or sleptons are in principle allowed.
When {\small LSP} is light
the only open channels are those involving light fermions in the final 
state. Then the annihilation of {\small LSP}'s into neutrinos for instance, 
a channel which is always open, proceeds via the 
exchange of a $Z$-boson which is non-vanishing and dominates the reaction 
when $m_0$ is sufficiently large, due to the heaviness of sfermions. This  
puts a lower bound on 
$ {\sum}_{f} \vev{ v  \sigma  ( \lsp \, \lsp \goes f \,\bar{f} ) }$ and hence 
an upper 
bound on $ \Omega_{\lsp} h_0^2 $ which can be within 
the experimental limits quoted in the introduction. 
On these grounds one would expect that by increasing 
$m_0$, while keeping $M_{1/2} $ fixed and low, there are regions in which  
the relic density stays 
below its upper experimental limit. 
Although such corridors of low $M_{1/2}$ and large
$m_0$ values\footnote{ These corridors 
of low $M_{1/2}$ and large $m_0$ values 
have been also presented in Ref.~\cite{recent}.} are
cosmologically acceptable they are ruled out by the
recent bound put on the chargino mass. Hence
the possibility of having a light {\small LSP} and a heavy
sfermion spectrum is excluded.

\begin{figure}[t] 
\begin{center} 
\epsfig{file=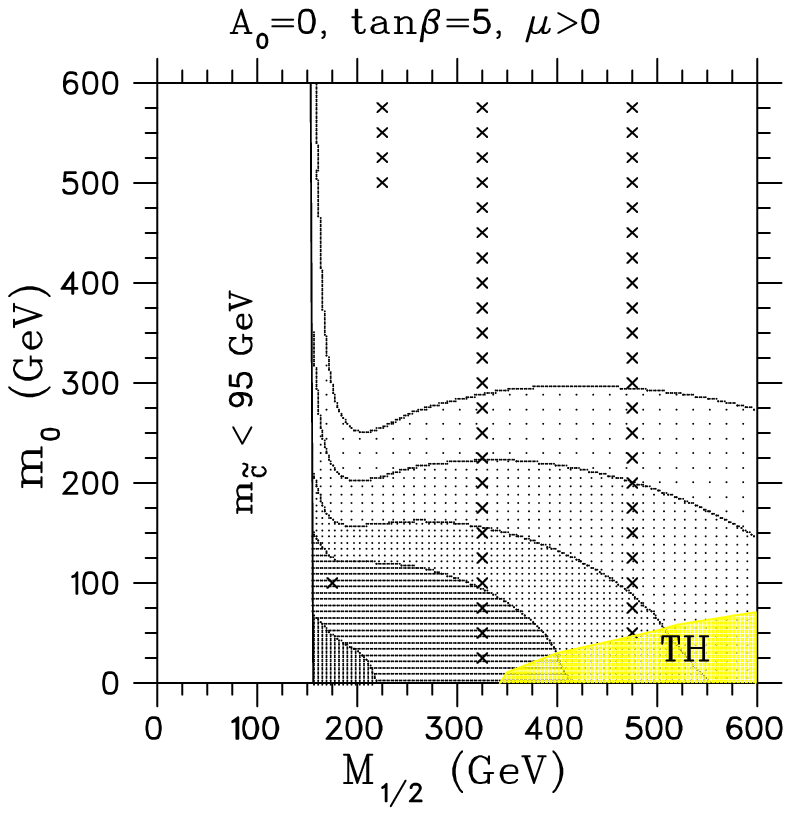,height=7.5cm,width=7.5cm} 
\hspace{.3cm}
\epsfig{file=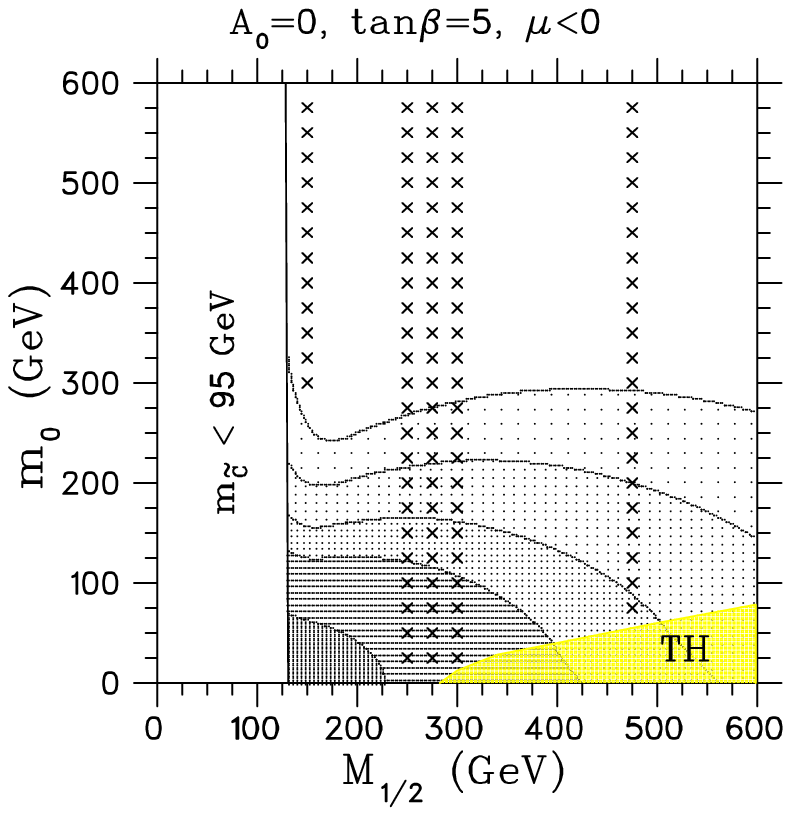,height=7.5cm,width=7.5cm}

\vspace{1cm}
\epsfig{file=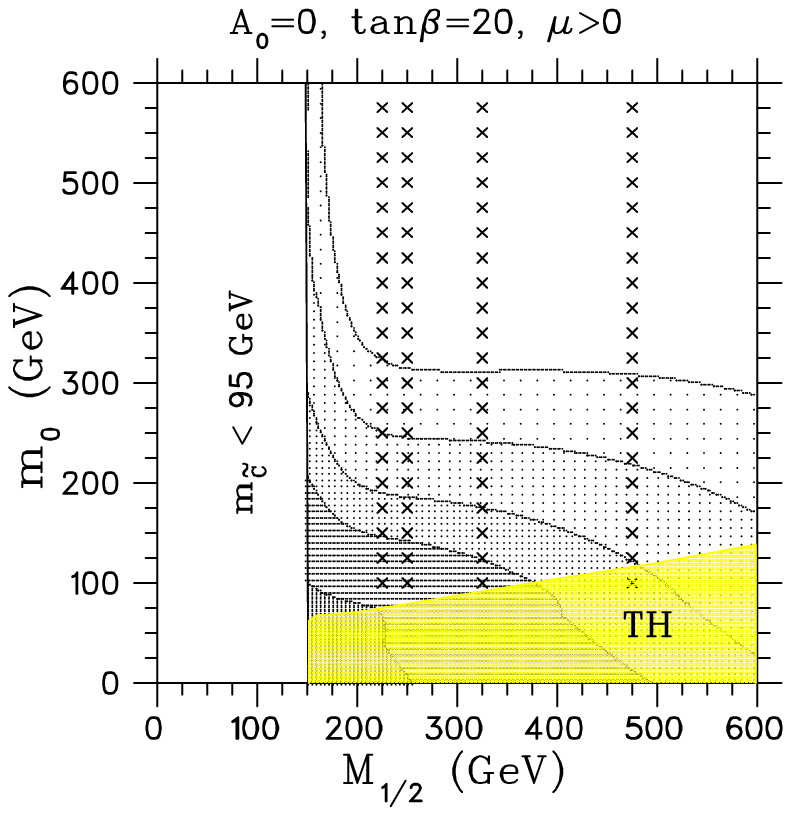,height=7.5cm,width=7.5cm} 
\hspace{.3cm}
\epsfig{file=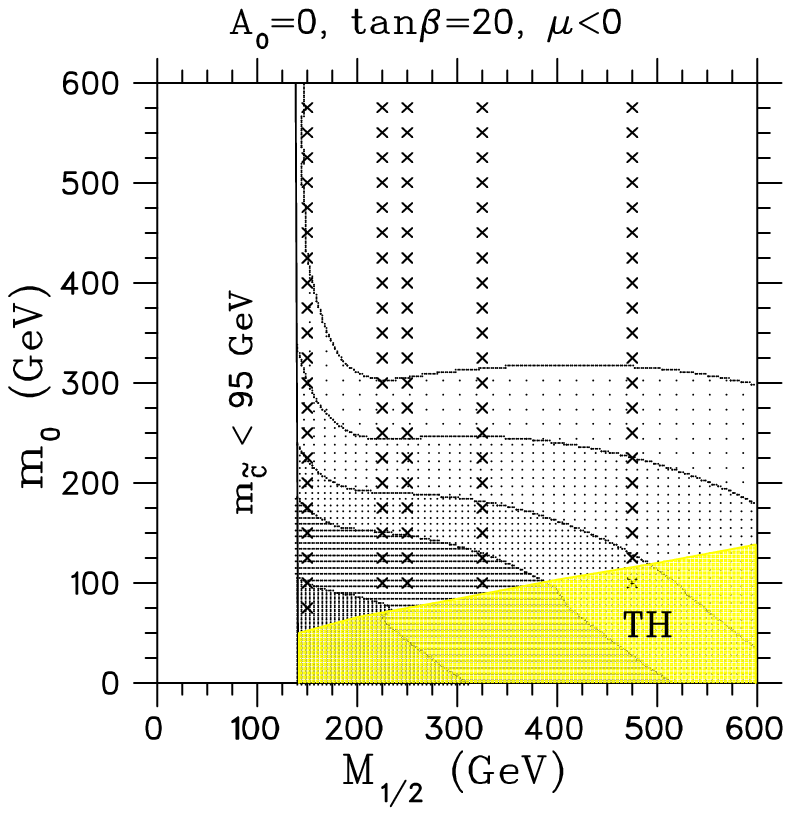,height=7.5cm,width=7.5cm} 

\begin{minipage}[t]{14.cm}  
\caption[]{
The {\small LSP} relic density
$\relic$ in the ($m_0$,$M_{1/2}$) plane for
given values of $A_0$, $\tan \beta$ and sign of $\mu$. 
Grey tone regions, from darker to lighter,
designate areas in which the {\small LSP} relic density
takes values in the intervals:
$0.00 -0.08$, $0.08-0.22$,
$0.22-0.35$, $0.35-0.60$ and $0.60-1.00$ respectively.
In the blanc area $\relic > 1.0$. The area marked by ``{\small TH}" is
theoretically excluded (see main text). The area labelled
by $m_{\tilde {C}}<95\GeV$ is excluded by chargino searches.
%% The recent experimental data $m_{\tilde {C}}>95\GeV$ shift
%% the boundary of this area to the thick vertical line.
Crosses
 denote points for which thresholds or poles are
encountered.}

\label{fig1}  
\end{minipage}  
\end{center}  
\end{figure}

We have scanned the parameter space for values of
$ m_0$, $M_{1/2}$, $A_0$ up to $1 \TeV$ and  $ \tan \beta$
from around 1.8 to 40 for both positive and negative values of $\mu$.
The top quark mass is taken $175 \GeV$. In
figure~\ref{fig1} we display
representative outputs in the ($M_{1/2},m_0$)
plane for fixed values of $A_0$ and $\tan \beta$. Both signs of the parameter
$\mu$ are considered. In the displayed figure 
$A_0 = 0 $ and $\tan \beta = 5, \;20$. The five different grey tone
regions met as we move from bottom left to right up, correspond to regions in
which $ \relic$ takes values in the intervals
$0.00 -0.08$,
$0.08-0.22$, $0.22-0.35$, $0.35 - 0.60$ and 
$0.60 - 1.00$ respectively\footnote{These regions are 
chosen in accord with new and old bounds
on $\relic$ which have been cited in the literature.}.
In the blanc area covering the right up region, 
the relic density
is found to be larger than unity.
The boundary of the area excluded by chargino searches, 
designated by $m_{\tilde C} <95 \GeV$, 
refers to the bound quoted in the beginning of this section.

In these figures whenever a
cross appears it designates that we are near either a pole or a threshold,
according to the criteria given in the previous section. In these cases
the non-relativistic expansions used 
are untrustworthy and no safe conclusions can be drawn. For low values
of $M_{1/2}$ crosses correspond to mainly poles,which are either a $Z$-boson
or a light Higgs, while for higher values, where 
{\small LSP} is heavier and hence more
channels are open, these correspond to thresholds.
The grey area at the bottom labelled by ``{\small TH}", which usually occurs for
low values $m_0 \leq  150\GeV$, is excluded mainly
because it
includes points for which the {\small LSP} is not a neutralino.
In a lesser extend some of these correspond to points which are theoretically
excluded in the sense that 
either radiative breaking of the {\small EW} symmetry does not occur
and/or other unwanted minima, breaking color or lepton number, are developed.
 From these figures it is seen that as  $\tan \beta$ increases from
$\tan \beta = 5$ to $\tan \beta = 20$
the region for which the
{\small LSP} is not a neutralino is enlarged. This is due to the fact that by
increasing $\tan \beta$ the stau sfermion $\tilde{{\tau}}_{R}$ becomes
lighter, since its mass, as do 
the masses of all the third generations sfermions, depends rather strongly 
on $\tan \beta$ (and also on $A_0$). Although not displayed, similar is the
case when one increases the value of the parameter $A_0$.

For fixed $M_{1/2} > 150 \GeV $ the relic density $ \relic $
increases, with increasing $m_0$, 
due to the fact that cross sections involving sfermion
exchanges decrease. Thus the area corresponding to
$ \relic < 0.22 $ concentrates to the left bottom of the figure. In this
region $ m_0 < 200 \GeV$.
For fixed $m_0 $ the relic density $ \relic < 0.22 $ also decreases with
increasing $M_{1/2}$, since an increase in
$M_{1/2}$ enlarges squark and slepton masses as well 
yielding smaller cross sections.
If $M_{1/2} $ is further increased
the {\small LSP} will eventually cease to be a neutralino.

\begin{figure}[t]  
\begin{center}  
\epsfig{file=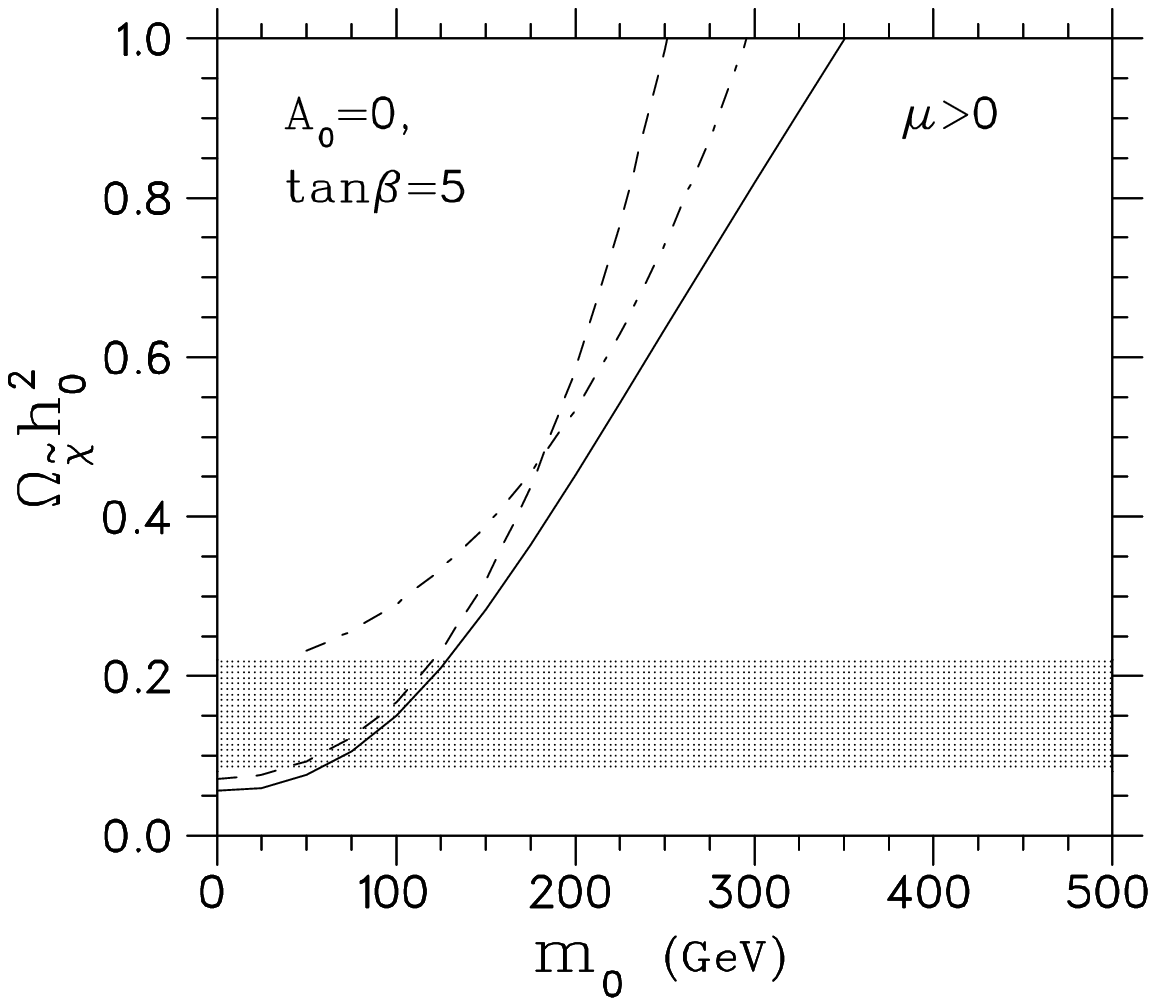,height=7.5cm,width=7.5cm} 
\hspace{.3cm}
\epsfig{file=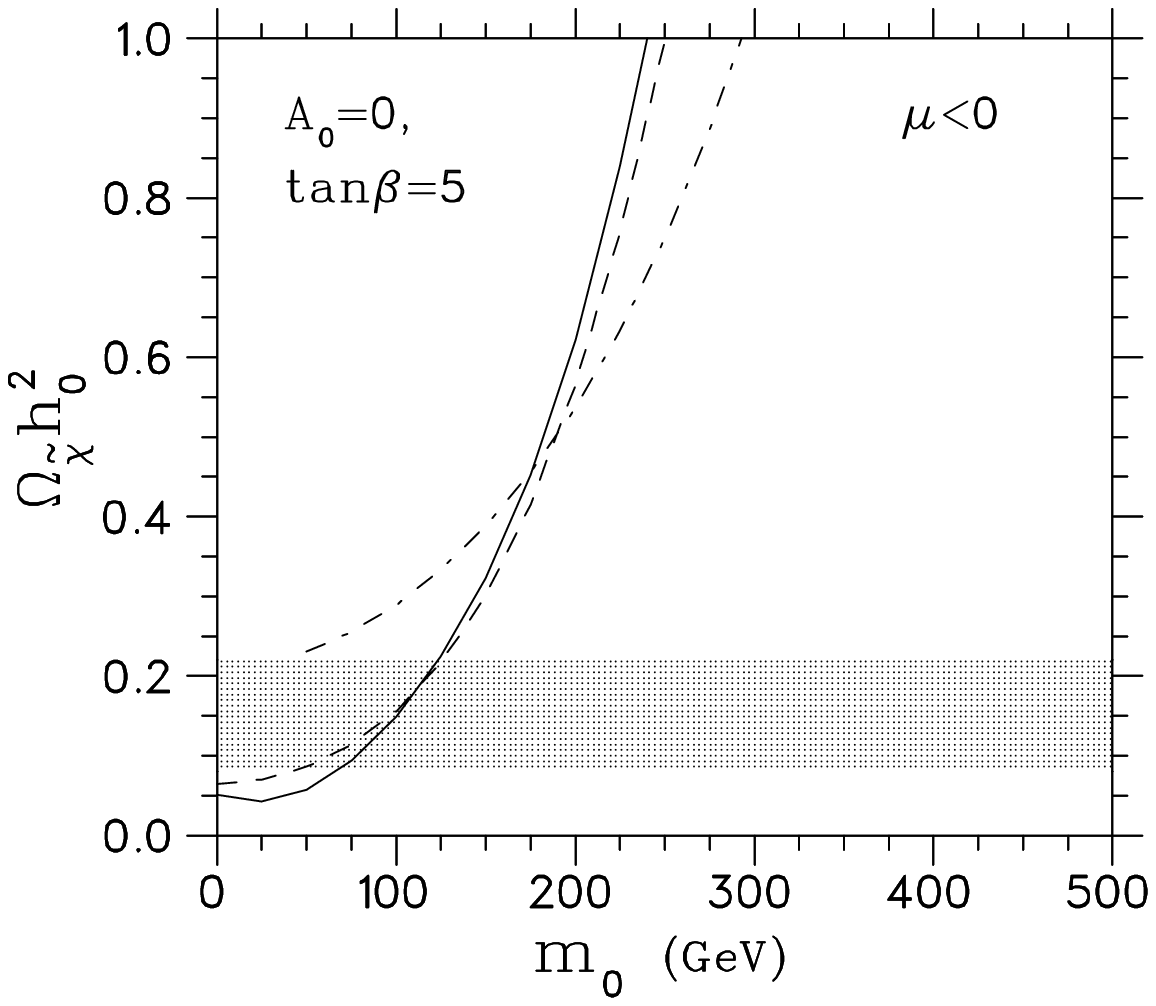,height=7.5cm,width=7.5cm}

\vspace{1cm}
\epsfig{file=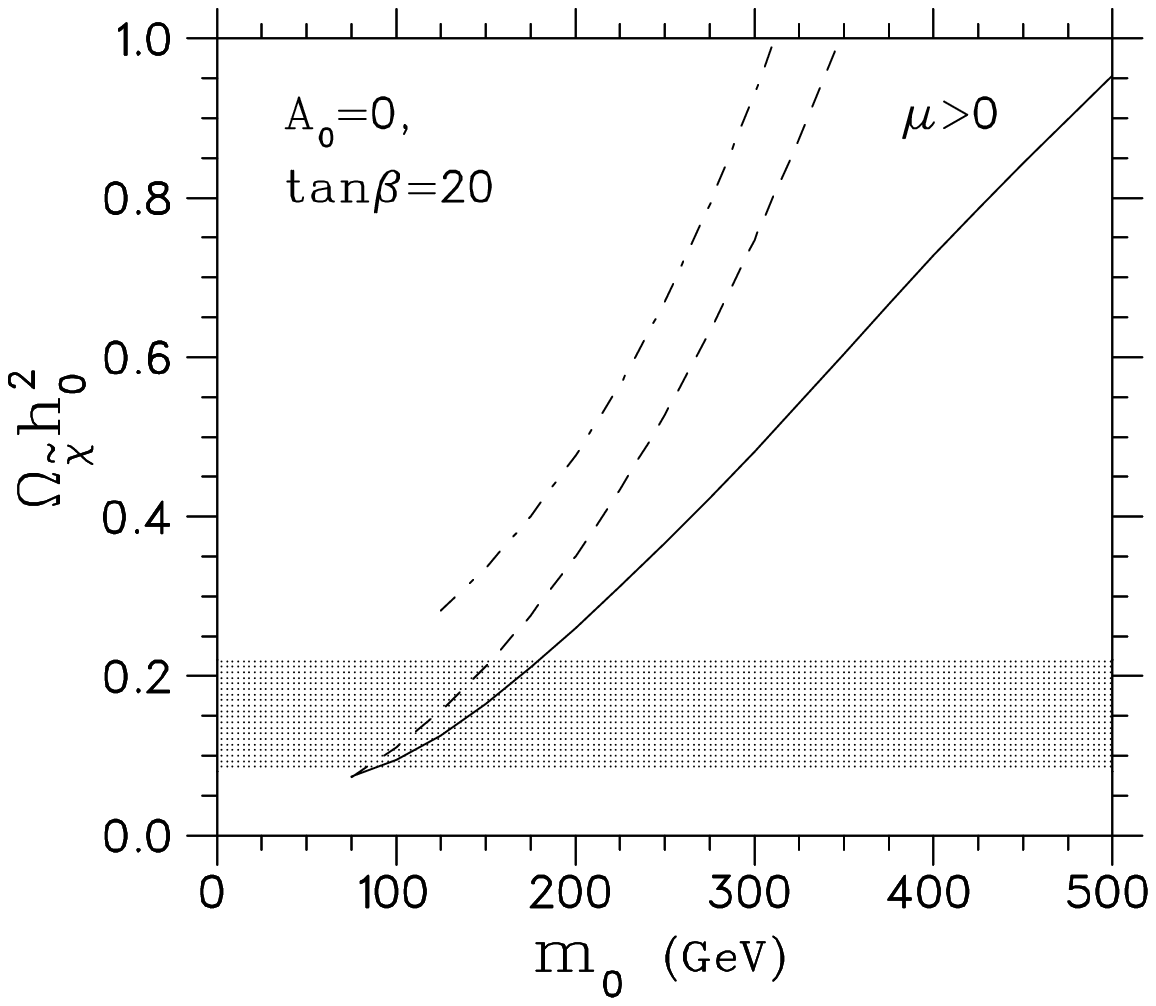,height=7.5cm,width=7.5cm} 
\hspace{.3cm}
\epsfig{file=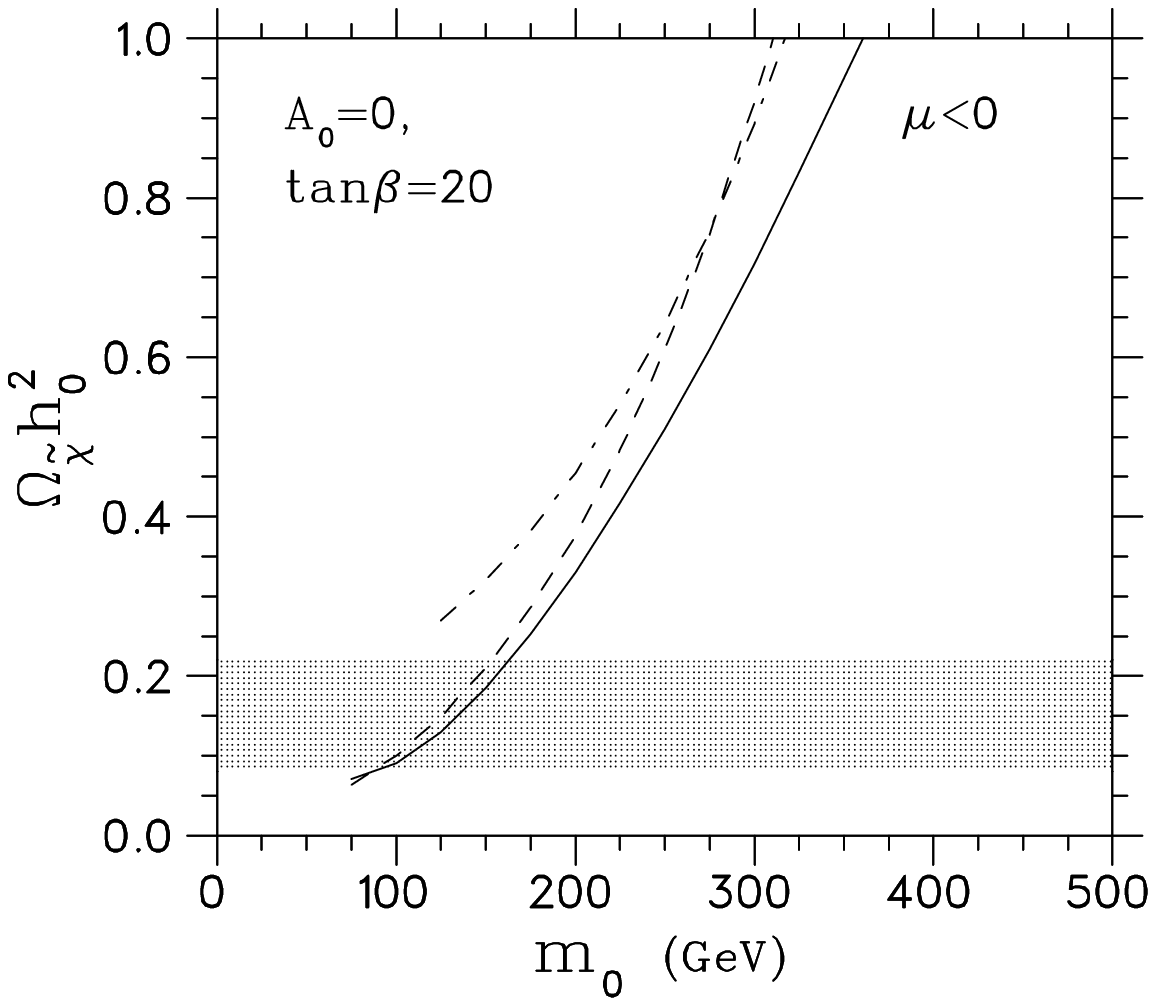,height=7.5cm,width=7.5cm}

\begin{minipage}[t]{14.cm} 
\caption[]{The relic density as function of $m_0$ for
fixed values of the remaining parameters.
The solid, dashed and dot-dashed lines correspond to
$M_{1/2}=170$, $200$ and $400\GeV$ respectively.}
\label{fig2} 
\end{minipage} 
\end{center} 
\end{figure}

In figure~\ref{fig2} and for fixed values of the parameter $A_0$
and $\tan \beta$ we
plot the {\small LSP} relic density as function of the soft scalar mass $m_0$ for
values of $M_{1/2} = 170$, $200$, $400 \GeV$ respectively.
The value $M_{1/2} = 170\GeV$ has been chosen close to the lowest
allowed by the recent chargino searches, and avoids poles or thresholds.
It is obvious from this figure that
for higher $ \tan \beta$ values $ \Omega_{\lsp} h_0^2 $  gets lowered, 
for fixed $m_0$, leaving more room for larger 
$m_0$ and hence for sfermion masses.
The abrupt stop in some of the displayed lines, towards their left endings, 
is due to the fact that the {\small LSP} ceases to be a neutralino for sufficiently
low values of $m_0$. 

\begin{figure}[t] 
\begin{center}

\epsfig{file=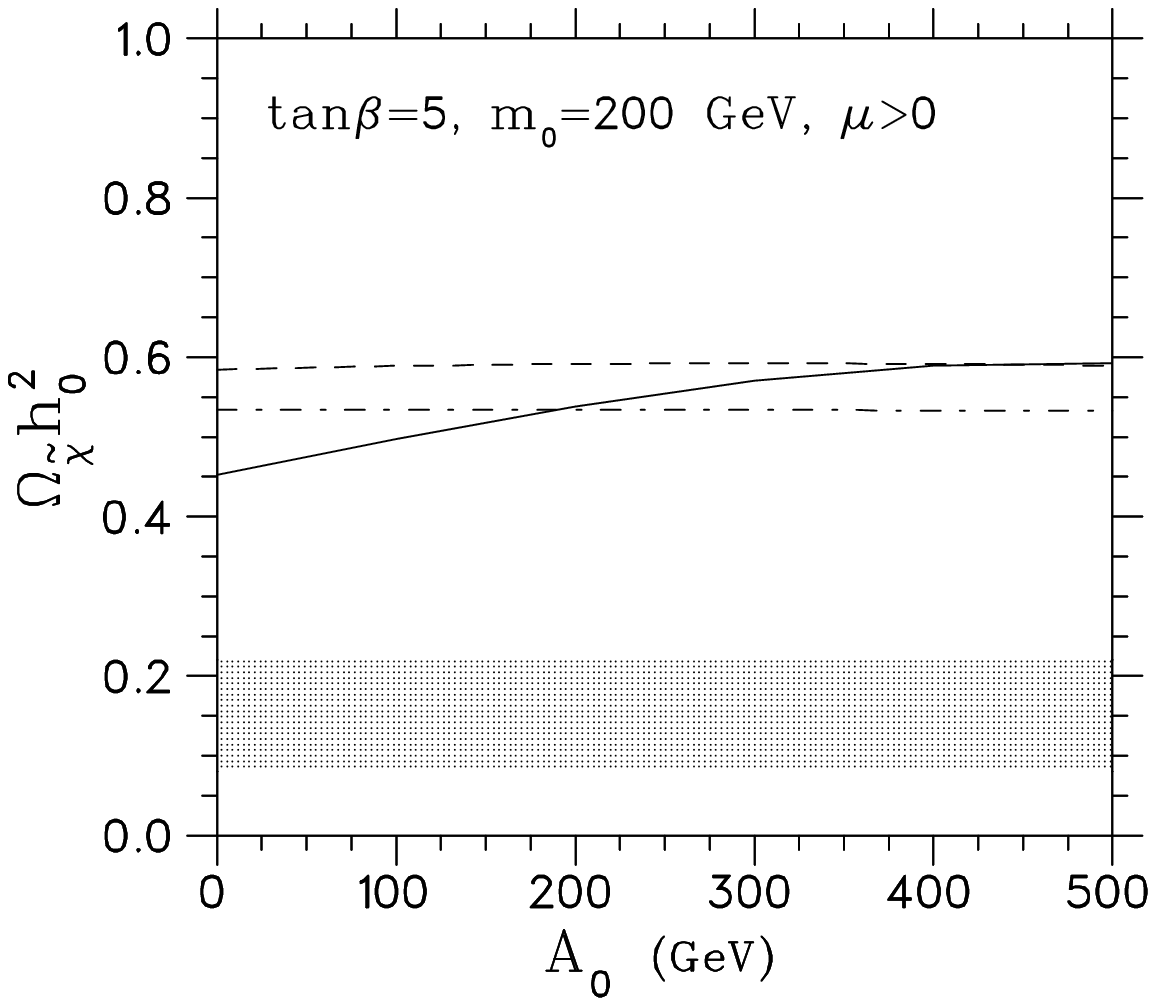,height=7.5cm,width=7.5cm} 
\hspace{.3cm}
\epsfig{file=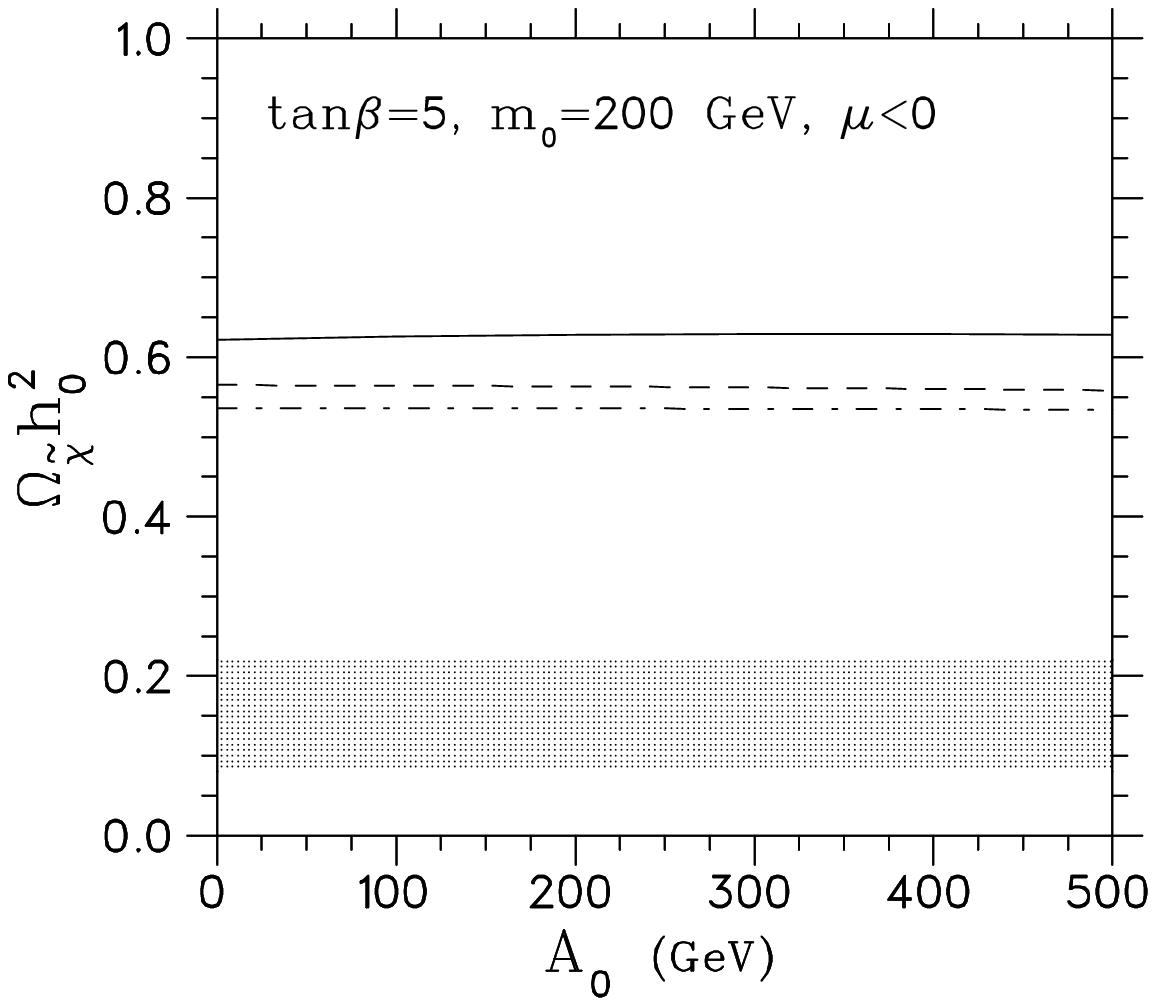,height=7.5cm,width=7.5cm}

\vspace{1cm}  
\epsfig{file=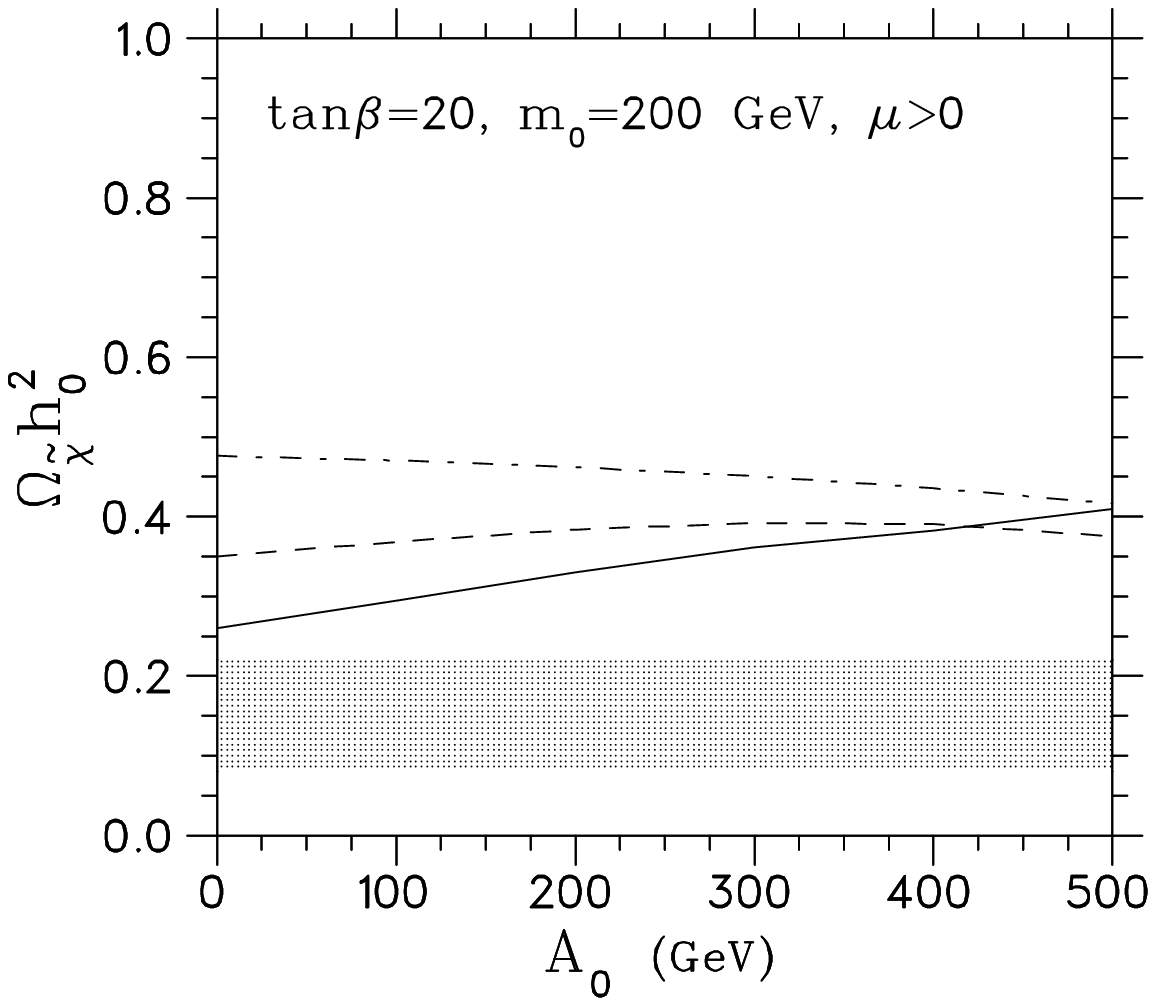,height=7.5cm,width=7.5cm} 
\hspace{.3cm}
\epsfig{file=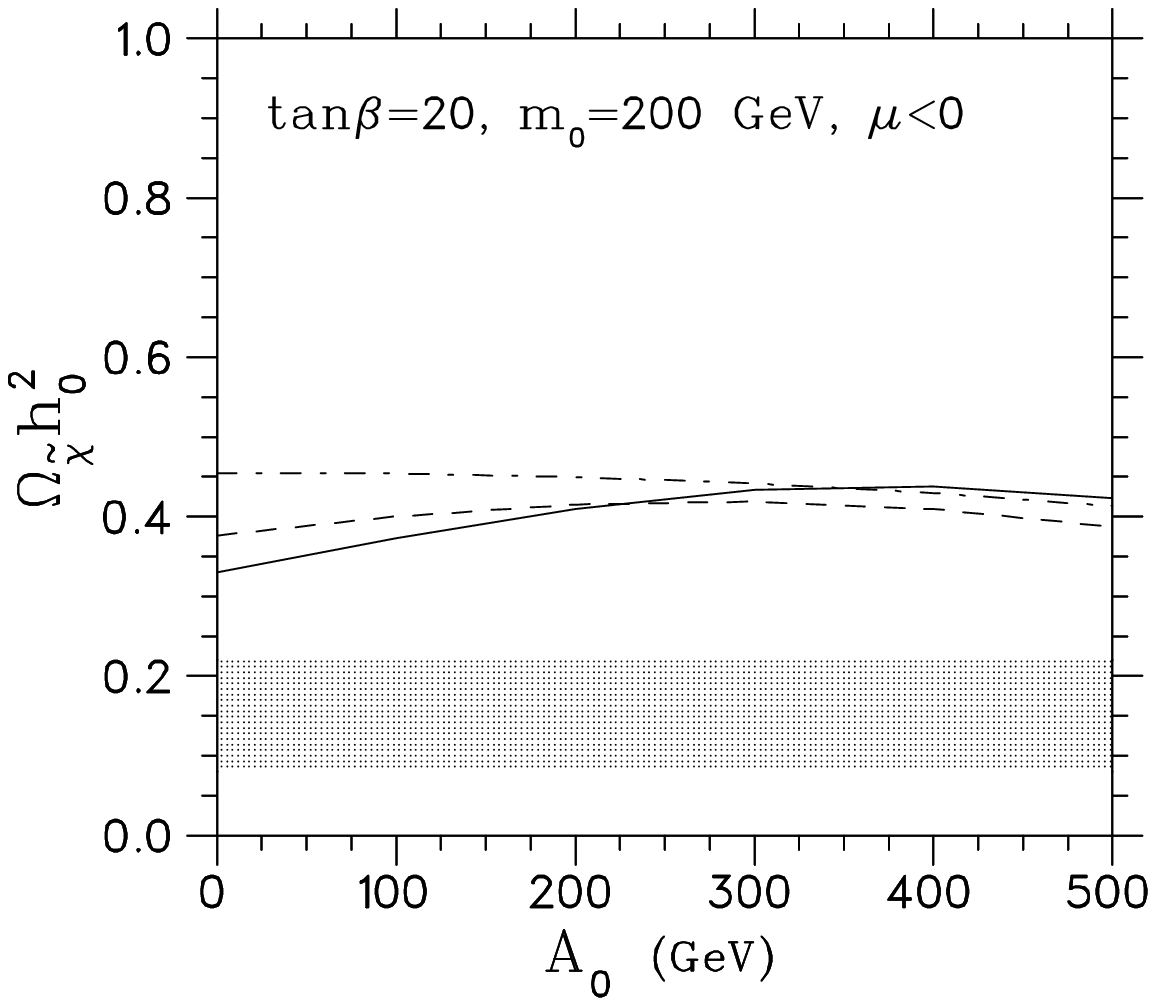,height=7.5cm,width=7.5cm}

\begin{minipage}[t]{14.cm} 
\caption[]{ The relic density as function of $A_0$. 
The lines are as in figure~\ref{fig2}.} 
\label{fig3} 
\end{minipage} 
\end{center} 
\end{figure}

\begin{figure}[t] 
\begin{center} 
\epsfig{file=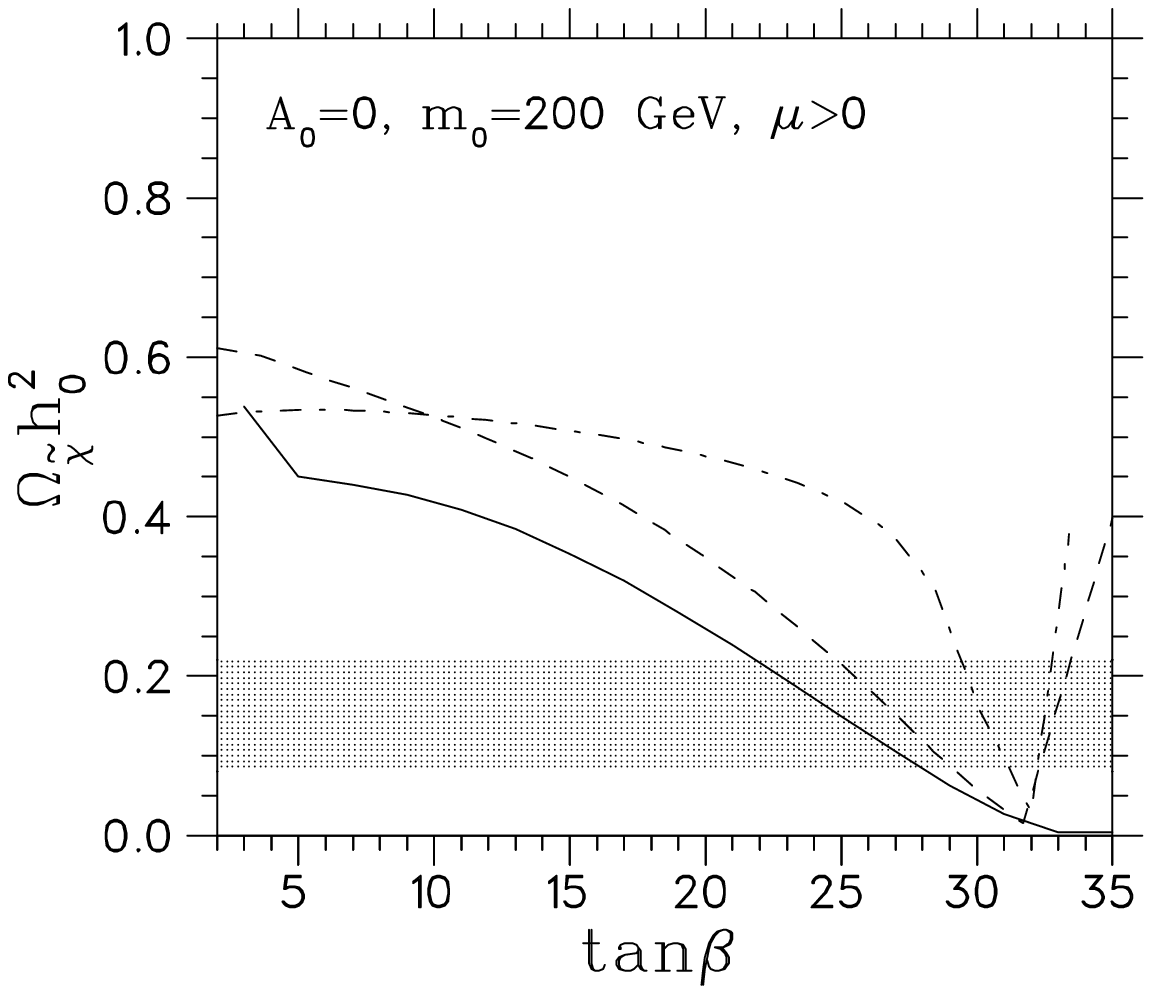,height=7.5cm,width=7.5cm} 
\hspace{.3cm}
\epsfig{file=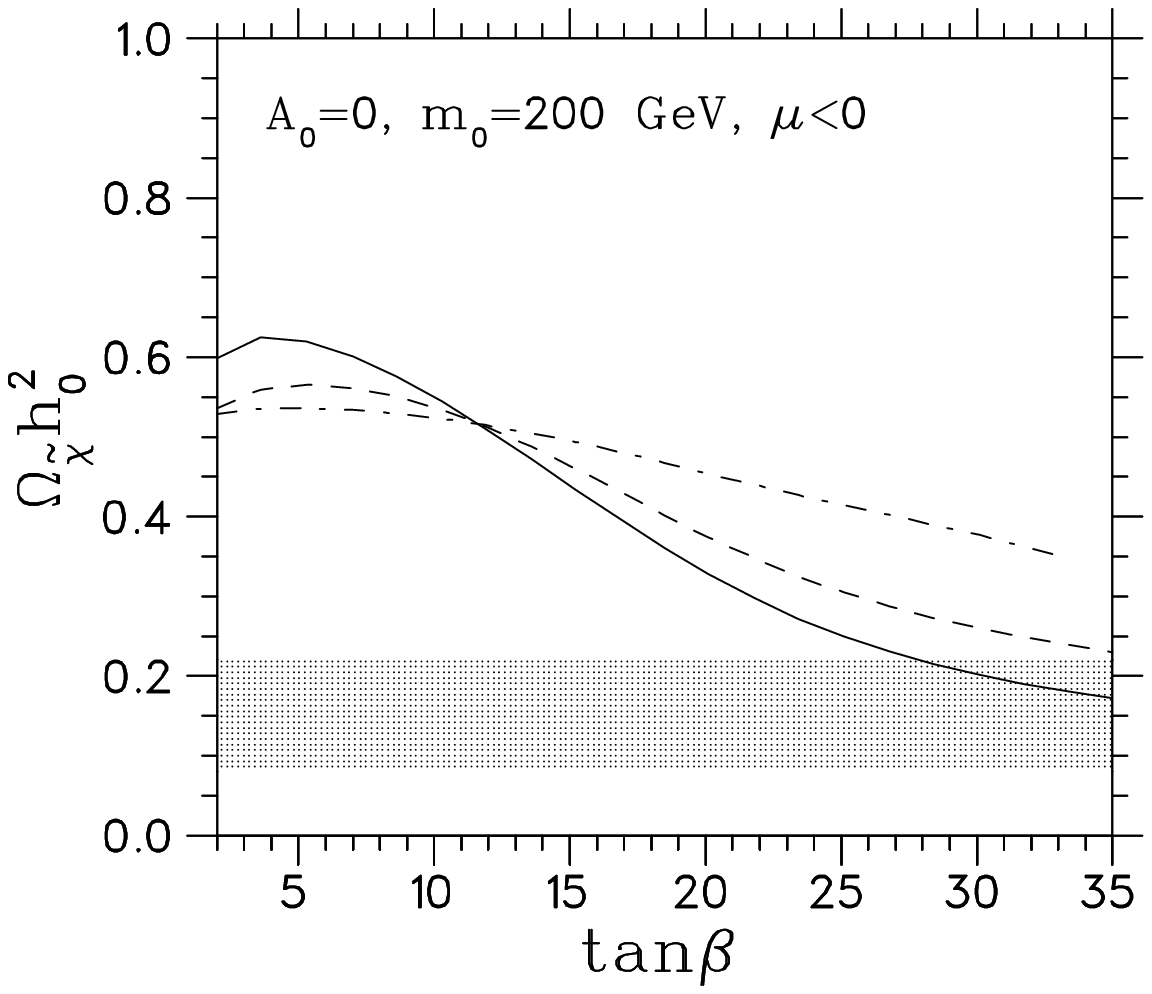,height=7.5cm,width=7.5cm}

\vspace{1cm}
\epsfig{file=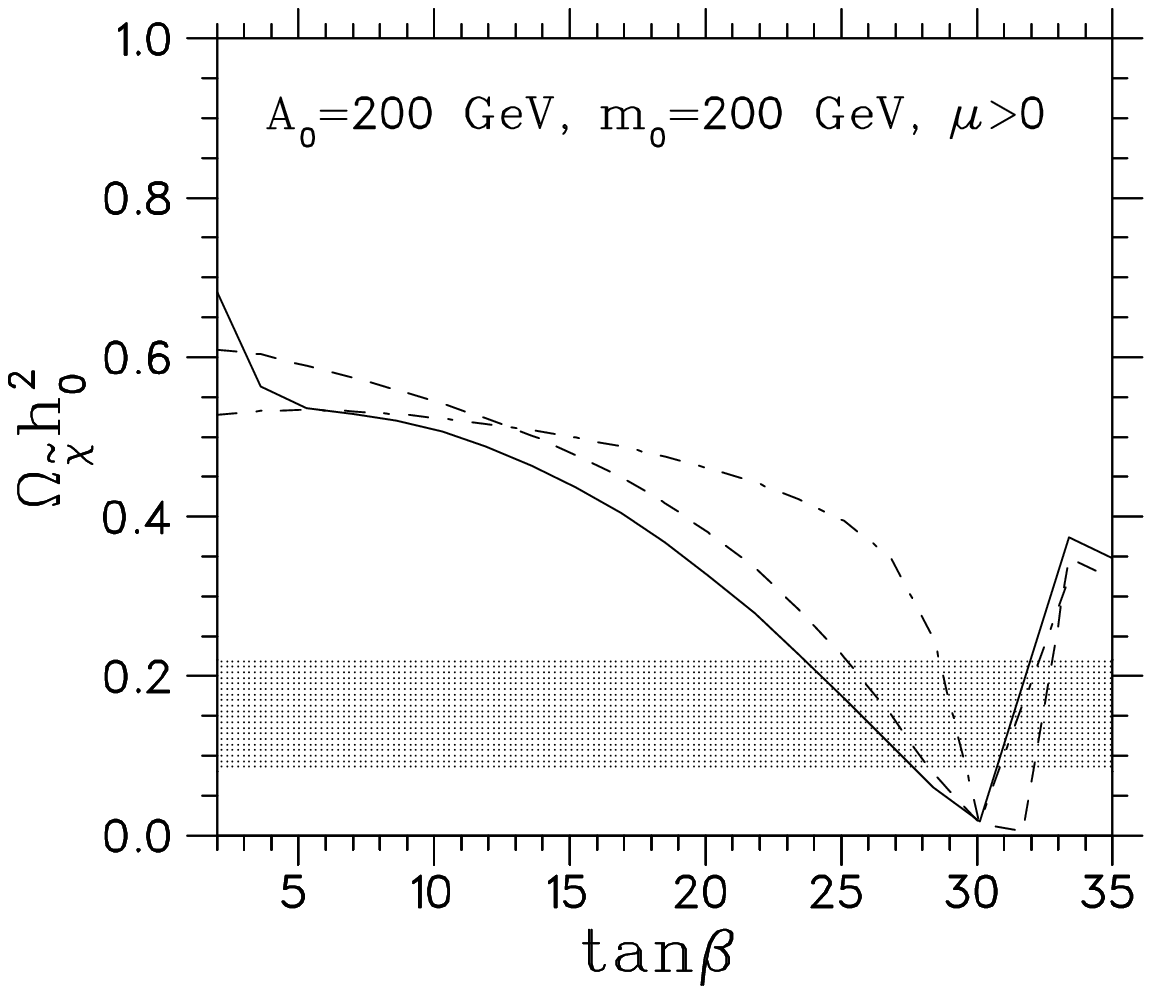,height=7.5cm,width=7.5cm} 
\hspace{.3cm}
\epsfig{file=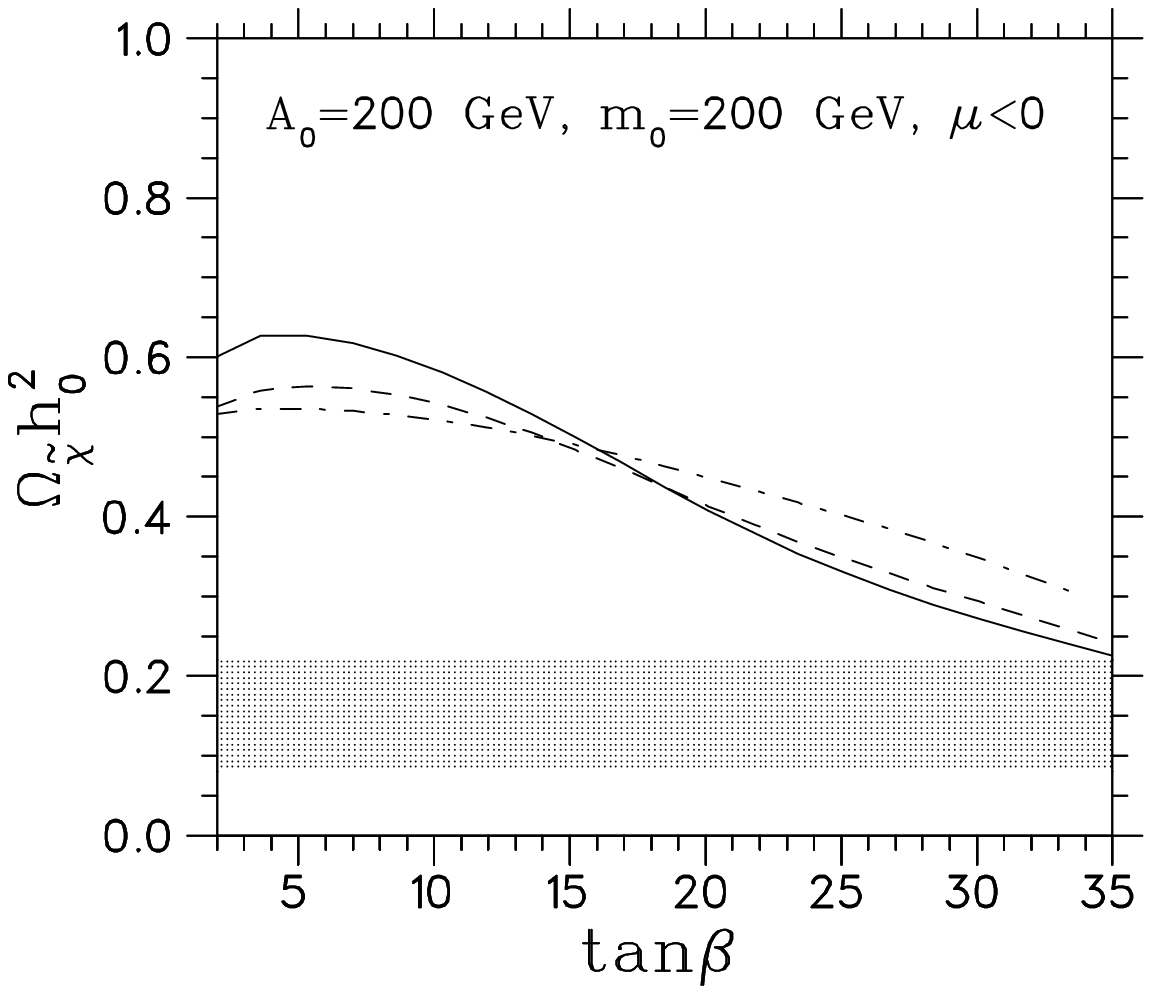,height=7.5cm,width=7.5cm} 

\begin{minipage}[t]{14.cm} 
\caption[]{The relic density as function of $\tan \beta$.
The lines are as in figure~\ref{fig2}.}
\label{fig4} 
\end{minipage} 
\end{center} 
\end{figure}

In figures~\ref{fig3} and \ref{fig4} we plot the {\small LSP} relic density as
function of the parameters
$A_0$ and $\tan \beta $ respectively by keeping, in each case, the other
parameters fixed. 
In figure~\ref{fig3} we see that for a relatively large value of the
parameter $m_0 = 200 \GeV$, and for all cases shown,  
the relic density takes unacceptably large values.
Although we have only depicted the case $m_0 = 200 \GeV $ this holds true
even for larger values of $m_0$, provided that $M_{1/2}$ stays
larger than about $150 \GeV$.

The behaviour of $\relic$, as the parameter $\tan \beta$
varies from 2 to 35, is depicted in figure~\ref{fig4}.
Keeping the parameters
$A_0$, $m_0$ fixed one observes that for large values of
$\tan \beta \geq 30$, 
$ \relic$ gets smaller falling below 0.22 even for large values of
the soft parameter $M_{1/2}$. The reason of getting small relic densities
for such large values of $\tan \beta$ is due to the fact that in these cases
the pseudoscalar Higgs boson $A$ has a mass close to $2\;m_{\lsp}$, and thus
its exchange dominates in the production of a fermion--antifermion 
pair in the final state.
This, along with the fact that $A \tau {\bar \tau}$
and $A b {\bar b}$ vertices are proportional to $\tan \beta$, enhances the
relevant cross sections, resulting to small relic densities within
the allowed cosmological limits.
This behaviour
agrees at least qualitatively with the findings of Ref.~\cite{Drees}
(see figure 4 in that reference).

\begin{figure}[t] 
\begin{center} 
\epsfig{file=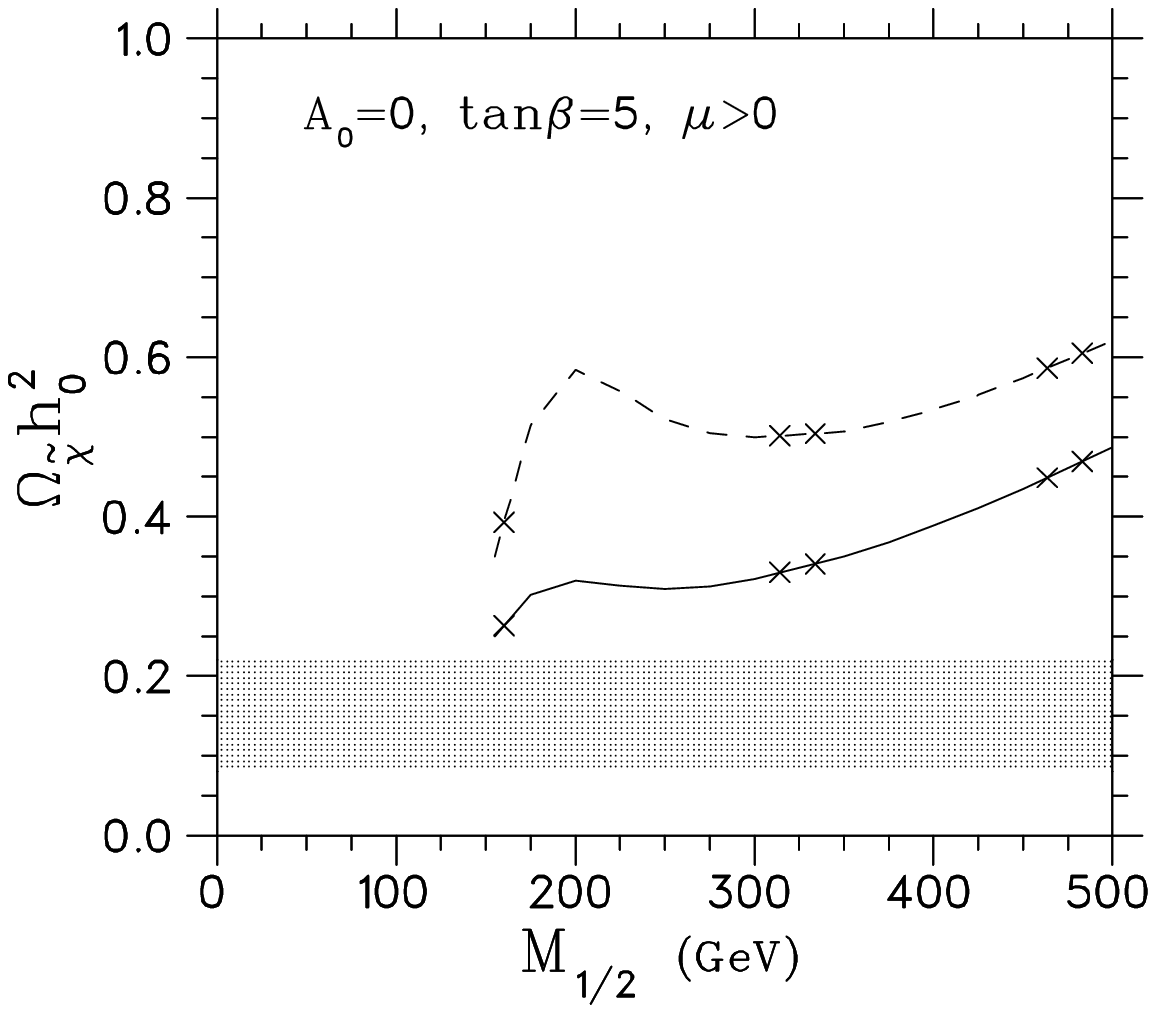,height=7.5cm,width=7.5cm} 
\hspace{.3cm}
\epsfig{file=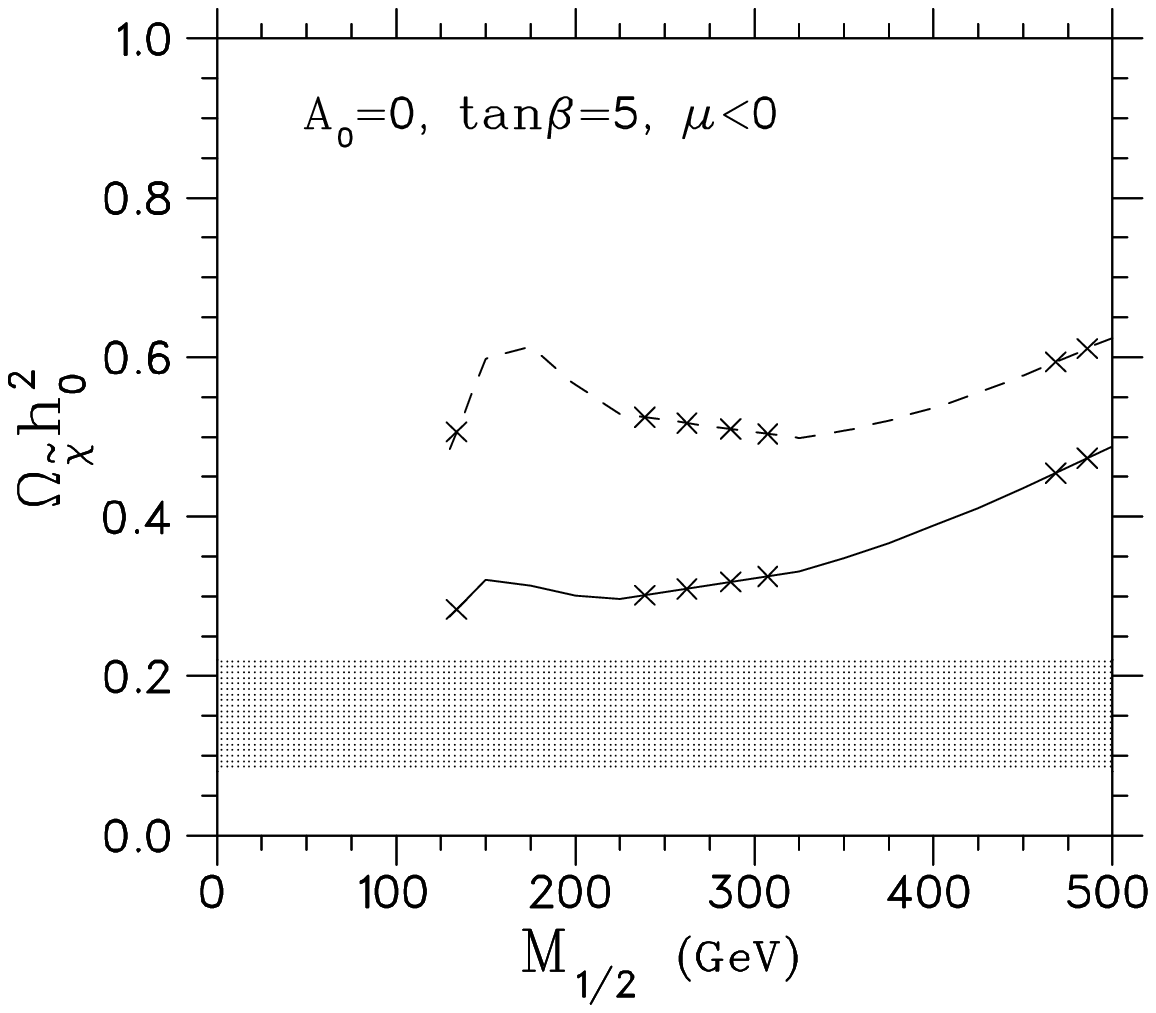,height=7.5cm,width=7.5cm}

\vspace{1cm}
\epsfig{file=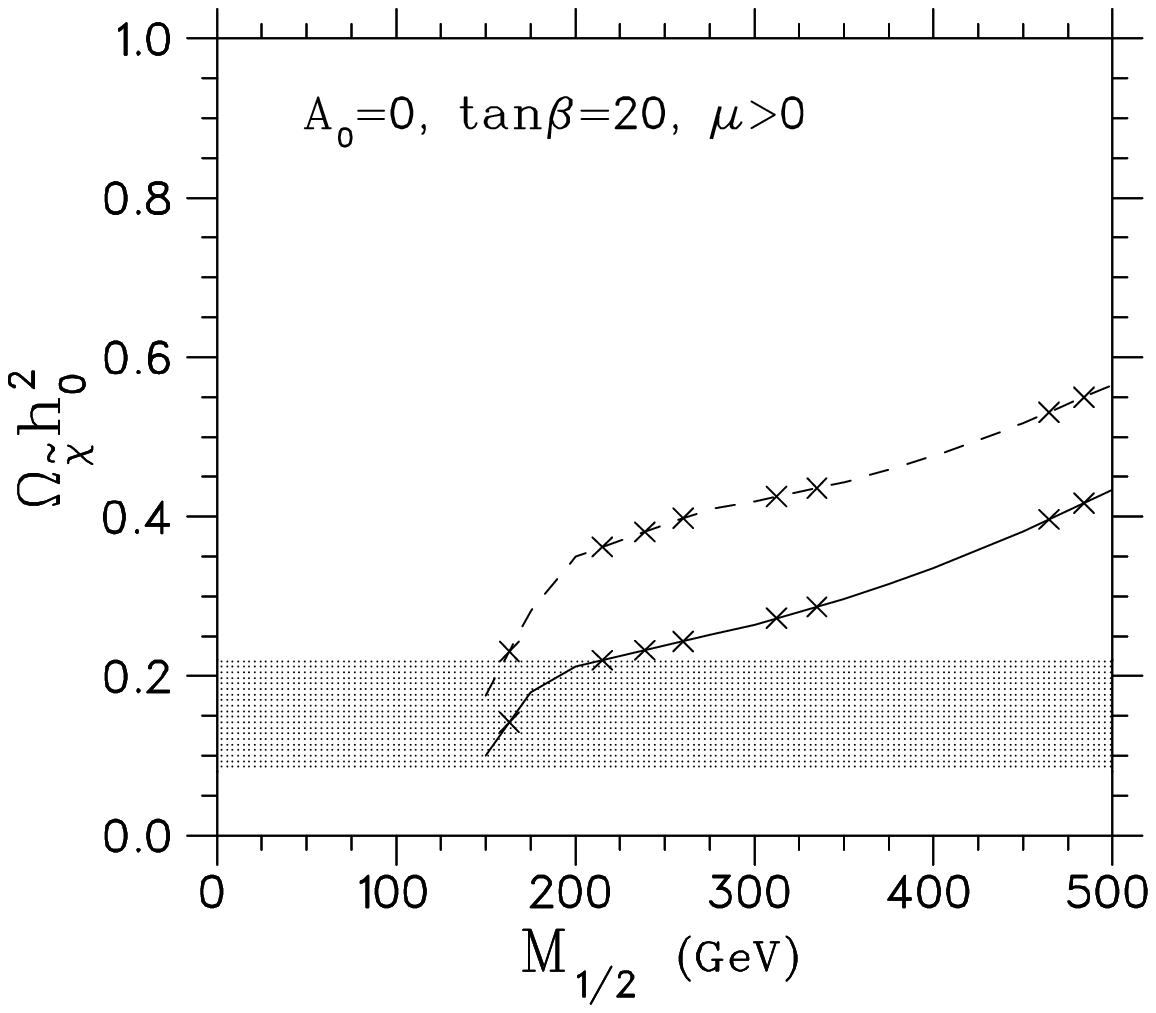,height=7.5cm,width=7.5cm} 
\hspace{.3cm}
\epsfig{file=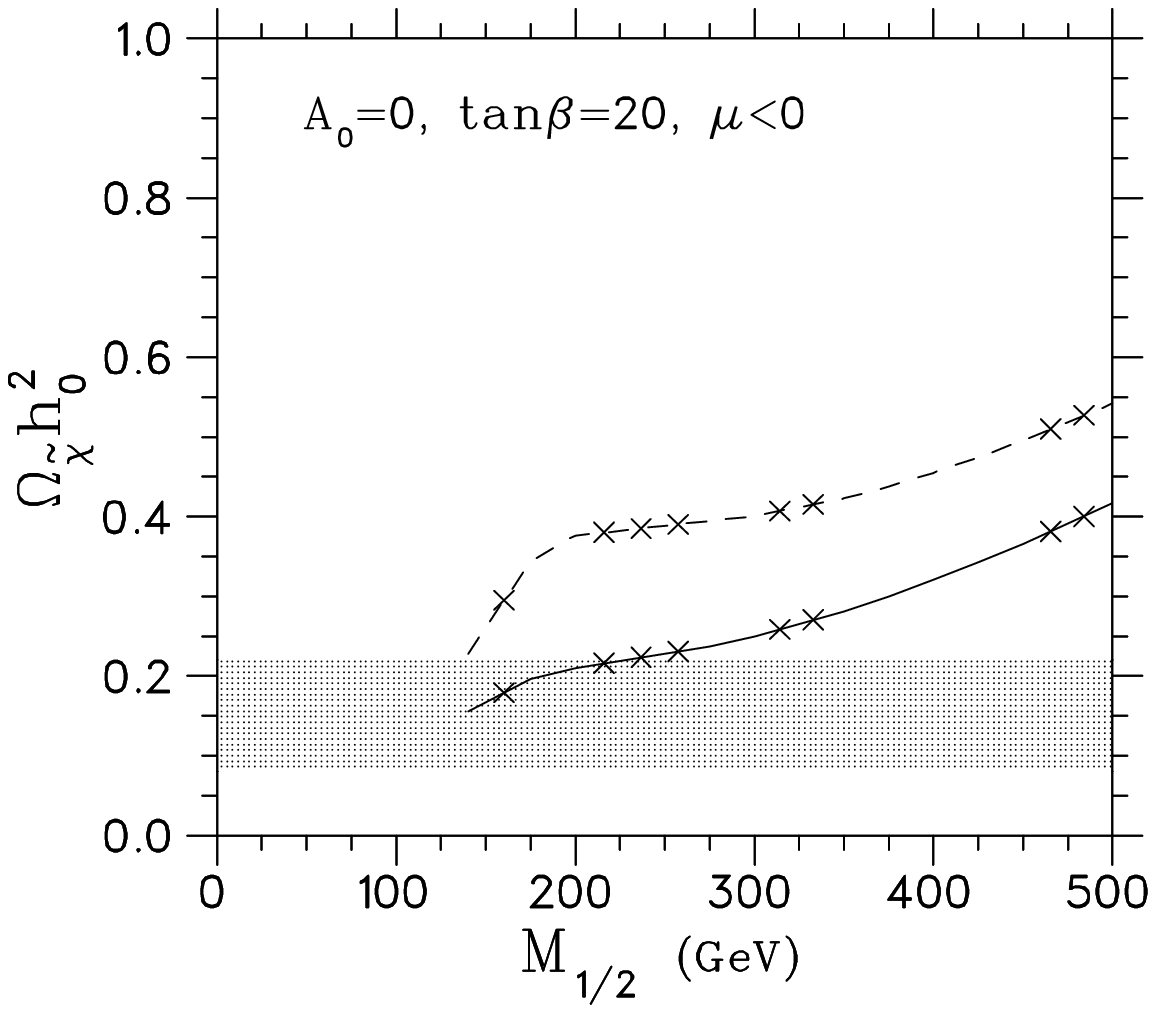,height=7.5cm,width=7.5cm} 

\begin{minipage}[t]{14.cm} 
\caption[]{The relic density as function of $M_{1/2}$.
Crosses denote points that  
are near thresholds or poles. The solid (dashed) line
corresponds to $m_0=150 \; (200) \GeV$.}
\label{fig5} 
\end{minipage} 
\end{center} 
\end{figure}

In figure~\ref{fig5} the {\small LSP} relic density is plotted as a
function of the
parameter $M_{1/2}$ for values of $A_0$, $\tan \beta$ shown on the figures, and
for $m_0 =150  \GeV$ (solid line) and $m_0 =200  \GeV$
(dashed line). 
The crosses denote points for which poles or thresholds are encountered.
It is obvious in these figures the tendency for the {\small LSP} relic density to
increase as $M_{1/2}$ increases especially for values $M_{1/2}>300\GeV$.
In this region, and for fixed
$M_{1/2}$, we observe that $ \relic$ decreases as
$\tan \beta$ is increased from $\tan \beta=5$ to $\tan \beta=20$.

So far in our analysis we have not 
studied neutralino--stau coannihilation effects,  
which if included can lower the values of the neutralino relic density
in some regions of the parameter space.
However as we shall see, even in those 
cases our calculation of relic density can be used to estimate with fair
accuracy the actual relic density by using the results of
Ref.~\cite{Falk1}.

These coannihilation processes are of relevance 
for values of the parameters for
which $m_{\lsp}<m_{{\tilde{\tau}}_R} \lesssim 1.2  m_{\lsp} $, that is 
near the edge where $\lsp$ and
$\tilde{\tau}_{R}$ are almost degenerate in mass.
Since so far in our analysis 
we have neglected such coannihilation effects, the conclusions reached are
actually valid outside the stripe
$m_{\lsp}<m_{{\tilde{\tau}}_R} \lesssim 1.2  m_{\lsp} $.
Inside this
band $\lsp-{\tilde{\tau}}$ coannihilations, and also
${\tilde{l}} - {\tilde{l}}$ annihilations, dominate the cross sections,
decreasing $\lsp$ relic densities leaving corridors of opportunity to
high $M_{1/2}$ and $m_0$ values as emphasized in other studies \cite{Falk1}.
Thus depending on the inputs $m_0$, $M_{1/2}$, $A_0$, $\tan \beta$ and
the sign of $\mu$ we can distinguish two cases:
\newcounter{biba}
\begin{list} 
{(\roman{biba})}{\usecounter{biba}} 
\item  $ 1.25 \; m_{\lsp} \leq m_{{\tilde{\tau}}_R} $ and    
\item $ m_{\lsp}<m_{{\tilde{\tau}}_R} < 1.25 \;  m_{\lsp} $,  
\end{list}
which are both compatible with having {\small LSP} as one of the neutralino states.
In region (ii) the stau ${\tilde{\tau}}_R$ is nearly degenerate in mass with
$\lsp$ and $\lsp-{\tilde{\tau}}$ coannihilation effects, and to a lesser
extend $\tilde{\tau} - \tilde{\tau}$, 
$\tilde{e} - \tilde{e}$ and $\tilde{\mu} - \tilde{\mu}$
annihilations, play an important role \cite{Falk1}.
We shall call this ``coannihilation" region
to be distincted from region (i) which will be designated hereafter as
``coannihilation free" region.

\begin{figure}[t] 
\begin{center} 
\epsfig{file=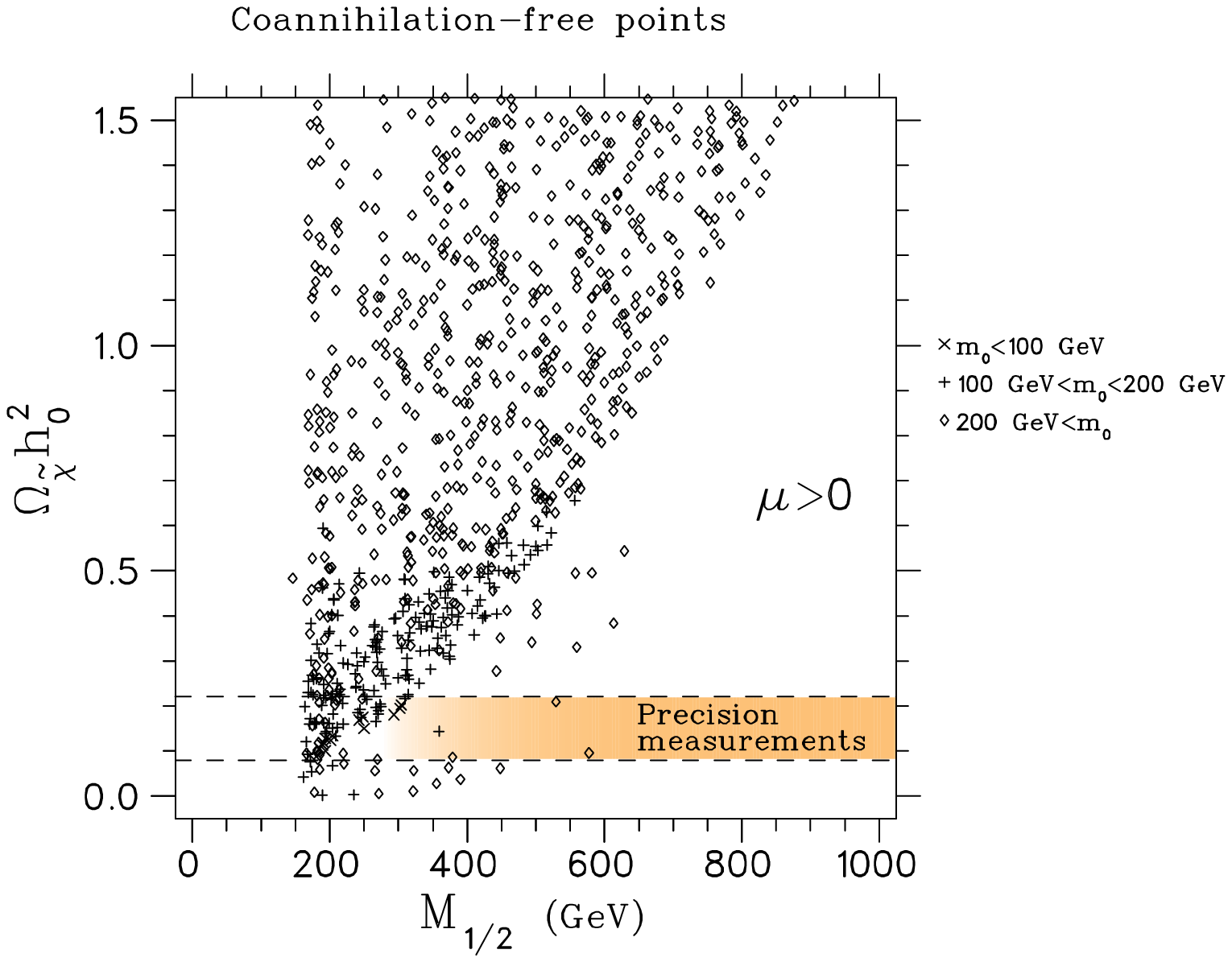,height=8.9cm,width=12cm}

\vspace{.5cm}
\epsfig{file=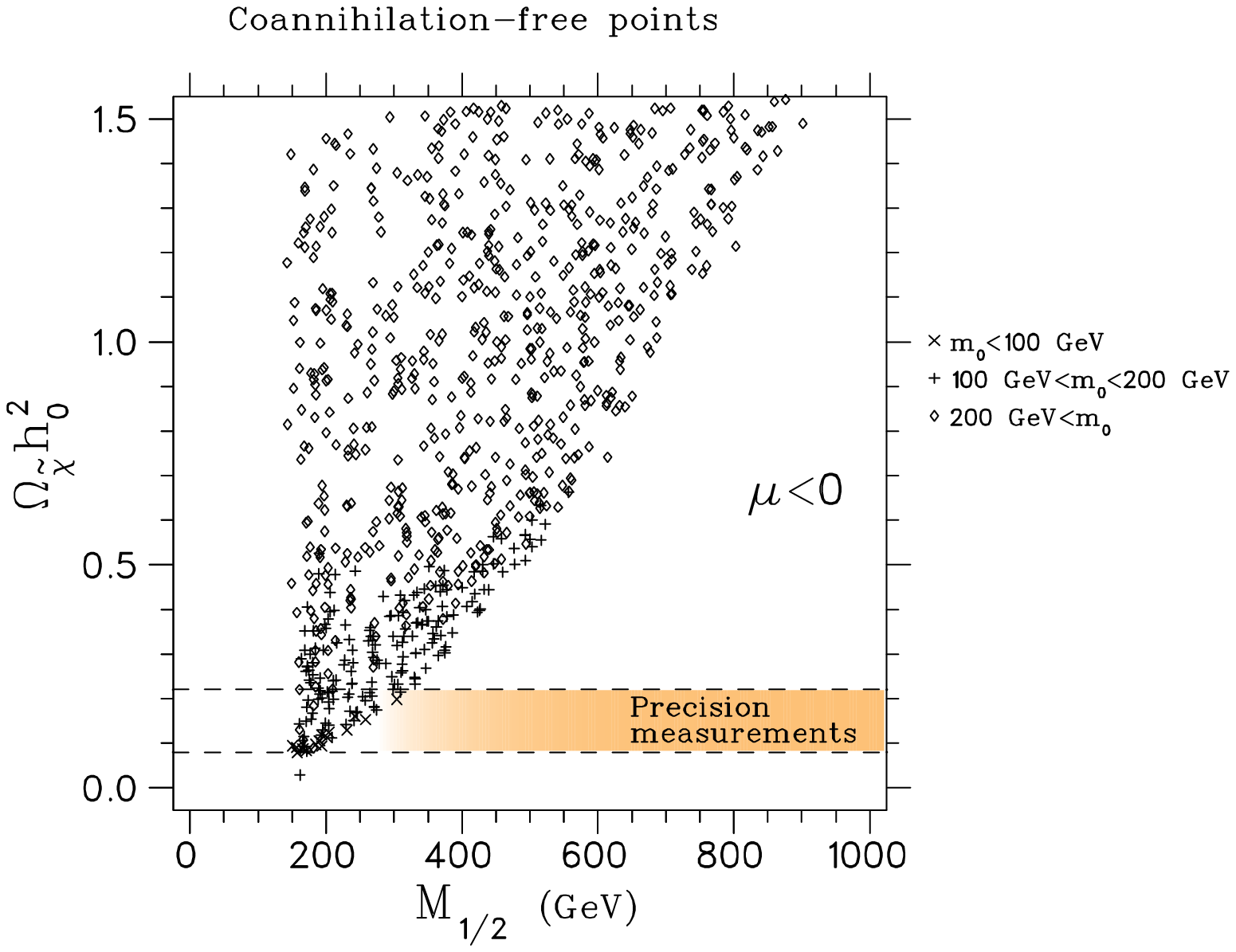,height=8.9cm,width=12cm}
\begin{minipage}[t]{14.cm}
\caption[]{Scattered plot of the relic density versus $M_{1/2}$ from
a sample of 4000 random points in the parameter space.
Low $M_{1/2}$ values are excluded by chargino searches.
All points shown
are in the coannihilation free region. Only the points with relic density
less than 1.5 are shown. The grey tone region within the cosmologically
allowed stripe designates the region which 
agrees with {\small EW} precision data (see main text).
The horizontal dashed lines mark the limits
$0.08 < \Omega_{\lsp} h_0^2 <0.22$
 .}
\label{fig6} 
\end{minipage} 
\end{center} 
\end{figure}

We shall first discuss the region 
(i) in which such effects are negligible and the
ordinary way of calculating $\lsp$ relic densities, with the omission
of the coannihilation processes, is very accurate and
reliable. In the coannihilation free region upper limits on $M_{1/2}$ and
$m_0$ can be established by imposing the cosmological constraint
$0.08< \relic < 0.22$, which are more strict than those discussed so far.
In fact within the coannihilation free region we find that for low and
moderate $\tan \beta$ the upper bounds on these parameters are
$M_{1/2} \lesssim 340\GeV$, $m_0\lesssim 200\GeV$. The upper limit set on
$m_0$ is correlated to the value of $M_{1/2}$, and is almost insensitive to
the value of the parameter $A_0$ and $\tan \beta$ as long as the latter does
not get values larger than about $\simeq 30$.
For instance the upper bound
$\simeq 200 \GeV$ on $m_0$ is reached when $M_{1/2} \sim 140 \GeV$, 
the lowest allowed by chargino searches, but
it is lowered to $\simeq 130 \GeV$ when $M_{1/2} \simeq 340 \GeV$.
%%%%%%%%%INSERT2
This behaviour is very clearly seen in the 
scattered plots shown in figure~\ref{fig6}. 
The sample consists of 4000 random points that cover the most interesting
part of the parameter space, which is within the limits:
$1.8<\tan\beta<40$, $150\GeV<M_{1/2}<1\TeV$, $|A_0|<500 \GeV$ and
$m_0<500 \GeV$\footnote{Higher values for $m_0$ are of relevance only
for low $M_{1/2}$ values, already
ruled out by the recent experimental bounds on chargino masses.
Also since $\relic$ does not depend strongly on $A_0$ for $M_{1/2}>150\GeV$,
as it can be realized from figure~\ref{fig3}, 
it suffices to focus on values $|A_0|< 500\GeV$.}.
 From the given sample
only points which lie entirely within the coannihilation free region are
shown. Also points  
which lead to relics larger than 1.5 are not displayed in the figure. 
The experimental bounds discussed before, restrict by about
$40 \%$ the values of the allowed points. 
The points shown are struck by a cross ($\times$) when
$m_0 < 100 \GeV$, by a plus (+) when  $100\GeV < m_0 < 200 \GeV$ 
and by a diamond ($\diamond$) when $m_0$ exceeds $200\GeV$. It is obvious 
the tendency to have $M_{1/2} \leq 340\GeV$ in the cosmologically interesting 
domain which lies in the stripe between the two lines at 0.08 and 0.22.
Actually 
except for a few isolated cases, which correspond to large $\tan \beta$ as
we shall see, all allowed points are accumulated
to values $M_{1/2} \leq 340 \GeV$
and $m_0 < 200\GeV$ (crosses or pluses). 

As a side remark, we point out that
the coannihilation free region under discussion overlaps
with the color and charge breaking 
({\small CCB}) free region as long as the parameter
$M_{1/2}$ stays less than $\simeq 300 \GeV$ \cite{Falk1}.

\begin{figure}[t] 
\begin{center} 
\epsfig{file=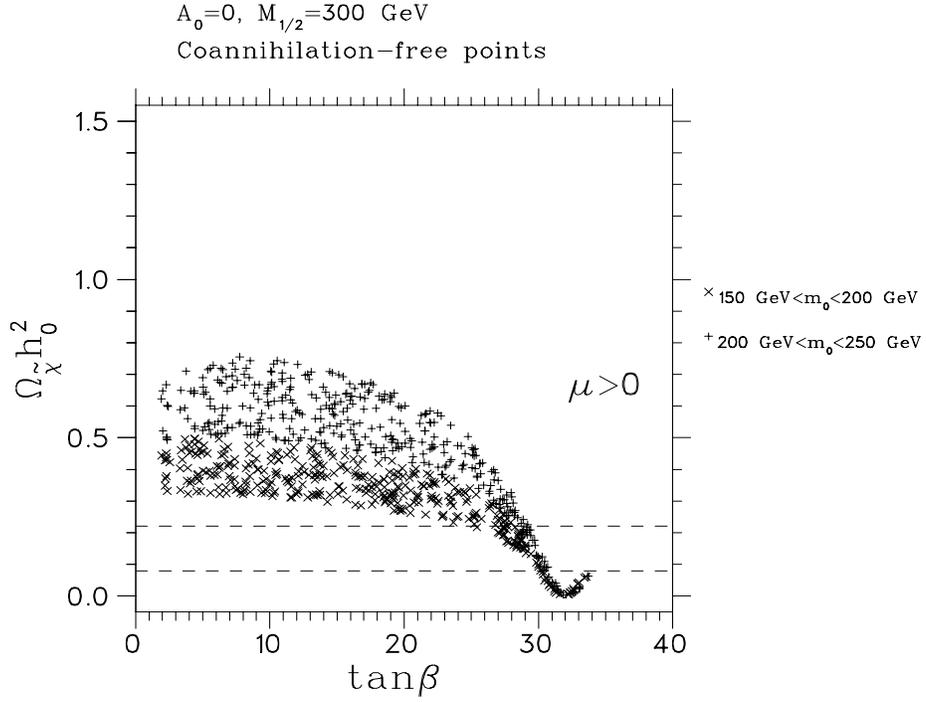,height=9.2cm,width=12.1cm} 

\vspace{1cm}
\epsfig{file=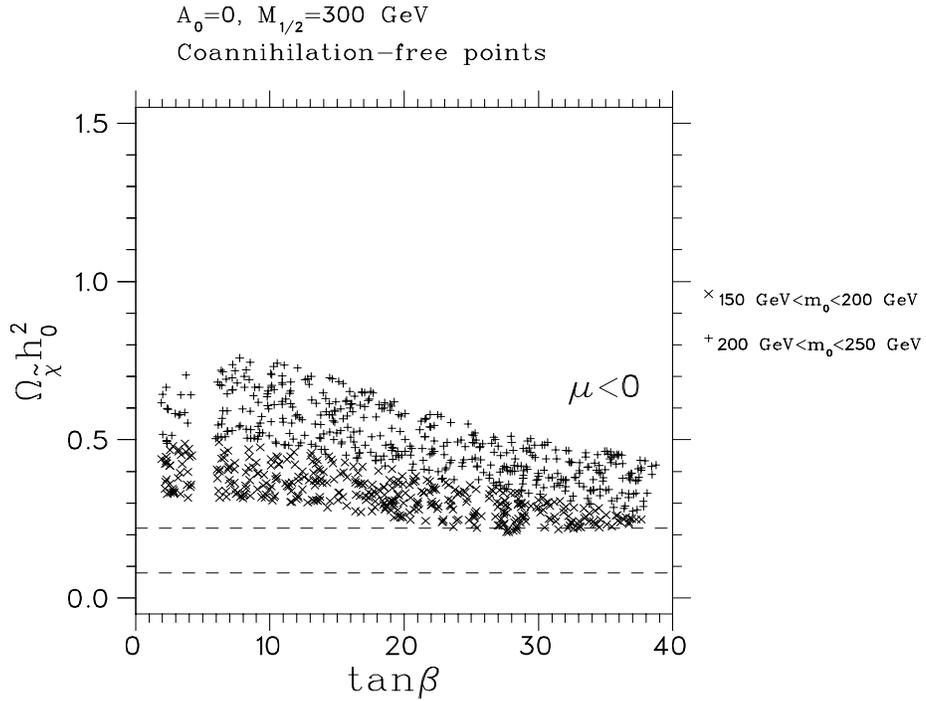,height=9.2cm,width=12.1cm} 

\begin{minipage}[t]{14.cm} 
\caption[]{Scattered plots of the relic density versus $\tan \beta$ from
a sample of random points with fixed $A_0 = 0\GeV$, 
$M_{1/2} = 300\GeV$.
The points shown fall within the coannihilation free region. 
The two horizontal dashed lines, as in figure~\ref{fig6}, 
mark the cosmologically allowed stripe. }

\label{fig7} 
\end{minipage} 
\end{center} 
\end{figure}

Anticipating a forthcoming discussion on {\small EW} precision data, 
we designate the region of $M_{1/2}$, which is rather favoured
by {\small EW} precision measurements. This is 
shaded in grey, which progressively becomes darker as we move to larger 
$M_{1/2}$ values, where the {\small SM} limit is attained.
In the coannihilation free region, these
upper limits set on $M_{1/2}$, $m_0 $ can be only evaded  
when $\tan \beta$ takes large values $\simeq 30$ and
$\mu$ is positive. Higher values of $\tan \beta$ can be also obtained at
the expense of changing the input value for the bottom pole mass as we
are discussing below.
In the aforementioned cases the pseudoscalar Higgs $A$ has a mass approaching
$2\;m_{\lsp}$,  and the $A \tau {\bar \tau}$, $A b {\bar b}$ couplings
are large as being proportional to $\tan \beta$. Both effects make 
the pseudoscalar Higgs exchange to dominate
the reactions $\lsp \;\lsp \goes \tau \; {\bar \tau}$ and
$\lsp \;\lsp \goes b \; {\bar b}$,
enhancing the corresponding cross sections
resulting to cosmologically acceptable relic densities
as already discussed
\footnote{ This requires the $A \lsp \lsp$ coupling to be non-vanishing.
This holds in regions of the parameter space where the {\small LSP} state
has a non-vanishing Higgsino component. 
}.
Such points allow for $M_{1/2}$, $m_0$ as large as $\simeq 450 \GeV$ 
and stay comfortably well as far as the process $b \goes s \;\gamma$
is concerned, which is not in conflict with large $\tan \beta$ values as
long as $\mu > 0$ \cite{recent,bsg}.
Since large values of $\tan \beta $ are compatible with Yukawa
coupling unification, the previous discussion shows that the possibility of
obtaining acceptable $\lsp$ relic densities in the coannihilation free
region is feasible in such schemes.
If Yukawa coupling unification is enforced, the input $b$-quark pole
mass should be lowered to values that are marginally consistent with the
experimental data. This has as an effect the increase of the value of  
$\tan \beta$. In fact by lowering the input value
$M_b^{pole} \ = \ 5 \GeV$, we were able to get relic
densities within the cosmologically allowed domain $0.08< \relic<0.22$ for 
$\tan \beta \approx 50$, without the need of invoking the
coannihilation mechanism as is done in Ref.~\cite{Lazarides}. 
Note the important role the pseudoscalar
Higgs boson plays in this case since it dominates the
$\lsp \; \lsp \goes \tau {\tilde \tau},b {\tilde {b}}$ reactions when the
{\small LSP}'s composition involves even a small Higgsino component. 
In figure~\ref{fig7}, and in order to exhibit the $\tan \beta$ behaviour of
the relic density, we display a scattered plot of random points, for fixed
$A_0$, $M_{1/2}$ and random values of $m_0 \geq 150 \GeV$, 
%%%%%and $ \tan \beta$,
as function of
$\tan \beta$, for both signs of $\mu$.
All points displayed refer to the coannihilation free region under discussion.
Actually for the $\mu>0$ case, only a few points of the given
sample are in the coannihilation region.
We see that only a small number of points with
$\mu < 0$ can marginally satisfy the cosmological constraints.
However for $\mu >0$ many such points 
exist for values of $\tan \beta $ which are around $\simeq 30$.
We recall that the bottom quark pole mass has been taken equal to $5\GeV$
which hardly allows for large values of $\tan \beta$. For this reason, 
and for the given sample, points beyond
${\tan \beta} \simeq 35 \ (38)$ for $\mu >0 \ (< 0)$ are absent in these
figures. In the bottom figure,
corresponding to $\mu < 0$ case, we do not display points in the gap around 
$\tan \beta \ = \ 5$, since we are close to a
two light Higgs threshold (see discussion in section III).

\begin{figure}[t] 
\begin{center} 
\epsfig{file=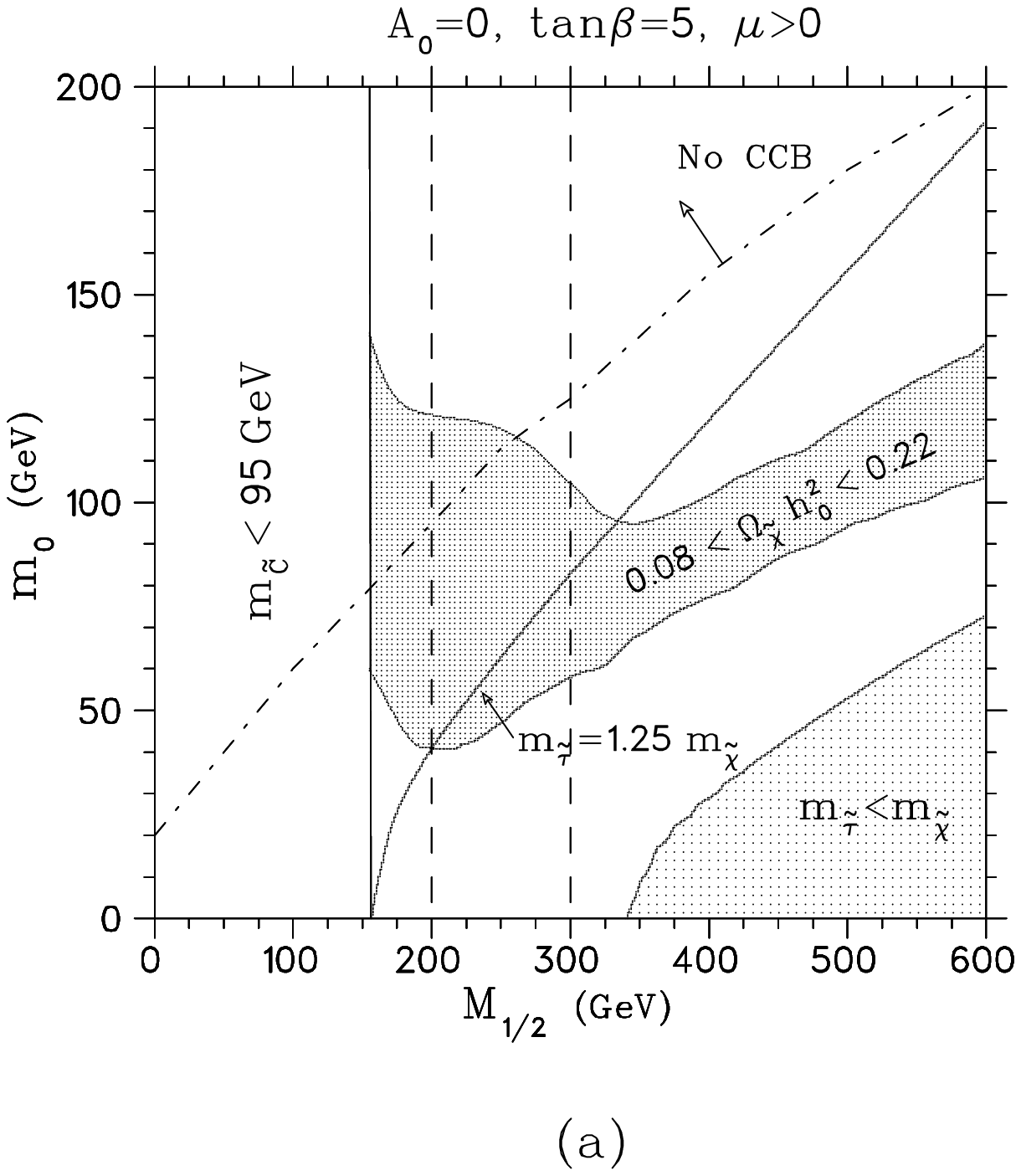,height=8.5cm,width=7.5cm} 
\vspace{.7cm} 

\epsfig{file=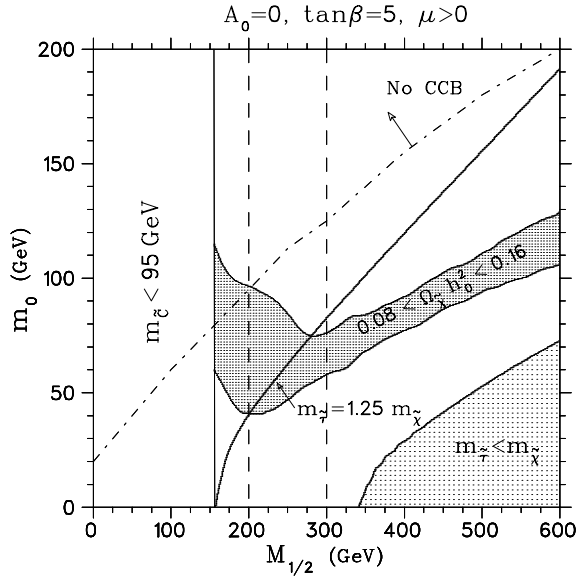,height=8.5cm,width=7.5cm}  

\begin{minipage}[t]{14.cm} 
\caption[]{The dark-shaded area in figure (a) (figure (b))
designates the cosmologically allowed region
$\Omega_{\lsp} h_0^2 \;=\;0.15 \pm 0.07\;(0.12 \pm 0.04) $. 
The boundary of the coannihilation free region is labelled by
$m_{\tilde{\tau}} =  1.25 \; m_{\tilde{\chi}}$. Also shown is the
region in which
$m_{\tilde{\tau}} < m_{\tilde{\chi}}$, shaded in light-grey tone.
The boundary of the region which is free of color and charged breaking
minima, marked as ``No {\small CCB}", 
is also shown. The vertical dashed lines represent
the boundaries of the regions $M_{1/2}>200 \GeV$ and $M_{1/2}>300 \GeV$.
}
\label{fig8} 
\end{minipage} 
\end{center} 
\end{figure}

{\small EW} precision data are in perfect agreement with the 
{\small SM} and hence also with supersymmetric 
extensions of the {\small SM} which are
characterized by a large supersymmetry breaking scale. In unconstrained
{\small SUSY} scenarios 
the bounds put on sparticle masses from the {\small EW} precision
data are not far from their lower experimental limits. In constrained
versions, 
such as the {\small CMSSM} 
which we study here,
lower bounds on $M_{1/2}$ can be established. In fact
phenomenological studies of the weak mixing angle ${\sin^{2}}\theta_{eff}$
restrict $M_{1/2}$ to lie in the region $M_{1/2} \geq 300\GeV$ if the
combined {small SLD} and {\small LEP} data are used 
for ${\sin^{2}}\theta_{eff}$\footnote{The SLD data alone leave more
freedom by allowing for lower $M_{1/2}$ values. On the contrary small LEP data
favour large $M_{1/2}$ values.}.
If in addition unification 
of gauge couplings at $M_{\mathrm{GUT}}$ is assumed then the
lower bound is shifted to higher values
(see Dedes {\etal} in Ref.~\cite{Lahanas}), in the absence
of high energy thresholds.
Therefore in the
context of the {\small CMSSM} it seems that 
{\small EW} precision data favour rather large
$M_{1/2}$ values in which case we are closer to the {\small SM} limit of
Supersymmetry. The higher the $M_{1/2}$ value the lower the $\chi^2$ is, and 
better agreement with the experimental data is obtained.
Adopting a lower bound of about $300\GeV$,
suggested by the above reasoning, can have a dramatic effect for the allowed
domain which lies entirely in the coannihilation free region. For low
$\tan \beta$ ($\lesssim 10$) the cosmologically allowed 
region is severely constrained almost
predicting the values of the soft masses. In fact $M_{1/2}$ is forced to
move within the rather tight limits
$M_{1/2} \simeq 300 \GeV - 340 \GeV$,
while at the same time $m_0 \simeq 80 \GeV - 110 \GeV$.
For higher values of $\tan\beta$ ($\sim 20$) the upper
bound on $m_0$ is sifted upwards by about $20 \GeV$ (see for
instance figure~\ref{fig1}).
This situation is depicted 
in figure~\ref{fig8}a where in the 
$(M_{1/2}$, $m_0)$ plane the dark-shaded
 area marks 
the cosmologically allowed region 
for values $A_0 = 0\GeV$, $\tan \beta = 5$. 
The coannihilation free region
under discussion lies above the line labelled by
$m_{{\tilde{\tau}}}  = 1.25 \; m_{\lsp} $. In this figure
it is seen that by enforcing a more relaxed lower bound,  
$M_{1/2} \geq 200\GeV$, not excluded by {\small SLD} data, 
a relatively large portion in the $(M_{1/2},m_0)$ plane is 
allowed which also overlaps
with regions in which neither color nor charge
are violated 
(marked as ``No {\small CCB}"\footnote{The alert 
reader may notice that the overlap between the
``No {\small CCB}" allowed region and the coannihilation region is of measure
zero, at least for $M_{1/2}$ less than $500 \GeV$. 
This trend may be very
suggestive in looking for the physically sound region in parameter
space.})\cite{Abel}.
However for $M_{1/2} \geq 300\GeV$ the allowed region, 
in the coannihilation free
domain, is shrunk to a small triangle.

%%%%%%% INSERTION %%%%%%%%%%%%%%%%%%%%%%%%%%%%%%%%%%%%%%%%%%%
The previously discussed 
bounds on $M_{1/2}, m_0$ affect the mass spectrum of supersymmetric
particles. For
$M_{1/2} = 300 \GeV - 340 \GeV$, $m_0 = 80 - 110 \GeV$,
$|A_0|<1 \TeV$ and
values of $\tan \beta < 10$, we have found the following bounds on the
masses of the {\small LSP} and 
the lighter of charginos, staus, stops and Higgs scalars:
%%%%%%%%%%%%%%%%%%%%
\bea  
m_{\mathrm{LSP}}\;&:&\; 115 \; (116) \GeV\;-\;130 \; (133)\GeV\,,\nonumber \\ 
m_{\tilde{C}}\;\;\;\;&:&\; 210 \; (218) \GeV\;-\;241 \; (250) \GeV\,,\nonumber \\ 
m_{{\tilde{\tau}}_R}\;\;\;&:&\;122 \; (130) \GeV\;-\;157 \; (158) \GeV,\nonumber \\ 
m_{{\tilde{t}}_1}\;\;\;&:&\;401 \; (403) \GeV\;-\;667 \; (687) \GeV\,,\nonumber \\ 
m_{{h}_0}\;\;\;&:&\;96 \; (87) \GeV\;-\;125 \; (122) \GeV\,.\nonumber  
\eea
%%%%%%%%%%%%%%%%%%%%%
These refer to the case $\mu>0$ ($\mu<0$). 

In order to see how the bound put on {\small SUSY} breaking parameters,
and hence the sparticle masses are affected, if the more
stringent cosmological limits quoted in Ref.~\cite{ostriker} are employed,
in the figure~\ref{fig8}b we have drawn 
the same situation as in figure~\ref{fig8}a
with $\relic=0.12 \pm 0.04$.
One notices that the decrease of the upper
bound on $\relic$ from 0.22 to 0.16 washes out the allowed  points in 
the coannihilation free region, if the lower bound $M_{1/2}>300\GeV$ is
enforced.

Within the {\small CMSSM} the only option to evade the stringent bounds put
on supersymmetry
breaking parameters, and hence on sparticle masses, remains
either to move to the large $\tan \beta$ regime, which we discussed
previously, or to go to the coannihilation region
$m_{\lsp}<m_{{\tilde{\tau}}_R} < 1.25 \; m_{\lsp} $ in which case $M_{1/2}$
is not actually bounded \cite{Falk1}. 

%%%%%%%%%%%%%% END INSERTION %%%%%%%%%%%%%%%%%%

\begin{figure}[t] 
\begin{center} 
\epsfig{file=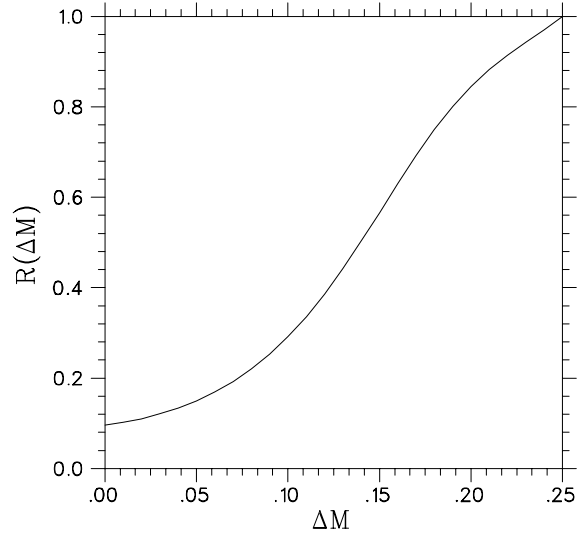,height=7cm,width=7.5cm} 

\begin{minipage}[t]{14.cm} 
\caption[]{The reduction factor $R(\Delta M)$ as function of
$\Delta M = (m_{{\tilde \tau}_R} - m_{\lsp})/m_{\lsp} $. }
\label{fig9} 
\end{minipage} 
\end{center} 
\end{figure}

\begin{figure}[t] 
\begin{center} 
\epsfig{file=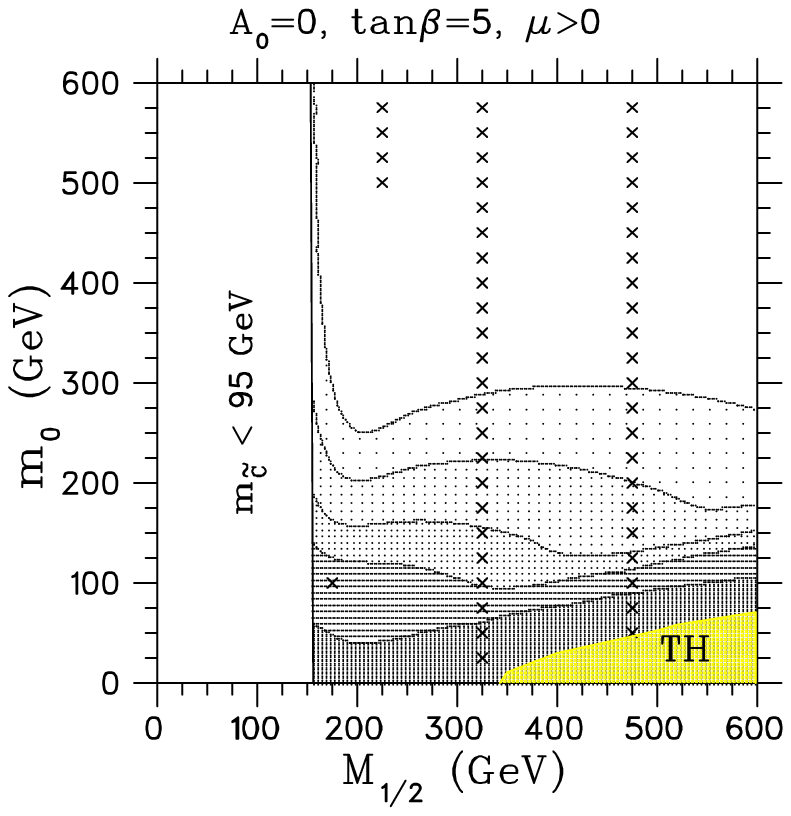,height=7.5cm,width=7.5cm} 
\hspace{.3cm}
\epsfig{file=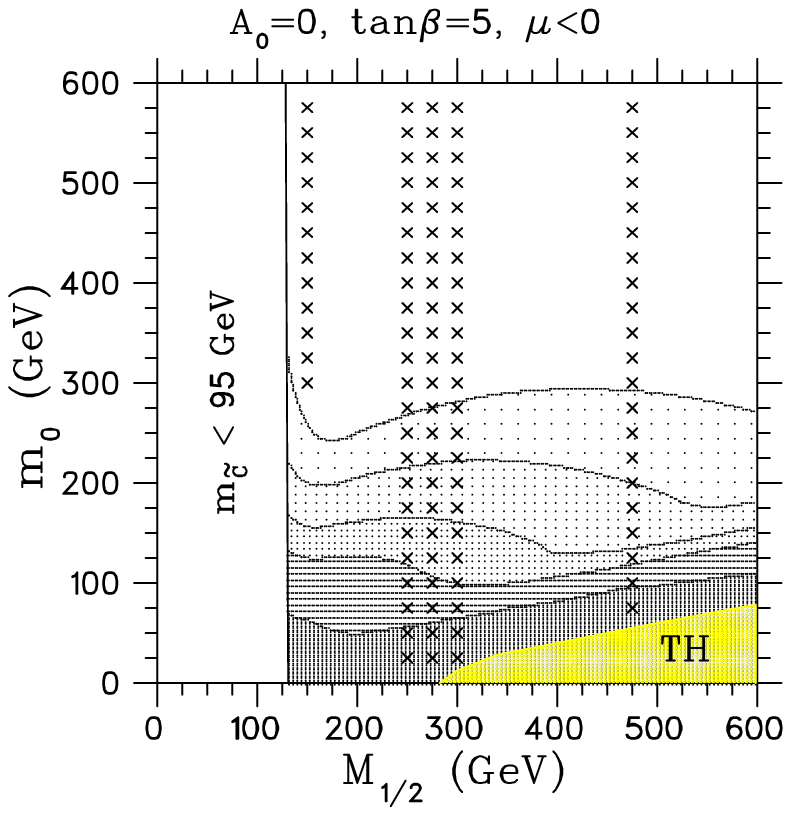,height=7.5cm,width=7.5cm}

\vspace{1cm}
\epsfig{file=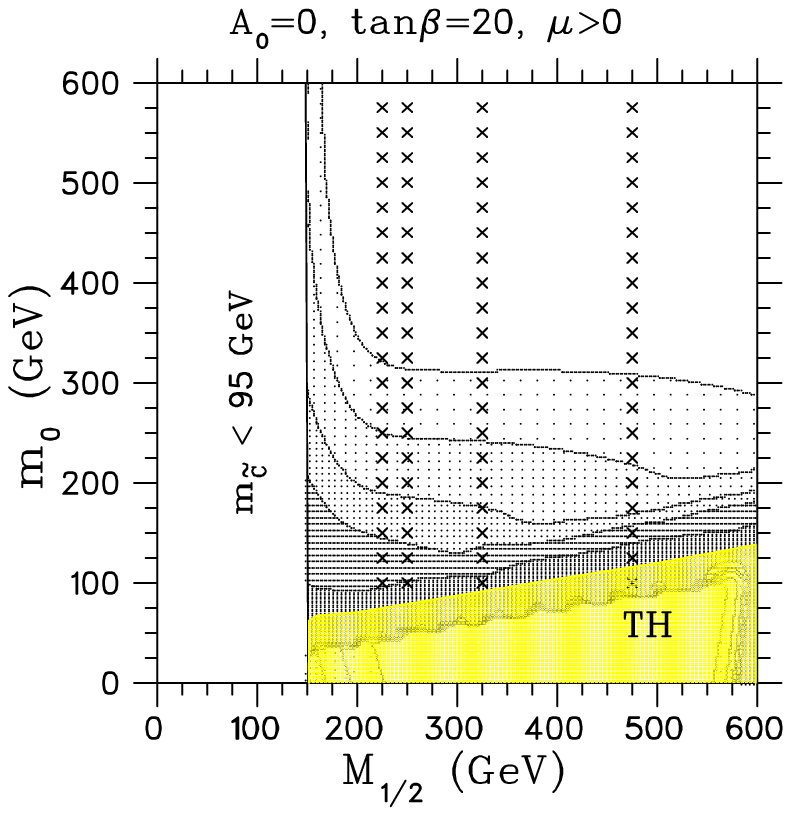,height=7.5cm,width=7.5cm} 
\hspace{.3cm}
\epsfig{file=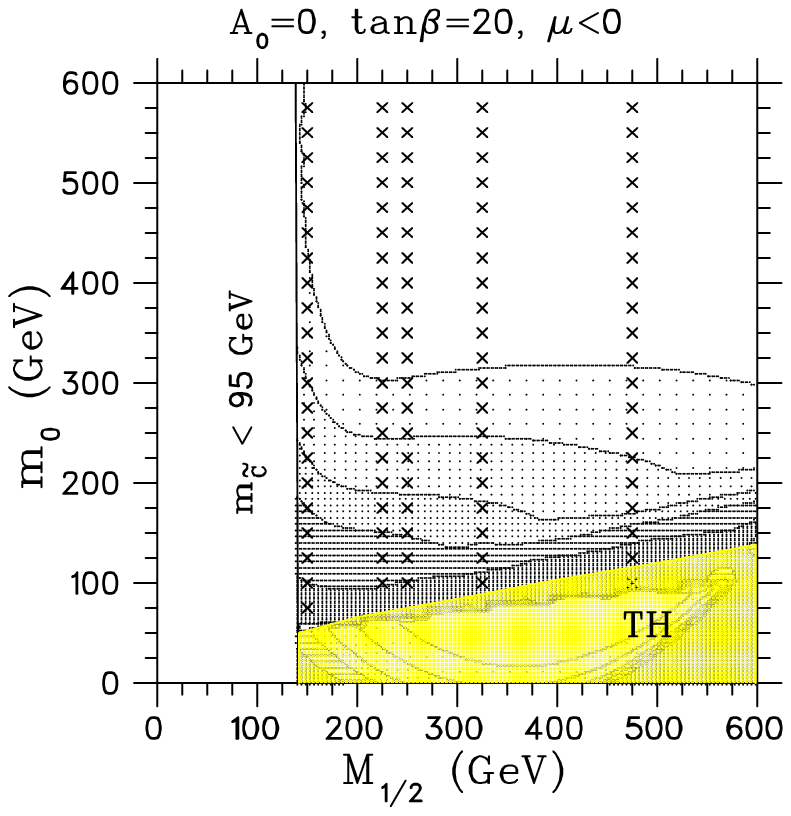,height=7.5cm,width=7.5cm} 

\begin{minipage}[t]{14.cm}  
\caption[]{
The {\small LSP} relic density
$\relic$ in the ($m_0$,$M_{1/2}$) plane for
given values of $A_0$, $\tan \beta$ and sign of $\mu$ when coannihilation
effects are taken into account. The inputs are the same as in
figure~\ref{fig1}. }

\label{fig10}  
\end{minipage}  
\end{center}  
\end{figure}

In the coannihilation region (ii) our results concerning the neutralino relic
density do
 not hold any more.
However the conclusions of our analysis and that presented in
Ref.~\cite{Falk1} can be both combined to infer information on the actual
relic density, $\relic$, from the one have calculated which we shall
hereafter denote by $\Omega_{\lsp}^0\;h_0^2$.
Using the findings of this reference we can express the actual relic density
as
%%%%%%%%%
\bea
\Omega_{\lsp} \;=\; R  (\Delta M) \;\Omega_{\lsp}^{0} \,,  \label{red}
\eea
%%%%%%%%%%%%%%%%
where the reduction factor $ R (\Delta M)$ depends on
$\Delta M \;=\; (m_{{\tilde {\tau}}_R} - m_{\lsp})/m_{\lsp} $ and
is plotted in figure~\ref{fig9}. It is seen that $ R (\Delta M)$ smoothly
interpolates between $\approx 0.1$ and 1.0 for values of 
$\Delta M $ in the range $0.00 - 0.25$.
The above equation is a handy device and reproduces 
the results cited in Ref.~\cite{Falk1}. The cosmologically allowed
domain shown in figures~\ref{fig8} has been actually drawn using this
equation.
In figure~\ref{fig10} we see how the
contours of figure~\ref{fig1} are distorted when Eq.~(\ref{red}) is implemented.
Notice the change of the shape at the bottom of the figure where the mass of 
${\tilde \tau}_R$ starts approaching that of the {\small LSP}.

\begin{figure}[t] 
\begin{center} 
\epsfig{file=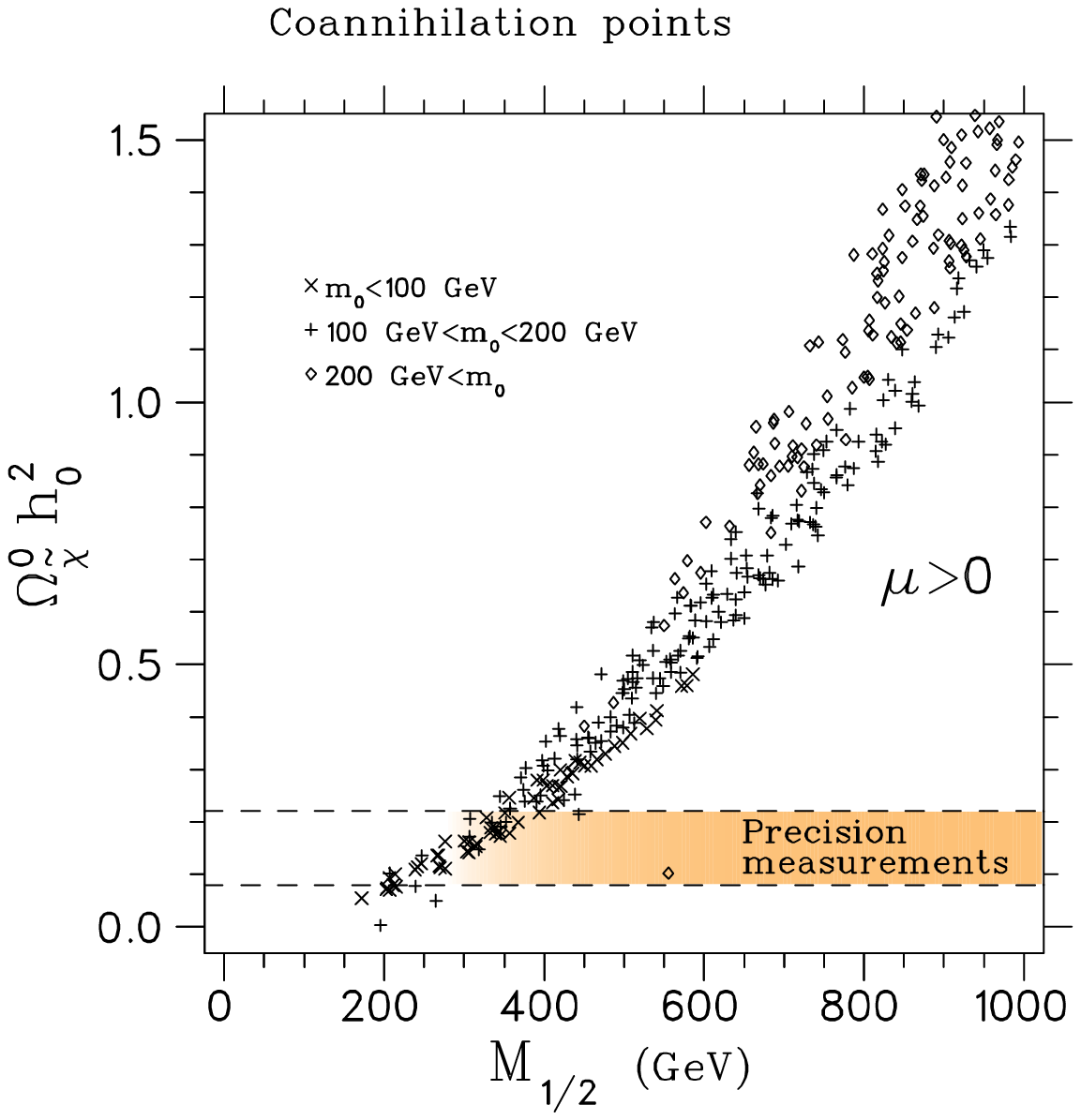,height=8cm,width=7.7cm} 
\hspace{.15cm}
\epsfig{file=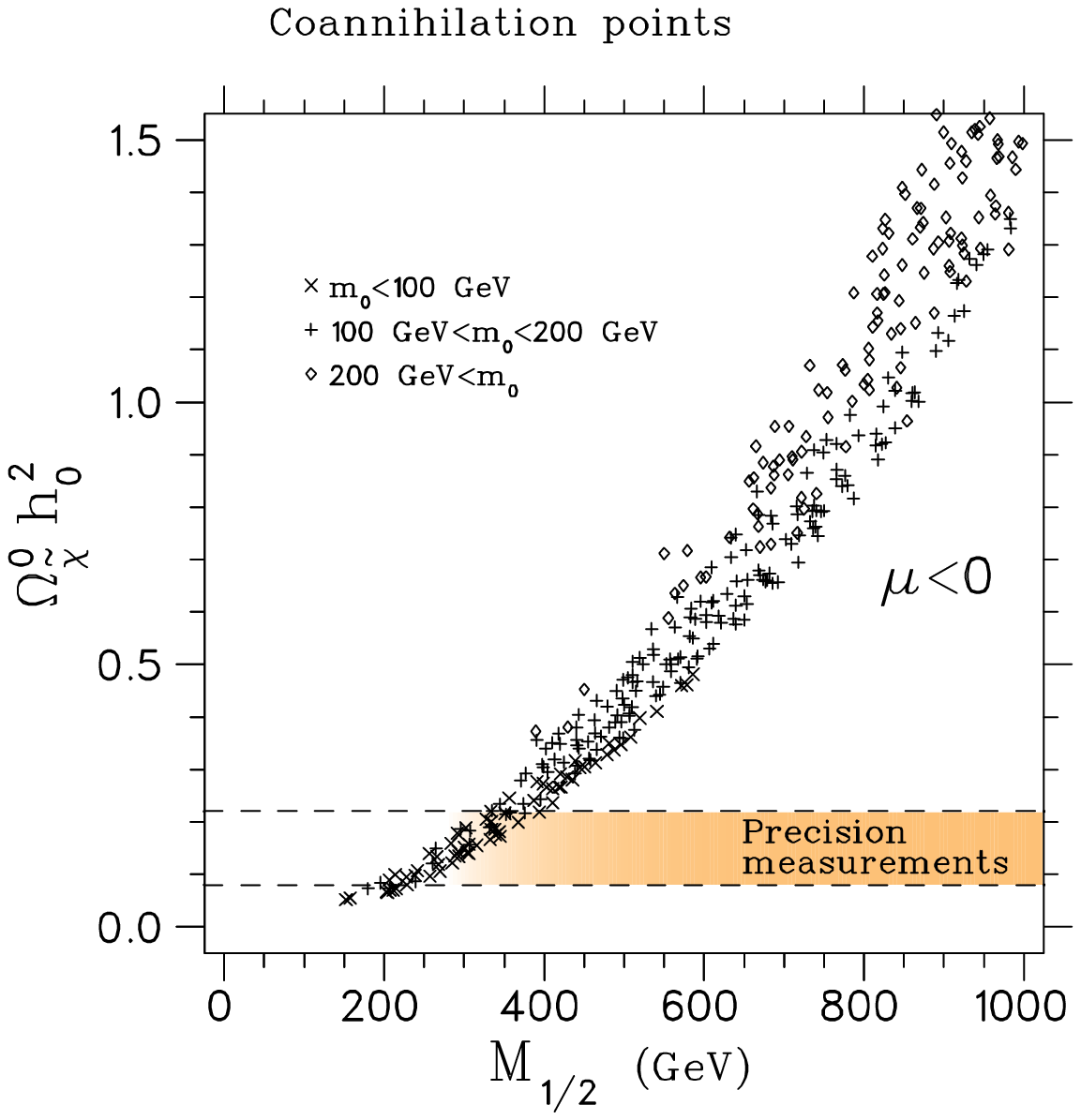,height=8cm,width=7.7cm}

\vspace{1cm}
\epsfig{file=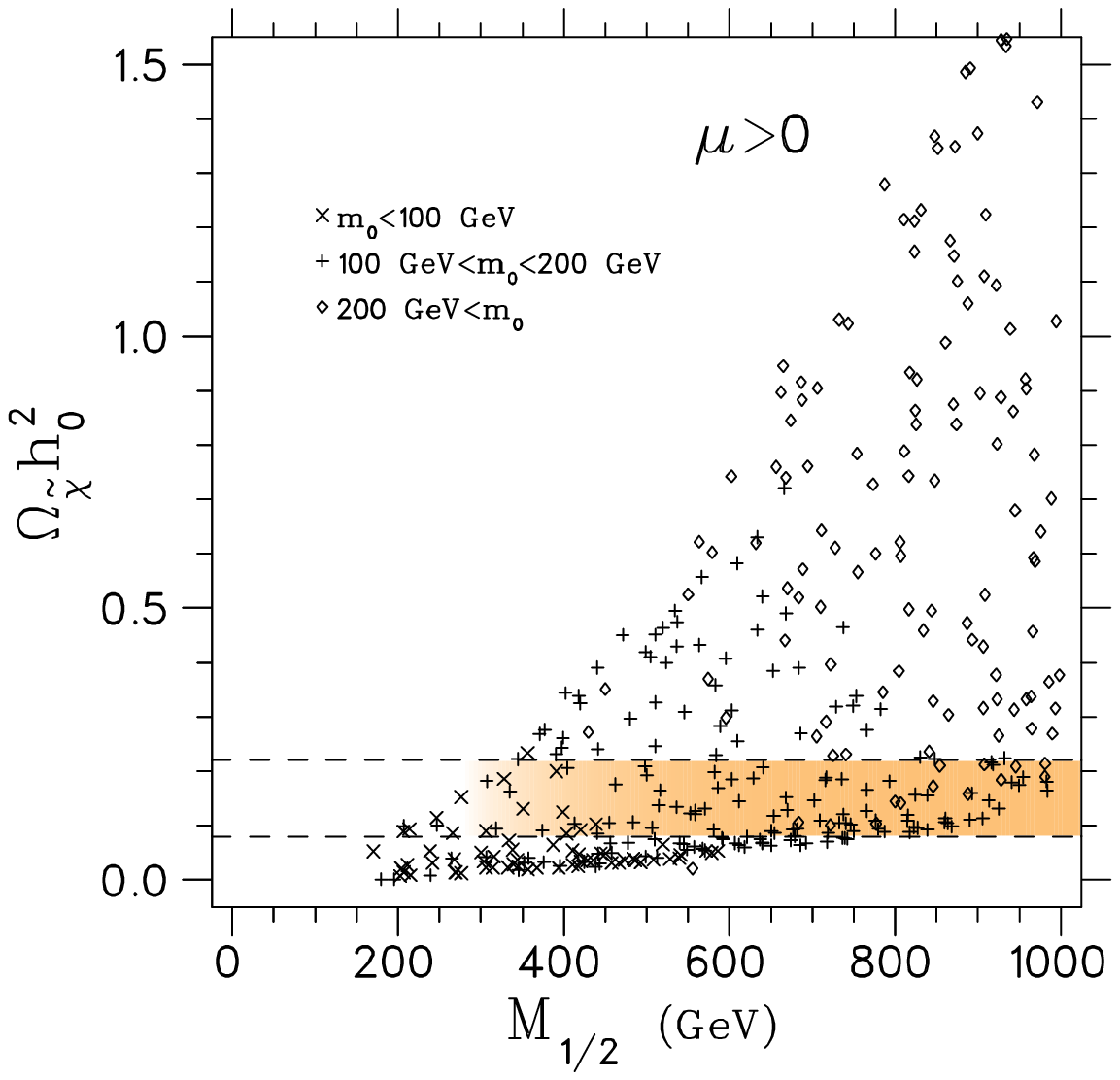,height=7.5cm,width=7.7cm} 
\hspace{.15cm}
\epsfig{file=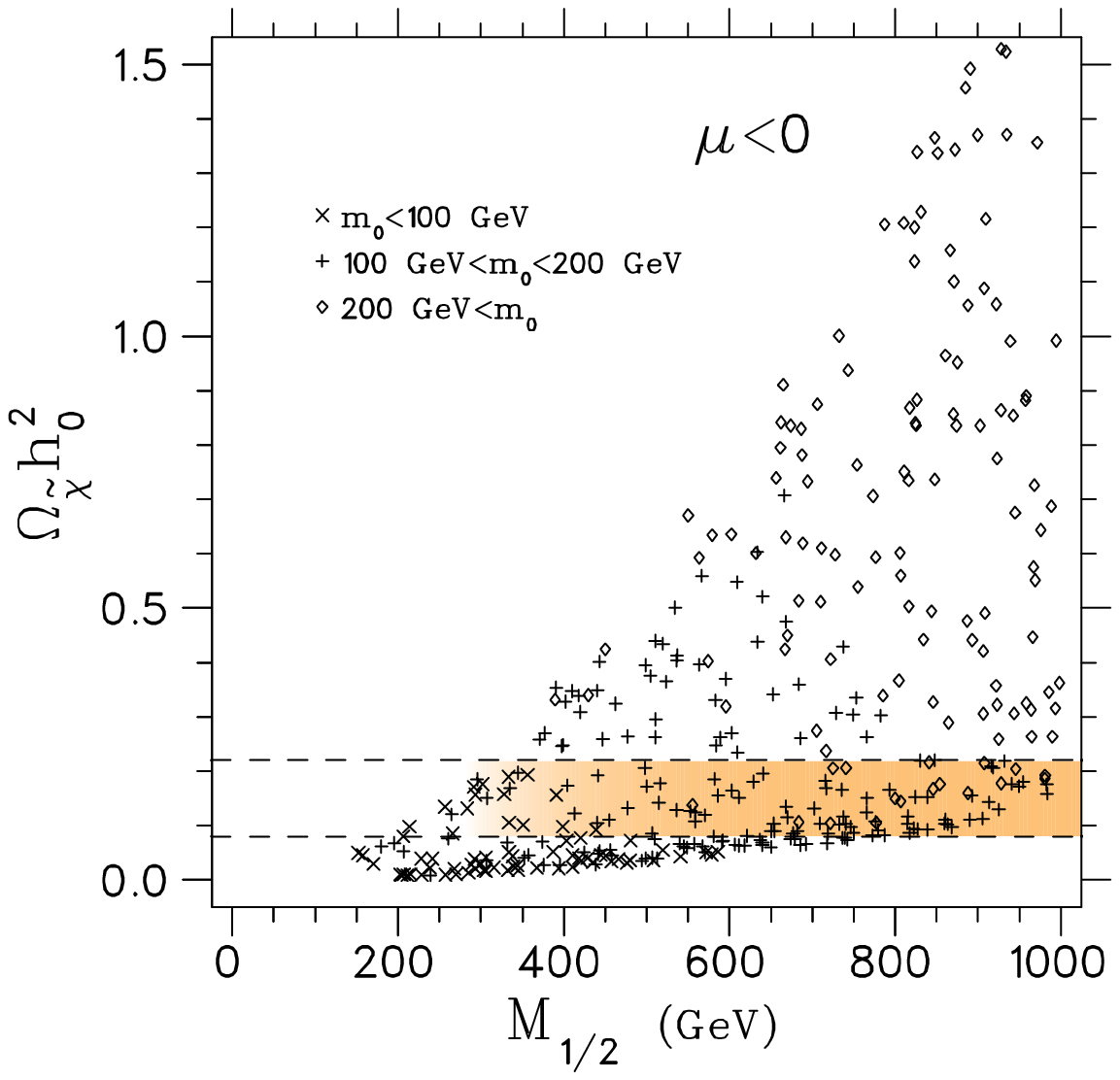,height=7.5cm,width=7.7cm} 

\begin{minipage}[t]{14.cm}  
\caption[]{The first two figures on the top show the points,  
from  a random sample of 4000 points, 
which lie entirely within the coannihilation region. The vertical
axis refers to $\Omega_{\lsp}^{0} \ h_0^2$ (see main text). The
figures at the bottom represent the same situation for the actual
relic density $\Omega_{\lsp} \ h_0^2$. }

\label{fig11}  
\end{minipage}  
\end{center}  
\end{figure}

In the scattered plot of figure~\ref{fig11} we show all points of the random
sample which were previously used for the production of figure~\ref{fig7}, 
which lie strictly within the coannihilation region.
These points were not displayed in the figure~\ref{fig7}.
In the figures at the top
the vertical axis refers to values of the relic density which is based
on our own calculation ($\Omega_{\lsp}^{0} \ h_0^2$). The second set of the
figures, at the bottom,  
shows how some of these points collapse, if the Eq.~(\ref{red}) is used,
falling within the cosmologically allowed stripe allowing for high
$M_{1/2}$ (and $m_0$), values. The vertical axis now refers to the actual
relic density ($\relic$).

%%%%%%%%%%%%%%%%%%%%%%%% SECTION CONCLUSIONS  %%%%%%%%%%%%%%%%%%%%%%%%%%
%%%%%%%%%%%%%%%%%%%%%%%%%%%%%%%%%%%%%%%%%%%%%%%%%%%%%%%%%%%%%%%%%%%%%%%%
\section{Conclusions}
In this paper we have evaluated the relic neutralino abundance 
in view of recent cosmological data which support evidence for 
a flat and accelerating Universe. The acceleration is mainly driven by a 
non-vanishing cosmological constant which weighs
about 2/3 of the total
matter-energy density of the Universe. Such a large contribution of the 
cosmological constant (vacuum energy) pushes the matter density,
and consequently the {\small CDM} density, to relatively small values 
$\Omega_{\mathrm{CDM}}\;h_0^2 \simeq 0.15 \pm 0.07$, 
constraining the theoretical predictions of supersymmetric extensions of
the {\small SM} model.

Supersymmetric theories, with $R$-parity conservation, offer 
a comprehensive theoretical framework
which provide us with
a good candidate for the Dark Matter particle, the {\small LSP}, which
turns out to be the lightest of the neutralinos. 
The bound $\relic \simeq 0.15 \pm 0.07$ shows preference towards low values
for the effective supersymmetry breaking scale $M_{\mathrm{SUSY}}$, which in
conjunction with electroweak precision measurements, pointing to the
opposite direction favoring rather large
values for $M_{\mathrm{SUSY}}$, put severe constraints affecting
supersymmetric predictions.

We have undertaken the calculation of the relic density in the context
of the {\small CMSSM}, with radiatively 
induced breaking of the electroweak symmetry
and universal boundary conditions for the soft supersymmetry breaking
parameters
in which the {\small LSP} plays the role of the Dark Matter particle.

Our analysis have revealed the following:

Although the cosmological data do not rule out
corridors in the $(m_0, M_{1/2})$ plane in which the {\small LSP}
is light, with substantial Higgsino mixing, with no bound put 
on sfermion masses, nevertheless such regions 
be excluded in view of the latest experimental data from chargino
searches. 

Towards the large $M_{1/2}$ regime we have found that
in the cosmologically interesting domain, $M_{1/2}$ cannot exceed
$\approx 340 \GeV$, while at the
same time $m_0  \lesssim 200 \GeV$. These bounds are obtained provided one
stays within the region  $1.25\;\mlsp \leq m_{{\tilde{\tau}}_R}$ where
coannihilation
processes do not play any significant role. Putting a lower bound on
$M_{1/2}$ suggested by {\small EW} precision 
data can have a dramatic effect on the
allowed $m_0, M_{1/2}$ values. If for instance, based on phenomenological
studies of the electroweak mixing angle, we impose $M_{1/2} \geq 300 \GeV$
then $m_0, M_{1/2}$ are restricted to lie within the tight limits 
$M_{1/2} \simeq 300 \GeV - 340 \GeV$, $m_0 \simeq 80 \GeV - 130 \GeV$.
These limits are insensitive to the choice of the parameter $A_0$
and hold as long as $\tan\beta < 30$.
If, as other analyses suggest, the more restrictive cosmological data
are imposed,
$\relic=0.12 \pm 0.04$, then 
there are no allowed points in the region 
$m_{{\tilde{\tau}}_R}>1.25 \, m_{\lsp}$ for $M_{1/2}>300\GeV$.

Within {\small CMSSM} there are two ways to reconcile the 
experimental information from {\small EW} and cosmological data
with values of $m_0$ and $M_{1/2}$ that
lie outside the strict bounds quoted above. We have either to go to the large 
$\tan\beta$ (with $\mu>0$) regime, while staying within
$1.25\;\mlsp \leq m_{{\tilde{\tau}}_R}$, or move inside the narrow band
$m_{\lsp}<m_{{\tilde{\tau}}_R} \lesssim 1.25  m_{\lsp} $ in which case 
$\tilde{\tau}_R$, the next to {\small LSP} sparticle, is almost degenerate in mass 
with the {\small LSP} and 
$ {{\tilde{\tau}}_R} - {\lsp}$ coannihilation processes 
are relevant to keep neutralinos
in equilibrium.

In the first case  the pseudoscalar Higgs boson $A$ plays an essential role.
Depending on the inputs its mass may 
approach $2\,\mlsp$ while the $A\tau\bar{\tau}$, $Ab\bar{b}$ 
couplings are large as being proportional to $\tan\beta$.
Both effects make the pseudoscalar Higgs exchange dominate the
reactions $\lsp \lsp \goes \tau \bar{\tau}, b \bar{b}$, and enhance
the corresponding cross sections, resulting to relic densities which are 
compatible with the cosmological data.
It is worth pointing out that large $\tan\beta$ values, for $\mu>0$, are
compatible with the {\small CLEO} data for the 
process $b \goes s \gamma$. In addition  
since large values of $\tan \beta $ are compatible with Yukawa
coupling unification, this mechanism may offer the possibility of obtaining
cosmologically acceptable $\lsp$ relic densities in the coannihilation free
region $1.25\;\mlsp \leq m_{{\tilde{\tau}}_R}$, in such unification schemes.  

The second possibility to make the recent astrophysical data
compatible with values of $m_0, M_{1/2}$ outside the narrow domain quoted 
above, is to move to the coannihilation region
$\mlsp<m_{{\tilde{\tau}}_R}<1.25\;\mlsp$.
In this region the ${{\tilde {\tau}}_R}-\lsp$ coannihilation effects 
enhance $\xsec$, lowering significantly 
the values of the neutralino relic
density. By using the findings of Ref.~\cite{Falk1} we have found a handy
way to relate the actual relic density $\Omega_{\lsp}$ to that calculated
using the traditional way, $\Omega_{\lsp}^{0}$, in which coannihilation
reactions are not counted for.
We find that in the
region of the parameter space in which the {\small LSP} is nearly degenerate with
the next to the {\small LSP} particle, namely ${{\tilde {\tau}}_R}$, 
no upper limit is imposed on the parameter $M_{1/2}$. Given a value for
$M_{1/2}$ the parameter $m_0$ is however constrained to lie within a narrow
band which is dictated by $\mlsp<m_{{\tilde{\tau}}_R}<1.25\;\mlsp$. 

\vspace{.5cm}
\acknowledgments 
A.B.L. acknowledges support from ERBFMRXCT--960090 TMR programme
and D.V.N. by D.O.E. grant DE-FG03-95-ER-40917.  
V.C.S. acknowledges an enlightening discussion with D.~Schwarz.

%%%%%%%%%%%%%%%%%%%%%%%%%%%%%%%%%%%%
%%%%%%%%%%%% TABLE %%%%%%%%%%%%%%%%%%%%%%%%
\begin{table}
\caption{Sample of values for the ratio
${q^{\prime}_0} /{2 \;q^2_0} $ for an {\small LSP} mass equal to $100\GeV$. 
The masses
of the remaining particles are as described in the main text. }
\end{table}

\begin{center}
\begin{tabular}{|c|c|}
\hline\hline
 $\hspace{.3cm} x \hspace{.3cm}$ & ${q^{\prime}_0} /{2 \;q^2_0} $
 \\ \hline\hline
$\hspace{.3cm} .08 \hspace{.3cm} $ &  $2.47 \times 10^8 $  \\ 
$\hspace{.3cm} .07 \hspace{.3cm} $ &  $1.62 \times 10^9 $  \\ 
$\hspace{.3cm} .06 \hspace{.3cm} $ &  $1.95 \times 10^{10} $ \\ 
$\hspace{.3cm} .05 \hspace{.3cm} $ &  $0.62\times 10^{12} $\\ 
$\hspace{.3cm} .04 \hspace{.3cm} $ &  $1.05 \times 10^{14} $\\ 
$\hspace{.3cm} .03 \hspace{.3cm} $ &  $0.51 \times 10^{18} $\\
\hline\hline
\end{tabular}
\end{center}
%%%%%%%%%%%%%%%%%%%%%%%%%%%%%%%%%%%%%%%

\newpage
{\noindent\bf Appendix: Supersymmetric conventions}
\vspace{.7cm}
\setcounter{equation}{0}
\renewcommand{\theequation}{A.\arabic{equation}}

The supersymmetric Lagrangian we are using in this paper has
a superpotential given by
\bea
{\cal W} \;=\; h_t \;H_2^T \; \epsilon\; Q\;U^c \;+\;
               h_b \;H_1^T \; \epsilon\; Q\;D^c \;+\;
               h_\tau \;H_2^T \; \epsilon\; L\;E^c \;+\;
               \mu \; H_2^T \; \epsilon\; H_1
\eea
where the elements of the antisymmetric $2 \times 2$ matrix $\epsilon$ are
given by $\; \epsilon_{12}\;=\;-\epsilon_{12}\;=\;1\;$. In the
superpotential above we have only shown the dominant Yukawa terms of the
third generation.

The scalar soft part of the Lagrangian is given by
\bea
{\cal L}_{scalar} \;=\; &-&\;{\sum_i}\;m_i^2 \; |{\phi_i}|^2\; \nonumber \\ 
&-&\;(A_t\;h_t \;H_2^T \; \epsilon\; Q\;U^c \;+\;
    A_b\;h_b \;H_1^T \; \epsilon\; Q\;D^c \;+\;
    A_\tau\;h_\tau \;H_2^T \; \epsilon\; L\;E^c \;+\;h.c\;) \nonumber \\
    &+&\;(\; m_3^2\;H_2^T \; \epsilon\; H_1\;+\;h.c\;)  \; ,
\eea
where the index $\;i\;$ in the sum in the equation above runs over all
scalar fields and all fields appearing denote scalar parts of the
supermultiplets involved.

The gaugino fields soft mass terms are given by
\bea
{\cal L}_{gaugino}\;=\; -{\frac{1}{2}}\;(\;M_1\;{\tilde B}\;{\tilde B}\;+\;
M_2\;{\tilde W}^{(i)}\;{\tilde W}^{(i)}\;+\;M_3\;{\tilde G}\;{\tilde G}\;
+\;h.c.) .
\eea
In this equation $\;{\tilde B},\;{\tilde W}^{(i)},\;{\tilde G}\;$ are the
gauge fermions corresponding to the $\;U(1),\;SU(2),\;$ and $\;SU(3)\;$ gauge
groups.

For comparison with other notations \cite{Drees,Haber}
it is perhaps useful to remark that covariant derivatives in this paper
are defined by
$$
D_{\mu}\;=\;\partial_{\mu} \;-\;i\;g\;T^{(k)}\;A^{(k)}_{\mu} \; .
$$
Thus there is a sign difference in the gauge couplings used in this
paper and in Refs.~\cite{Drees,Haber}.
Besides that, the gaugino fields we use through differ in sign from those
used in those papers and the parameters  
$\;M_i\;$ and $\;\mu\;$ are opposite in sign too.
These remarks set the rules of passing from one notation to the other.

In the $\tilde{B}$, $\tilde{W}^{(3)}$, 
$i \tilde{H}_{1}^0$, $i \tilde{H}_{2}^0$, basis 
the neutralino mass matrix is
%%%%%%%%%%%%%%%%%%%
%\begin{equation}
%{\cal M}_N \ =\ \left(\begin{array}{cccc} M_1 & 0 &
%M_Z \sg cos\beta &
%-M_Z \sg sin\beta
%\\[1mm] 0 & M_2 & -M_Z \cg cos\beta &
%M_Z \cg sin\beta
%\\[1mm] M_Z \sg cos\beta&
%-M_Z \cg cos\beta  & 0 & -\mu \\[1mm]
%-M_Z \sg sin\beta
%&M_Z \cg sin\beta & -\mu & 0
%\end{array} \right)\ .\label{mchi0}
%\end{equation}
%%%%%%%%%%%%%%
%%%%%%%%%%%%%%%%%%%
\begin{equation}
{\cal M}_N \ =\ \left(\begin{array}{cccc} M_1 & 0 &
{g^\prime v_1} /{\sqrt {2}}  &
-{g^\prime  v_2}/{\sqrt {2}}
\\[1mm] 0 & M_2 & -{g v_1}/{\sqrt {2}} &
{g v_2}/{\sqrt {2}}
\\[1mm] {g^\prime  v_1}/{\sqrt {2}}&
-{g v_1}/{\sqrt {2}} & 0 & -\mu \\[1mm]
-{g^\prime  v_2}/{\sqrt {2}}
&{g v_2}/{\sqrt {2}} & -\mu & 0
\end{array} \right)\ .\label{mchi0}
\end{equation}
%%%%%%%%%%%%%%

In this expression
%%$\;M_Z\;$ is the tree level Z - boson mass,
%%$\; \sg  \;$ is the tree level weak mixing angle while
the tangent of the
angle $\; \beta \;$ sets the ratio of the v.e.v's of the two Higgses
$\; \tan \beta \;=\; \vev{H_2}/\vev{H_1} \;$.

The mass eigenstates (${\tilde \chi}_{1,2,3,4}^0$)
of neutralino mass matrix ${\cal M}_N$ are
written as 
%%%%%%%%%%%%%%%%%%%
\begin{equation}
{\cal O} \,\,  \left ( \begin{array}{c} {\tilde \chi}_1^0 \\
{\tilde \chi}_2^0 \\ {\tilde \chi}_3^0 \\ {\tilde \chi}_4^0
\end{array} \right )
\ =\ \left ( \begin{array}{c} \tilde{B} \\ \tilde{W}^{(3)} \\
i \tilde{H}_{1}^0 \\ i \tilde{H}_{2}^0
\end{array} \right )\,.
\end{equation}
%%%%%%%%%%%%%%%%%%%
and 
%%%%%%%%%%%%
\begin{equation}
{\cal O}^T {\cal M}_N {\cal O}\ ={\rm Diag} \left (
m_{{\tilde \chi}^0_1},m_{{\tilde \chi}^0_2}, m_{{\tilde \chi}^0_3},
m_{{\tilde \chi}^0_4} \right ) \,,
\end{equation}
%%%%%%%%%%%%
where ${\cal O}$ is a real orthogonal matrix. Note that when 
electroweak breaking effects are ignored ${\cal O}$ can get
the form
%%%%%%%%%%%%%%%%%%%
\begin{equation}
{\cal O} \ =\ \left ( \begin{array}{cc} {\bf 1_2} & {\bf 0_2} \\[1mm]
{\bf 0_2} & \begin{array}{cc} \frac{1}{\sqrt{2}} & \frac{1}{\sqrt{2}} \\
-\frac{1}{\sqrt{2}} & \frac{1}{\sqrt{2}} \end{array} \end{array} 
\right ) \,.
\end{equation}
%%%%%%%%%%%%%%%%%

The chargino mass matrix can be obtained from the following
Lagrangian mass terms
%%%%%%%%%%%%
\begin{equation}
{\cal L}^{mass}_{charginos} \ =\ 
- \left ( \tilde{W}^- , {i \tilde{H}_{1}^-} \right ) {\cal M}_c 
\left (\begin{array}{c} \tilde{W}^+ \\ {i \tilde{H}_{2}^+} \end{array}
\right ) \, + \, (h.c) \,,
\end{equation}
%%%%%%%%
where we have defined $\tilde{W}^\pm \equiv \frac{ \tilde{W}^{(1)}
\mp i \tilde{W}^{(2)} }{\sqrt{2}}$ and 
%%%%%%%%%
%\begin{equation}
%{\cal M}_C \ =\ \left (\begin{array}{cc} M_2 &
% -{\sqrt{2}}\; M_W \; sin\beta \\[1mm]
%- {\sqrt{2}}\; M_W \;  cos\beta & \mu \end{array}
%\right )\; .
%\label{chmat}
%\end{equation}
%%%%%%%%%%%%
%%%%%%%%%
\begin{equation}
{\cal M}_C \ =\ \left (\begin{array}{cc} M_2 &
 -\;g\;v_2 \\[1mm]
-\;g\;v_1 & \mu \end{array}
\right )\; .
\label{chmat}
\end{equation}
%%%%%%%%%%%%

Diagonalization of this matrix gives
%%%%%%%%
\begin{equation}
U {\cal M}_c V^\dagger \ =\ \left (\begin{array}{cc} m_{\tilde{\chi}_1}
& 0 \\[1mm]
0 & m_{\tilde{\chi}_2} \end{array} \right ) \; .
\end{equation}
%%%%%%%%%%%%%%%%%%%
Thus,
%%%%%%%%%%%%
\begin{equation}
{\cal L}^{mass}_{charginos} \ =\ -m_{\tilde{\chi}_1} 
\bar{\tilde{\chi}_1} \tilde{\chi}_1 -  m_{\tilde{\chi}_2}
\bar{\tilde{\chi}_2} \tilde{\chi}_2 \; .
\end{equation}
%%%%%%%%%%%%%%
The Dirac chargino states $\tilde{\chi}_{1,2}$ are given by
%%%%%%%%%%%%%%%
\begin{equation}
\tilde{\chi}_1 \equiv \left (\begin{array}{c} \lambda_1^+ \\
\bar{\lambda}_1^- \end{array} \right ) \,\, , \,\,
\tilde{\chi}_2 \equiv \left (\begin{array}{c} \lambda_2^+ \\
\bar{\lambda}_2^- \end{array} \right ) \,.
\end{equation}
%%%%%%%%%%%%%%%%%
The two component Weyl spinors $\lambda^\pm_{1,2}$ are related
to $\tilde{W}^\pm$, ${i \tilde{H}_{1}^-}$, ${i \tilde{H}_{2}^+}$ by
%%%%%%%%%%%%%%%
\begin{equation}
V \left ( \begin{array}{c} \tilde{W}^+ \\ {i \tilde{H}_{2}^+}
\end{array} \right ) \equiv \left (\begin{array}{c}
\lambda_1^+ \\ \lambda_2^+ \end{array} \right ) \,\, , \,\,
\left ( \tilde{W}^- \, , \, {i \tilde{H}_{1}^-} \right ) U^\dagger
 \equiv \left ( \lambda_1^- \, , \, \lambda_2^- \right ) \,.
\end{equation}
%%%%%%%%%%%%%%%%

The gauge interactions of charginos and neutralinos can
be read from the following Lagrangian\footnote{
In our notation ${e} \equiv $electron's charge.} 
%%%%%%%%%%%%%%%
\begin{equation}
{\cal L} \ =\ {g} \left ( W^+_\mu J_-^\mu + W_\mu^- J_+^\mu 
\right ) + {e} A_\mu J_{em}^\mu +\frac{{e}}{{s} {c}}
Z_\mu J_Z^\mu \,.
\end{equation}
%%%%%%%%%%%%%%%%%
In the equation above $\;s\;=\; \sin \theta_W,\;c\;=\; \cos \theta_W\;$.
Also,
%%%%%%%%%%%%%
\begin{equation}
\left (\begin{array}{c} Z_\mu \\[1mm] A_\mu \end{array} \right )
\ =\ \left ( \begin{array}{cc} {c} & {s} \\[1mm]
-{s} & {c} \end{array} \right ) \, 
\left ( \begin{array}{c} W_\mu^{(3)} \\[1mm] B_\mu \end{array}
\right ) \,.
\end{equation}
%%%%%%%%%%%%%%%%%
The currents $J_+^\mu$, $J_{em}^\mu$ and $J_Z^\mu$ are given by
%%%%%%%%%%%%%%%%%
\begin{equation}
J_+^\mu \equiv \bar {{\tilde \chi}^0_a}  \gamma^\mu \left [
{\cal P}_L {\cal P}^L_{a i}+ 
{\cal P}_R {\cal P}^R_{a i}  \right ]
\tilde{\chi}_i \;\; a=1...4,\;\; i=1,2 \,,
\end{equation}
%%%%%%%%%%%%%%
where
${\cal P}_{L,R} = \frac{1 \mp \gamma_5 }{2}$ and
%%%%%%%%%%%%%%%
\begin{eqnarray}
& &{\cal P}^L_{a i}\equiv +\frac{1}{\sqrt{2}} {\cal O}_{4 a} V^{*}_{i 2}
- {\cal O}_{2 a} V^{*}_{i 1}\,,\nonumber \\[2mm]
& &{\cal P}^R_{a i}\equiv -\frac{1}{\sqrt{2}} {\cal O}_{3 a} U^{*}_{i 2}
- {\cal O}_{2 a} U^{*}_{i 1} \,.
\end{eqnarray}
%%%%%%%%%%%%%% 
The electromagnetic current $J_{em}^\mu$ is
%%%%%%%%%%%%
\begin{equation}
J_{em}^\mu \ =\ \bar{\tilde\chi}_1 \gamma^\mu \tilde{\chi}_1 +
\bar{\tilde\chi}_2 \gamma^\mu \tilde{\chi}_2 \,.
\end{equation}
%%%%%%%%%%%%%%%%%%%
Finally, the neutral current $J_Z^\mu$ can be read from
%%%%%%%%%%%%%%%
\begin{equation}
J^\mu_Z \equiv \bar  {{\tilde \chi}_i}  \gamma^\mu \left [
{\cal P}_L {\cal A}^L_{i j} + {\cal P}_R {\cal A}^R_{i j} \right ]
\tilde{\chi}_j + \frac{1}{2} \bar {{\tilde \chi}^0_a} \gamma^\mu \left [
{\cal P}_L {\cal B}^L_{a b} + {\cal P}_R {\cal B}^R_{a b} \right ]
{{\tilde \chi}^0_b} \,,
\end{equation}
%%%%%%%%%%%%%%%%%%%%%
with
%%%%%%%%%%%%%%
\begin{eqnarray}
{\cal A}^L_{i j} &=& {{c}^2} \delta_{i j} -\frac{1}{2} V_{i 2} V^{*}_{j 2}
\,,\nonumber \\
{\cal A}^R_{i j} &=& {{c}^2} \delta_{i j} -\frac{1}{2} U_{i 2} U^{*}_{j 2}
\,,\nonumber \\
{\cal B}^L_{a b} &=& \frac{1}{2} \left ( {\cal O}_{3 a} {\cal O}_{3 b} -
{\cal O}_{4 a} {\cal O}_{4 b} \right )
\,,\nonumber \\
{\cal B}^R_{a b} &=& - {\cal B}^L_{a b} \,.
\end{eqnarray}
%%%%%%%%%%%%%%%%%%
Note that since ${\cal B}^R_{a b} = - {\cal B}^L_{a b}$ the neutralino
contribution to $J_Z^\mu$ can be cast into the form
%%%%%%%%%%%%%%%%%%%
\begin{equation}
J_Z^\mu \ =\ -\frac{1}{2} {\cal B}^L_{a b} \left ( \bar{{\tilde \chi}^0_a}
\gamma^\mu \gamma^5 {{\tilde \chi}^0_b} \right ) \,.
\end{equation}
%%%%%%%%%%%%%

For the calculation of the $\lsp\;\lsp \goes f \;{\bar f}$ cross
sections we need know the
chargino and neutralino couplings to fermions and sfermions.
The relevant chargino couplings are given by the following
Lagrangian terms
%%%%%%%%%%%%%
\begin{equation}
{\cal L} = i \; {\bar{\tilde {\chi}}}_{i}^{c} \;
 ({\cal P}_L \, a^{ {f^{\prime}}   {\tilde f} }_{ij} +
{\cal P}_R \, b^{ {f^{\prime}}   {\tilde f} }_{ij}) \,  {f^\prime } \,
 {\tilde f}_{j}^{*}
\, + \,
i \; {\bar{\tilde {\chi}}}_{i} \;
 ({\cal P}_L \, a^{f {\tilde f}^{\prime} }_{ij} +
{\cal P}_R \, b^{f {\tilde f}^{\prime} }_{ij}) \,  {f} \,
 {\tilde f}_{j}^{\prime *} \, + \, (h.c) \,.
\end{equation}
In this, ${\chi}_{i} \; (i=1,2)$ are the positively charged charginos  and 
${\chi}_{i}^{c}$ the corresponding charge conjugate states having 
opposite charge. $f\, , \,{f}^{\prime}$ are ``up" and ``down"
fermions, quarks or leptons, while ${\tilde f}_i\, , \,{\tilde f}_i^{\prime}$
are the corresponding sfermion mass eigenstates.
The left and right-handed couplings appearing above are given by
%%%%%%%%%%%%%%%%%%%%%%
\begin{eqnarray}
a^{{f^{\prime}}{\tilde f} }_{ij} \ori &= \ori
g V_{i1}^{*} \, K^{\tilde f}_{j1} - h_{f} V_{i2}^{*}  K^{\tilde f}_{j2} \ori ,
\ori  &b^{ {f^{\prime}} {\tilde f} }_{ij} \ori =
 \ori -h_{f ^\prime} \,  U_{i2}^{*} K^{\tilde f}_{j1}  \,,  \nonumber  \\
a^{f {\tilde f}^{\prime} }_{ij} \ori &= \ori
g U_{i1} \, K^{{\tilde f}^{\prime}}_{j1} + h_{f^\prime} \, U_{i2}
  K^{{\tilde f}^{\prime}}_{j2} \ori , \ori
&b^{f {\tilde f}^{\prime} }_{ij} \ori = \ori
h_{f } \,  V_{i2} K^{{\tilde f}^{\prime}}_{j1} \,.  \nonumber
\end{eqnarray}
%%%%%%%%%%%%%%%%%%%%%%
In the equation above $h_{f } \, , \, h_{f ^\prime} $ are the Yukawa
couplings of the up and down fermions respectively. The matrices
$K^{\tilde{f},{\tilde f}^\prime} $ which diagonalize the sfermion mass matrices become the
unit matrices in the absence of left-right sfermion mixings.
For the electron and muon family the lepton masses are taken to be
vanishing in the case that mixings do not occur. In addition the 
right-handed couplings, are zero.  \newline 
The corresponding neutralino couplings are given by
%%%%%%%%%%%%%%%%%%%%%%%%%%%%%%%%%%%%
\begin{equation}
{\cal L} = i \; {\bar  {{\tilde \chi}^0_a}} \;
 ({\cal P}_L \, a^{ {f}   {\tilde f} }_{aj} +
{\cal P}_R \, b^{ {f}   {\tilde f} }_{aj}) \,  {f} \,
 {\tilde f}_{j}^{*}
\, + \,
i \; {\bar{{\tilde \chi}^0_a}} \;
 ({\cal P}_L \, a^{{f^{\prime}} {\tilde f}^{\prime} }_{aj} +
{\cal P}_R \, b^{{f^{\prime}} {\tilde f}^{\prime} }_{aj}) \, {f^{\prime}} \,
 {\tilde f}_{j}^{\prime *} \, + \, (h.c) \,.
\end{equation}
%%%%%%%%%%%%%%%%%%%%%%%
The left and right-handed couplings for the up fermions, sfermions
are given by
%%%%%%%%%%%%%%%%%%%%%%
\begin{eqnarray}
a^{{f}{\tilde f}}_{aj} \ori &= &\ori
{\sqrt{2}} \, ( g {T^{3}_f}{O_{2a}}+{g^{\prime}}{\frac{Y_f}{2}} \,{O_{1a}})
\, {K^{f}_{j1}} \ori + \ori h_{f} \, {O_{4a}}\,  {K^{f}_{j2}}  \ori , \ori
\nonumber \\
b^{{f}{\tilde f} }_{aj} \ori & = & \ori
{\sqrt{2}} \, (-{g^{\prime}}{\frac{Y_{f^c}}{2}} \,{O_{1a}}) \, {K^{f}_{j2}}
\ori - \ori h_{f} \, {O_{4a}}\,  {K^{f}_{j1}} \,,  \nonumber  
\end{eqnarray}
%%%%%%%%%%%%%%%%%%%%%%
while those for the down fermions and sfermions are given by
%%%%%%%%%%%%%%%%%%%%%%%%%%%%%
\begin{eqnarray}
a^{{\fp}{\tilde \fp}}_{aj} \ori &= &\ori
{\sqrt{2}} \,
( g {T^{3}_{\fp}}{O_{2a}}+{g^{\prime}}{\frac{Y_{\fp}}{2}} \,{O_{1a}})
\, {K^{\fp}_{j1}} \ori - \ori h_{\fp} \, {O_{3a}}\,  {K^{\fp}_{j2}}  \ori , \ori
\nonumber \\
b^{{\fp}{\tilde \fp} }_{aj} \ori & = & \ori
{\sqrt{2}} \, (-{g^{\prime}} \frac{Y_{ {\fp}^c }}{2} \,{O_{1a}})
 \, {K^{\fp}_{j2}}
\ori + \ori h_{\fp} \, {O_{3a}}\,  {K^{\fp}_{j1}}\,.   \nonumber  
\end{eqnarray}

%%%%%%%%%bottom of article

%%%%%%%%%%%%%%%%%% FIGURES %%%%%%%%%%%%%%%%%%%%%%%%%%
\newpage
%%%%%%%%%%%%%%%%%%%%%%%%%% Figure 1 %%%%%%%%%%%%%%%%%%%%%%%%%%%%%%%% 

\newpage
%%%%%%%%%%%%%%%%%%%%%%%%%% Figure 2 %%%%%%%%%%%%%%%%%%%%%%%%%%%%%%%%  
%%%%%%%%%%%%%%%%%%%%%%%%%%%%%%%%%%%%%%%%%%%%

\newpage
%%%%%%%%%%%%%%%%%%%%%%%%%% Figure 3 %%%%%%%%%%%%%%%%%%%%%%%%%%%%%%%% 

%%%%%%%%%%%%%%%%%%%%%%%%%%%%%%%%%%%%%%%%%%

\newpage
%%%%%%%%%%%%%%%%%%%%%%%%%% Figure 4 %%%%%%%%%%%%%%%%%%%%%%%%%%%%%%%% 
%%%%%%%%%%%%%%%%%%%%%%%%%%%%%%%%%

\newpage
%%%%%%%%%%%%%%%%%%%%%%%%%% Figure 5 %%%%%%%%%%%%%%%%%%%%%%%%%%%%%%%% 
%%%%%%%%%%%%%%%%%%%%%%%%%%%%%%%%%

\newpage
%%%%%%%%%%%%%%%%%%%%%%%%%% Figure 6 %%%%%%%%%%%%%%%%%%%%%%%%%%%%%%%% 
%%%%%%%%%%%%%%%%%%%%%%%%%%%%%%%%%%%%%%%%%%%%%%%%%%%%%%%%%%%%

\newpage
%%%%%%%%%%%%%%%%%%%%%%%%%% Figure 7 %%%%%%%%%%%%%%%%%%%%%%%%%%%%%%%% 
%%%%%%%%%%%%%%%%%%%%%%%%%%%%%%%%%%%%%%%%%%%%%%%%%%%%%%%%%%%%

\newpage
%%%%%%%%%%%%%%%%%%%%%%%%%% Figure 8 %%%%%%%%%%%%%%%%%%%%%%%%%%%%%%%% 
%%%%%%%%%%%%%%%%%%%%%%%%%%%%%%%%%%%%%%%%%%%%%%%%%%%%%%%%%%%%

\newpage
%%%%%%%%%%%%%%%%%%%%%%%%%% Figure 9 %%%%%%%%%%%%%%%%%%%%%%%%%%%%%%%% 
%%%%%%%%%%%%%%%%%%%%%%%%%%%%%%%%%%%%%%%%%%%%%%%%%%%%%%%%%%%%

\newpage
%%%%%%%%%%%%%%%%%%%%%%%%%% Figure 10 %%%%%%%%%%%%%%%%%%%%%%%%%%%%%%%% 

%%%%%%%%%%%%%%%%%%%%%%%%%% Figure 11 %%%%%%%%%%%%%%%%%%%%%%%%%%%%%%%% 

\end{document}